\documentclass[twocolumn, twocolappendix]{aastex63}
\usepackage{hyperref}
\usepackage{amsmath,amstext}
\usepackage[T1]{fontenc}
\usepackage{graphicx}
\usepackage[figure,figure*]{hypcap}
\usepackage{multirow}
\usepackage{comment}
\usepackage{nicefrac}

\usepackage[]{algorithm2e} 

\def\SPARK{\texttt{SPARK}}
\def\TARDIS{\texttt{TARDIS}}
\def\AUTOSTRUCTURE{\texttt{AUTOSTRUCTURE}}
\def\approxposterior{\texttt{approxposterior}}

\usepackage{soul}

\def\sun{$_{\odot}$}

\def\f28{${f}_{2-8{\rm keV}}$}

\def\sun{$_{\odot}$}

\def\ergscm2{erg s$^{-1}$ cm$^{-2}$}
\def\yr-1{yr$^{-1}$}

\def\eg{{\it e.g.}}

\def\ie{{\it i.e.}}

\def\a{$\&$}
\def\x{$\times$}

\def\w{$\omega$}

\def\asec{\ifmmode^{\prime\prime}\else$^{\prime\prime}$\fi}

\hypersetup{linkcolor=magenta}

\received{(ApJ) September 14, 2022}
\accepted{on December 22, 2022}

\shorttitle{inferring $r$-process abundances of the GW170817 kilonova} 
\shortauthors{Vieira {\it et al.}}

\begin{document}

\title{Spectroscopic $r$-Process Abundance Retrieval for Kilonovae I: \\ The Inferred Abundance Pattern of Early Emission from GW170817}

\correspondingauthor{Nicholas~Vieira}
\email{nicholas.vieira@mail.mcgill.ca}

\author[0000-0001-7815-7604]{Nicholas~Vieira}
\affil{McGill Space Institute and Department of Physics, McGill University, 3600 rue University, Montreal, Qu{\'e}bec, H3A 2T8, Canada}

\author[0000-0001-8665-5523]{John~J.~Ruan}
\affil{Department of Physics and Astronomy, Bishop's University, 2600 rue College, Sherbrooke, Qu{\'e}bec, J1M 1Z7, Canada}

\author[0000-0001-6803-2138]{Daryl Haggard}
\affil{McGill Space Institute and Department of Physics, McGill University, 3600 rue University, Montreal, Qu{\'e}bec, H3A 2T8, Canada}

\author[0000-0001-8921-3624]{Nicole Ford}
\affil{McGill Space Institute and Department of Physics, McGill University, 3600 rue University, Montreal, Qu{\'e}bec, H3A 2T8, Canada}

\author[0000-0001-7081-0082]{Maria~R.~Drout}
\affil{Department of Astronomy and Astrophysics, University of Toronto, 50 St. George St., Toronto, Ontario, M5S 3H4, Canada}

\author[0000-0003-4619-339X]{Rodrigo Fern{\'a}ndez}
\affil{Department of Physics, University of Alberta, Edmonton, Alberta, T6G 2E1, Canada}

\author{N.~R.~Badnell}
\affil{Department of Physics, University of Strathclyde, Glasgow, G4 0NG, UK}

\begin{abstract}
Freshly-synthesized $r$-process elements in kilonovae ejecta imprint absorption features on optical spectra, as observed in the GW170817 binary neutron star merger. These spectral features encode insights into the physical conditions of the $r$-process and the origins of the ejected material, but associating features with particular elements and inferring the resultant abundance pattern is computationally challenging. We introduce Spectroscopic $r$-Process Abundance Retrieval for Kilonovae (\SPARK), a modular framework to perform Bayesian inference on kilonova spectra with the goals of inferring elemental abundance patterns and identifying absorption features at early times. \SPARK~inputs an atomic line list and abundance patterns from reaction network calculations into the \TARDIS~radiative transfer code. It then performs fast Bayesian inference on observed kilonova spectra by training a Gaussian process surrogate for the approximate posteriors of kilonova ejecta parameters, via active learning. We use the spectrum of GW170817 at 1.4 days to perform the first inference on a kilonova spectrum, and recover a complete abundance pattern. Our inference shows that this ejecta was generated by an $r$-process with either (1) high electron fraction $Y_e \sim 0.35$ and high entropy $s / k_{\mathrm{B}} \sim 25$, or, (2) a more moderate $Y_e \sim 0.30$ and $s / k_{\mathrm{B}} \sim 14$. These parameters are consistent with a shocked, polar dynamical component, and a viscously-driven outflow from a remnant accretion disk, respectively. We also recover previous identifications of strontium absorption at $\sim$8000~\AA, and tentatively identify yttrium and/or zirconium at $\lesssim$4500~\AA. Our approach will enable computationally-tractable inference on the spectra of future kilonovae discovered through multi-messenger observations.
\end{abstract}
\keywords{nuclear abundances, r-process, radiative transfer simulations, spectral line identification}


\section{Introduction}\label{sec:intro}

More than half a century ago, rapid neutron capture in astrophysical settings was identified as a likely source for the heaviest elements (\citealt{burbidge57, cameron57}). This so called `$r$-process' nucleosynthesis is now thought to contribute significantly to cosmic abundances of almost half of all known elements (see \citealt{cowan21} for a recent review). Mergers of two neutron stars (NS-NS) or a neutron star and black hole (NS-BH) have long been suspected as a site of this \textit{r}-process (\citealt{lattimer74, symbalisty82, eichler89, freiburghaus99, goriely11, korobkin12, bauswein13}). However, a major outstanding question is whether these mergers alone can robustly produce the observed abundances of \textit{r}-process elements. Recently, the measurement of large $r$-enhancement in the ultra-faint dwarf galaxy \ion{Reticulum}{2} was ascribed to a single high-yield event such as a NS-NS or NS-BH merger (\citealt{ji16, roederer16}), although the scenario of one or more core-collapse supernovae cannot be ruled out (\citealt{beniamini16}). Similarly, in the Solar System, very low meteoritic and deep-sea abundances of the actinides ($Z = 89-103$, but specifically ${}^{244}$Pu and ${}^{247}$Cm, \citealt{wallner15, bartos19}) argue for a single, rare actinide production site, consistent with mergers. Yet, isotopic ratios of ${}^{247}$Cm / ${}^{129}$I in meteorites are also consistent with both merger accretion disks and magnetorotationally-driven supernovae (\citealt{cote21}). Adding to the confusion, so-called `actinide-boost' stars can be reproduced by the dynamical ejecta from mergers (\citealt{lai08, roederer09, holmbeck18}), but their actinide-poor counterparts (\citealt{holmbeck19a}) imply either intrinsic variation in actinide yield from the same class of sources (as might be replicated in post-merger ejecta, \citealt{holmbeck19b}) or additional actinide production sites. Finally, mergers have difficulty producing observed trends in [Eu/Fe] abundance ratios in the disk of the Milky Way (\citealt{cote19}), while collapsars at sufficiently high accretion rates (\citealt{siegel19}) may produce these trends (\citealt{brauer20, yamazaki21}, but see \citealt{wanajo21}). Identifying the dominant site(s) of the $r$-process in the Universe will thus require further, more constraining observations.

One avenue for tackling this question is the direct measurement of the $r$-process abundances at the different candidate sites. The landmark NS-NS merger GW170817, detected first in gravitational waves (GWs) by the LIGO and Virgo observatories and then across the electromagnetic spectrum (\citealt{abbottLIGO17a, abbottLIGO17b}), offered the first opportunity to directly test these mergers' ability to synthesize the \textit{r}-process elements. Observations in the ultraviolet (UV), optical, and infra-red (IR), including both photometry (\citealt{andreoni17, arcavi17, coulter17, diaz17, drout17, evans17, hu17, kasliwal17, lipunov17, tanvir17, troja17, utsumi17, valenti17})\footnote{See \cite{villar17} for a compilation of this photometry considering inter-instrument variation.} and spectroscopy (\citealt{chornock17, kasen17, pian17, shappee17, smartt17, cote18}), broadly matched expectations for a transient kilonova powered by the radioactive decay of freshly-synthesized \textit{r}-process elements (see \citealt{metzger19, margutti21} for reviews). It was thus confirmed that NS-NS mergers must be the source of \textit{some} fraction of the \textit{r}-process elements in the Universe. However, whether the abundances of the heaviest $r$-process elements (\eg, the lanthanides $Z = 57 - 71$, and beyond) match those of the universal \textit{r}-process remains unclear. Indeed, \cite{ji19} reviewed the lanthanide mass fraction $X_{\mathrm{lan}}$ inferred for this event across the literature, and broadly found that the lanthanides were under-produced relative to the \textit{r}-process abundances measured in Galactic halo stars. They concluded that some fraction of future kilonovae would need to be significantly more lanthanide-rich for NS-NS mergers to remain a viable candidate for the dominant source of \textit{r}-process elements in the Milky Way.

To quantify the $r$-process yield of GW170817, various studies have attempted to associate features in the spectrum of the kilonova with specific elements through forward-modeling. This process is complicated by both the presence of up to billions of lines in the spectrum (\citealt{tanaka20}), and the broadening of these lines due to the high velocities of the kilonova ejecta. Nonetheless, a broad absorption feature (relative to a thermal continuum) at $\sim$8000~\text{\AA} has received much attention. \cite{smartt17} attributed this feature to a combination of absorption at $7000-7500$~\text{\AA} and $8000-8500$~\text{\AA} from \ion{Cs}{1} and \ion{Te}{1}, respectively. \cite{watson19} argued instead that \ion{Sr}{2} was responsible for the absorption at both $\sim$8000~\text{\AA} and potentially $\sim$4000~\text{\AA}. They noted that at the temperatures required for the presence of sufficient \ion{Cs}{1} and \ion{Te}{1}, singly-ionized species such as \ion{La}{2}, \ion{Gd}{2}, and \ion{Eu}{2} would also be highly populated. These lanthanide elements would themselves imprint absorption features on the spectra, but were not observed. Using updated atomic data, \cite{domoto21} similarly attributed the feature at $\sim$8000~\AA~to \ion{Sr}{2}. \cite{gillanders21} searched the spectrum for evidence of gold and platinum, and found no such evidence. Recently, using a combination of observed and theoretical atomic data, \cite{gillanders22} recovered the \ion{Sr}{2} feature and also found evidence for absorption from \ion{Y}{2} and \ion{Zr}{2} at wavelengths $\lesssim$5000~\text{\AA}. Similarly, \cite{domoto22} used a combination of observed and theoretical lines to argue for absorption from \ion{La}{3} and \ion{Ce}{3} at $\sim$ $12000 - 14000$~\AA, as well as recovering the aforementioned \ion{Sr}{2}, \ion{Y}{2}, and \ion{Zr}{2} features.

\begin{figure*}[!ht]
    \includegraphics[width=0.99\textwidth]{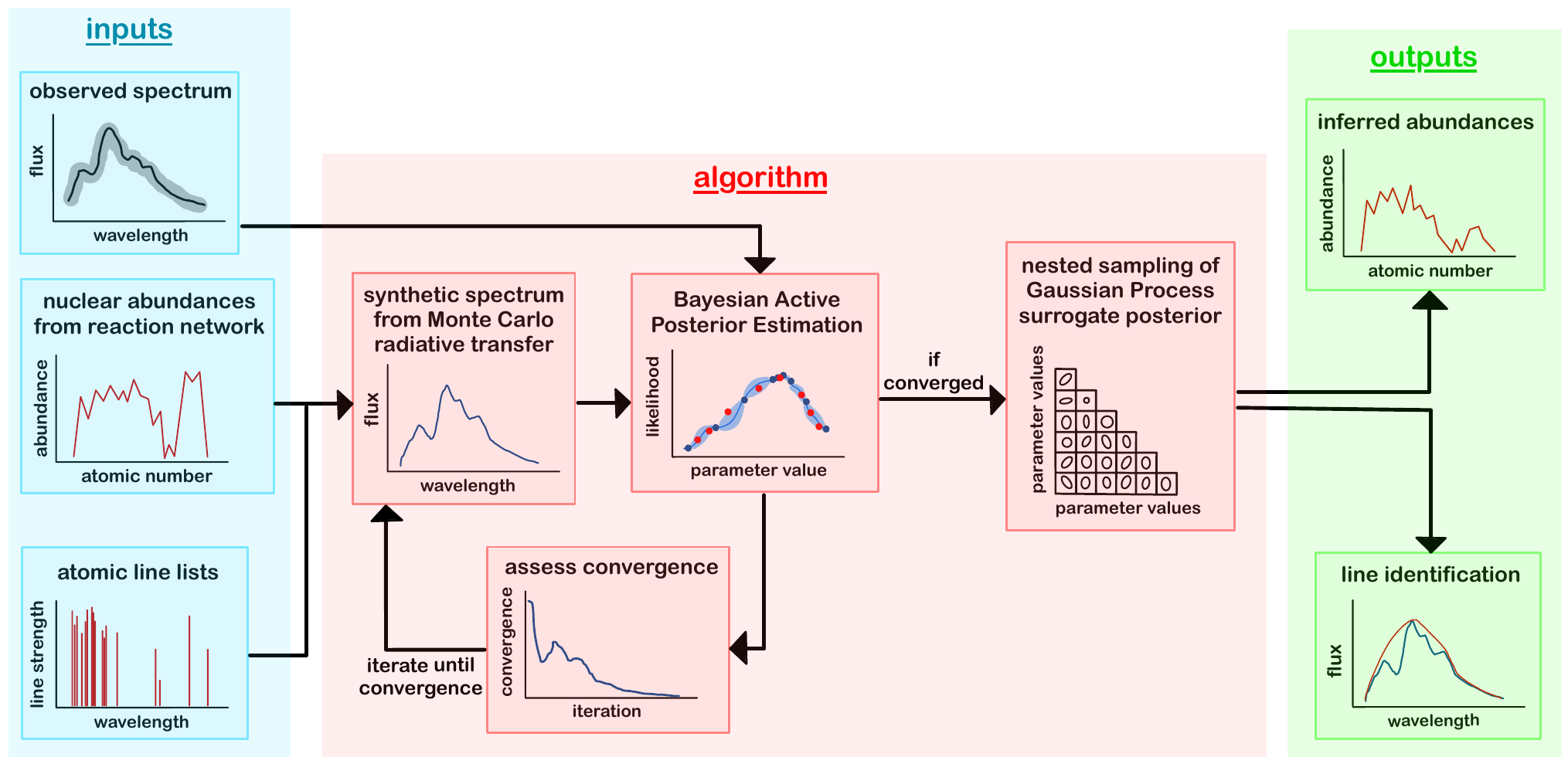}
    \figcaption{\textbf{Schematic of the steps in a \SPARK~run.} We take as inputs the observed spectrum of some kilonova, an atomic line list (Section \ref{ssc:linelist}), and the abundances from a nuclear reaction network (Section \ref{ssc:W18abunds}). The line list and abundances are used to generate synthetic spectra, via Monte Carlo radiative transfer with \TARDIS~(Section \ref{ssc:TARDIS}). We aim to construct a posterior distribution given a parametrization of the kilonova ejecta (Section \ref{ssc:params}) and find the best fit to the observed spectrum. Because this radiative transfer is computationally expensive, we use Bayesian Active Posterior Estimation (BAPE) to determine where in parameter-space to perform the radiative transfer (Section~\ref{ssc:inference}). In BAPE, a Gaussian process (GP) surrogate for the posterior actively learns the posterior as a training set of synthetic spectra is accumulated. As this training set grows, we periodically assess the convergence of the posterior. Once convergence is achieved, we produce our final approximate posterior using nested sampling on the GP surrogate. We extract the best fit and uncertainties for parameters of the kilonova, and use the relevant parameters to obtain a best-fit abundance pattern. Finally, we iteratively remove one element at a time from our best fit model, re-perform radiative transfer, and compare this new spectrum to the best fit to assess the impact of each individual element. In the end, we obtain the full abundance pattern and robustly associate features in the spectrum of the kilonova with specific species.}
    \label{fig:schema}
\end{figure*}

While the detection of these individual elements can be used to constrain their abundances, it has not been possible to systematically infer the full, element-by-element abundance pattern of the ejecta. First, due to the large number of heavy elements which may be co-produced in the ejecta and the large opacities of these elements under the conditions in the ejecta (as large as $\kappa \approx 10 - 100 ~\mathrm{cm^2~g^{-1}}$ for the lanthanides and actinides; \citealt{kasen13, kasen17, even20, fontes20, tanaka20, silva22}), spectral synthesis for comparison to the observed spectra is computationally expensive. Second, kilonova parameter-space is high-dimensional, and the dependence of the emergent spectrum on the elemental abundances is highly non-linear. The net effect is that a thorough exploration of abundance-space, and the impact of all of the elements on the spectrum, has not yet been possible.

Here, we solve this problem and perform the first inference on the spectrum of the GW170817 kilonova in the early-time optically-thick phase, to retrieve posteriors for parameters of the ejecta material. We introduce Spectroscopic $r$-Process Abundance Retrieval for Kilonovae (\SPARK), a framework for performing Bayesian inference of key kilonova parameters using optical spectra with the dual goals of (1) inferring the entire $r$-process elemental abundance pattern, and (2) robustly associating features in the observed spectra with specific $r$-process elements. Using a simple parametrization of the kilonova ejecta, we input a line list constructed from the Vienna Atomic Line Database (VALD; \citealt{ryabchikova15, pakhomov19}) and abundance patterns from the nuclear reaction network calculations of \cite{wanajo18} into the \TARDIS~(\citealt{kerzendorf14}) radiative transfer code. We then couple \TARDIS~to the approximate Bayesian inference framework of \approxposterior~(\citealt{fleming18, fleming20}), which uses a Gaussian process surrogate model to approximate a posterior distribution through active learning when forward modelling is computationally expensive. This yields a modular inference engine which is capable of inferring the properties of the kilonova ejecta with few forward model evaluations.

The organization of this paper is as follows. In Section~\ref{sec:methods}, we describe our line list, radiative transfer, kilonova parametrization, abundances, and inference techniques. In Section~\ref{sec:results}, we apply \SPARK~to the spectrum of the GW170817 kilonova at $t = 1.4$~days post-merger and present approximate posterior distributions and inferred abundances for this early, optically thick component of the ejecta. We also associate features in the spectrum with particular species. In Sections~\ref{sec:disco}~and~\ref{sec:conclusions}, we discuss the implications of our results, including the physical origins of the ejecta, and conclude.


\section{Methods}\label{sec:methods}

We present a general schematic for \SPARK~in Figure~\ref{fig:schema}. We discuss each of the steps in detail in the following sections.

\subsection{Building a Line List}\label{ssc:linelist}

In problems of spectroscopic inference, the accuracy of the inference is highly dependent on the quality of the employed atomic data. These atomic data or `line lists' include, at minimum: the energy levels, transition wavelengths, and the strengths of the transitions for all ions of relevance. Aside from these line lists, we also require the masses and ionization potentials of different elements. This poses a challenge for performing inference on kilonovae because a wide range of heavy elements, up to and including the actinides, are expected to be present. For many of these elements, and in particular ionized species, observed atomic data is limited or non-existent. Theoretical atomic data can be used, but may not be sufficiently accurate in wavelength (\eg, \citealt{tanaka20}). We therefore use the most complete source of atomic data currently available: the Vienna Atomic Line Database (VALD; \citealt{ryabchikova15, pakhomov19})\footnote{{\href{http://vald.astro.uu.se/~vald/php/vald.php}{http://vald.astro.uu.se/~vald/php/vald.php}}}. For elements from $Z = 1-92$ (with some missing), neutral to doubly ionized, VALD provides approximately 1.6  million \textit{observed} lines. We do not include any theoretical or semi-empirical lines at this time. Because of their relevance to kilonovae, we also use a small number of astrophysically measured lines from the APOGEE survey (\citealt{majewski17}). These include 10 \ion{Nd}{2} lines (\citealt{hasselquist16}) and 9 \ion{Ce}{2} lines (\citealt{cunha17}), in the $\sim$ $15,000 - 17,000$~\text{\AA} range. We convert these line lists to a format compatible with our radiative transfer using the \texttt{carsus}\footnote{\href{https://github.com/tardis-sn/carsus}{https://github.com/tardis-sn/carsus}} package. The final tally of lines for each ion is provided in Appendix~\ref{app:linelist}.

\subsection{Radiative Transfer with \textsc{TARDIS}}\label{ssc:TARDIS}

To generate synthetic spectra, we use the Monte Carlo radiative transfer code \TARDIS\footnote{\href{https://tardis-sn.github.io/tardis/}{https://tardis-sn.github.io/tardis/}}~(Temperature And Radiative Diffusion In Supernovae; \citealt{kerzendorf14}). \TARDIS~employs the indivisible photon packet Monte Carlo scheme formalized in \cite{mazzali93, lucy99, lucy02, lucy03} to produce a synthetic spectrum at a single point in time.\footnote{See \cite{noebauer19} for a review of Monte Carlo radiative transfer techniques.} In this scheme, photon packets are generated at a user-specified inner computational boundary (the photosphere\footnote{{Note, however, that this is a simplification: the photosphere is fundamentally a wavelength-dependent quantity (\eg, \citealt{fontes20}).}) and allowed to propagate until they either escape through the user-set outer boundary or are re-absorbed and lost at the inner boundary. As they propagate, packets may undergo bound-bound line interactions in the ambient ejecta and/or electron scattering. Over several iterations, the parameters of the radiation field are updated until a steady plasma state is achieved. During the run, interaction histories of all packets are tracked. At each interaction, $N_{\mathrm{v}} = 5$ `virtual' packets are generated, escape the ejecta without further interaction, and are recorded. This virtual packet technique (\citealt{long02}) yields a synthetic virtual packet spectrum in which the impact of Monte Carlo noise is reduced relative to a `real' packet spectrum.} For our purposes, we further smooth the spectrum with a Savitzky-Golay filter (\citealt{savitzky64}), as in \cite{vogl20}.

For details of the different modes available within \TARDIS, we refer the reader to \cite{kerzendorf14} and its extensive documentation. Briefly, we use the \texttt{macroatom} (\citealt{lucy02}) scheme to handle radiation-matter interactions. Note that \texttt{macroatom} does not consider auto-ionizing lines, and indeed all bound-free processes are neglected. Bremsstrahlung (free-free) and synchrotron processes are also neglected, as they are subdominant due to the much larger contribution from bound-bound processes to the opacities of the ejecta at all relevant wavelengths (\eg, \citealt{kasen13}).\footnote{Indeed,~\TARDIS~currently cannot model bound-free or free-free processes. In future work, it may be useful to take advantage of the modularity of \SPARK~to swap out \TARDIS~for some equivalent code which can include these processes, to assess their importance.} Ionization fractions and level populations in the plasma are computed assuming local thermodynamic equilibrium (LTE). This approximation is valid at the early times modelled here, but note that departures from LTE may be significant at later times ($\gtrsim$ $3-4$~days), when the ejecta becomes optically thin and enters a nebular phase (\eg, \citealt{gillanders21, hotokezaka21, pognan22a, pognan22b}).

At present, \TARDIS~can consider only 1D, spherically symmetric ejecta. Multi-dimensional general relativistic magnetohydrodynamics (GRMHD) simulations of NS-NS/NS-BH mergers indicate that the ejecta does not necessarily follow such a distribution; it may have a faster tidal component confined mostly to the equator, shocked ejecta from the collision interface in a NS-NS merger, and/or slower disk outflows emitted more isotropically or even more prominently toward polar regions (see, \eg, \citealt{fernandez16, baiotti17, shibata19, radice20} for reviews). These components are also characterized by different elemental compositions and opacities (\eg, \citealt{wanajo14, just15, mendoza-temis15, wu16}). As a result, the spectra and light curves of these events should be viewing-angle (and time) dependent, and axisymmetric profiles (\eg, \citealt{wollaeger18, bulla19, darbha20, kawaguchi20, heinzel21, korobkin21, wollaeger21}) would be more physically-motivated. However, an increase in spatial dimensions would significantly increase radiative transfer computation times and the complexity of our model. For now, we accept this limitation, and employ a simple power-law profile for the density $\rho$:

\begin{equation}\label{eqn:plaw_dens}
    \rho(v, t) = \rho_0 (t / t_0)^{-3} (v / v_0)^{n} , 
\end{equation}

\noindent where $\rho_0$, $v_0$, and $t_0$ are normalization terms.  We treat $\rho_0$ as a free parameter and fix $v_0 = 0.1c$ and $t_0 = 1.4$~days across all runs. Under the assumption that the ejecta is expanding homologously, which is accurate as of $\sim 10^2-10^3 $ seconds post-merger (\citealt{metzger10, kasen13, rosswog14, grossman14}), the velocity follows $v = r/t$ at all radii $r$ and can thus be interpreted as a radial coordinate at some time $t$. We fix $n=-3$, which agrees with hydrodynamical simulations (\eg, \citealt{kasen17, tanaka17, watson19}), although this choice is somewhat arbitrary and the emission is not particularly sensitive to the choice of density profile (\citealt{kasen17}). We fix the velocity at the outer computational boundary to a fiducial value of $v_{\mathrm{outer}} = 0.35c$, sufficiently large to allow propagating photon packets to redshift and interact with all relevant atomic lines (\citealt{gillanders22}). We find that our inference yields similar results for $v_{\mathrm{outer}}$ in the range $0.35c - 0.38c$. Finally, to simplify our model, ensure that there is an analytic mapping between our fit parameters and the synthesized spectra, and reduce run times, we use a single shell in the ejecta and perform a single \TARDIS~MC iteration. In this single-shell configuration, the plasma is described by a single temperature, mass density, and abundance pattern. These quantities influence the opacity of the ejecta and the features in the emergent spectrum.

Importantly, we use the fully relativistic implementation of \TARDIS~(\citealt{vogl19, vogl20}). Some previous works applying \TARDIS~to kilonovae (\citealt{smartt17, watson19}) did not include this full treatment, which was not available at the time.

\subsection{Parametrization of a Kilonova}\label{ssc:params}

The inputs required by the radiative transfer and our goal of inferring elemental abundances motivate our parametrization of a kilonova. We use as free parameters:

\begin{align}\label{eqn:params_inference}
\begin{split}
   \theta&=\{\log_{10}(L_{\mathrm{outer}}/L_{\odot}),~\log_{10}(\rho_0/\mathrm{g~cm^{-3}}),\\
   &~~~~~~v_{\mathrm{inner}},~v_{\mathrm{exp}},~Y_e,~s\}, 
\end{split}
\end{align}

\noindent corresponding to the luminosity at the outer computational boundary ($L_{\mathrm{outer}}$, normalized by solar luminosities $L_{\odot}$), the normalization in the density power law ($\rho_0$; Equation (\ref{eqn:plaw_dens})), the inner computational boundary velocity ($v_{\mathrm{inner}}$; also the photospheric velocity), and three parameters related to the abundances: the expansion velocity of the ejecta ($v_{\mathrm{exp}}$), the electron fraction of the ejecta ($Y_e$), and the specific entropy (per nucleon) of the ejecta ($s$). We note that the expansion velocity $v_{\mathrm{exp}}$ describes the conditions during $r$-process nucleosynthesis, which occurs in just the first second(s) post-merger, and is thus not necessarily equivalent to the velocities in the ejecta ($v_{\mathrm{inner}},v_{\mathrm{outer}}$) at later times; we discuss this detail further in Section~\ref{ssc:W18abunds}.

The luminosity at the outer boundary is used in \TARDIS~to set the initial guess for the blackbody temperature at the inner boundary, $T_{\mathrm{inner}}$ (\citealt{kerzendorf14}):

\begin{equation}
    T_{\mathrm{inner}} = \Big( \frac{L_{\mathrm{outer}}}{4 \pi r_{\mathrm{inner}}^2 \sigma} \Big)^{1/4}
\end{equation}

\noindent where $r_{\mathrm{inner}} = v_{\mathrm{inner}} t$ is the radius at the inner boundary and $\sigma$ is the Stefan-Boltzmann constant. Because we perform a single MC iteration within \TARDIS, $T_{\mathrm{inner}}$ is not iterated and keeps this relation to $L_{\mathrm{outer}}$; $L_{\mathrm{outer}}$ can thus be understood as an analog for the temperature of the photosphere in this configuration. The inner boundary velocity $v_{\mathrm{inner}}$ denotes the velocity of the ejecta at the photosphere. Given $\rho_0$, $v_{\mathrm{inner}}$, and our fixed $v_{\mathrm{outer}} = 0.35c$, we can compute the mass above the photosphere, and thus place a lower limit on the total mass of the ejecta. Finally, the electron fraction $Y_e$, expansion velocity $v_{\mathrm{exp}}$ and entropy $s$ are used to parametrize the abundances in the ejecta, as described in Section~\ref{ssc:W18abunds}.

This parametrization describes a single epoch of a single-component kilonova. In reality, kilonova ejecta are likely composed of multiple components of different masses, velocity profiles, spatial distributions, and compositions. These components interact with each other, leading to viewing-angle- and time-dependent effects such as lanthanide curtaining (\citealt{kasen15, wollaeger18, darbha20, nativi21}) and reprocessing where different components overlap in space (\citealt{kawaguchi20, korobkin21}). However, including additional components would significantly increase the dimensionality of our parametrization and the run time of the radiative transfer. Furthermore, the early-time spectrum (1.4 days) may be dominated by a single component, compared to later epochs which are better described with multiple components (\eg, \citealt{kasen17}). It is thus worth investigating if this simple single-component model can adequately reproduce the spectra. Given our Bayesian approach, it is also possible to compute the evidence of this model and compare it to that of multi-component models to determine whether additional component(s) are needed, which we will explore in future work.

\begin{figure*} [!ht]
    \includegraphics[width=0.98\textwidth]{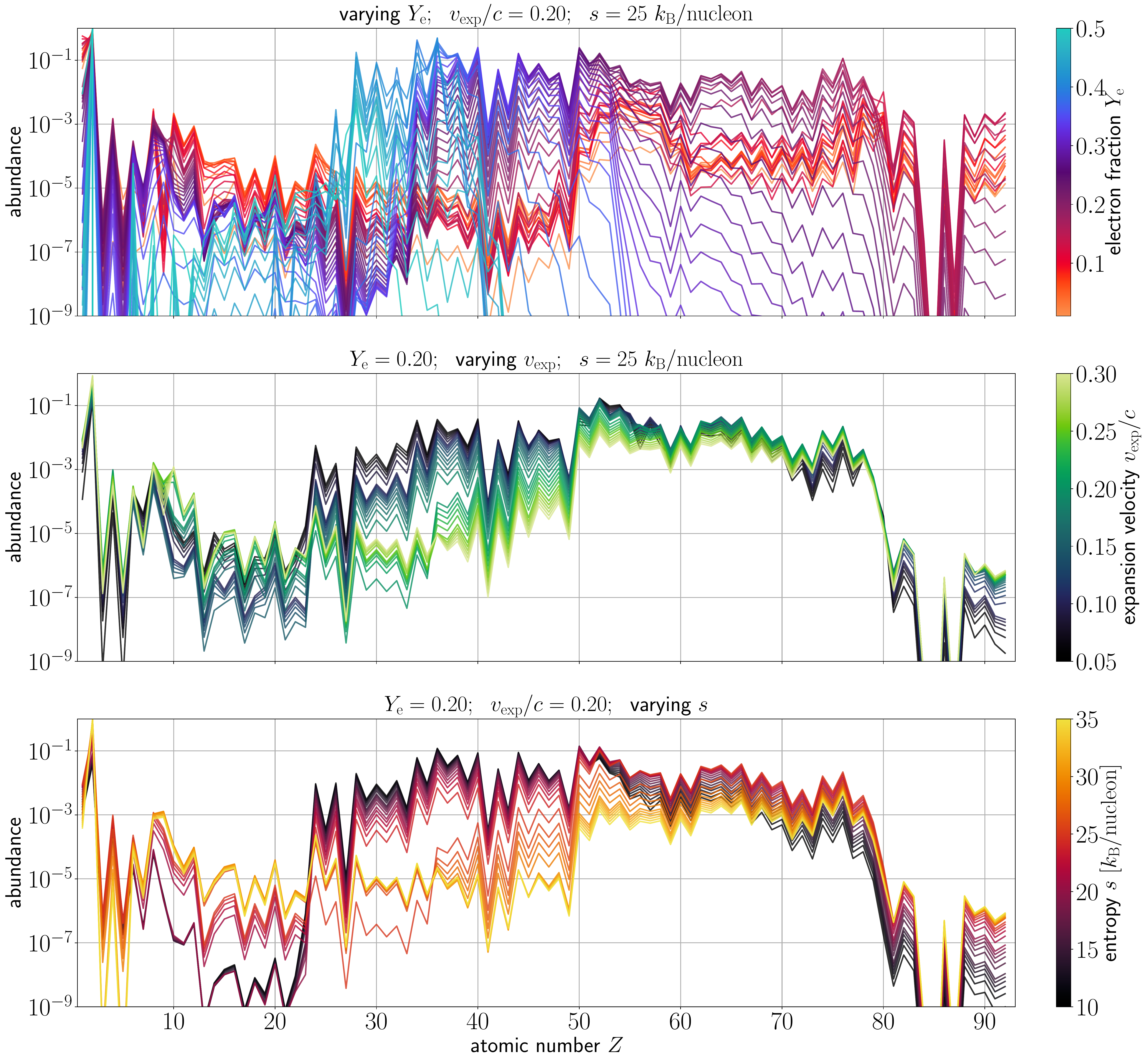}
    \figcaption{\textbf{Elemental abundances for $Z = 1-92$ as a function of the initial electron fraction $Y_e$, expansion velocity $v_{\mathrm{exp}}$, and entropy $s$, derived from \cite{wanajo18}, at $t=1.4$ days post-merger}. \cite{wanajo18} performed time-dependent reaction network calculations for electron fractions $Y_e = (0.01, 0.02, \ldots, 0.50)$, expansion velocities $v_{\mathrm{exp}}/c = (0.05, 0.10, ..., 0.30)$, and specific entropies $s = (10, 15, ..., 35)~k_{\mathrm{B}}/\mathrm{nucleon}$. For a smoother mapping of these parameters to abundances, we linearly interpolate $v_{\mathrm{exp}}/c$ in steps of 0.01 and $s$ in steps of 1~$k_{\mathrm{B}}/\mathrm{nucleon}$. \textbf{\textit{Top:}} Variation due to the electron fraction, for fixed $v_{\mathrm{exp}} = 0.25c$ and $s = 25~k_{\mathrm{B}}/\mathrm{nucleon}$. Abundance patterns are color-coded, with those corresponding to lower $Y_e$ and thus redder kilonovae colored red, and those of higher $Y_e$ and bluer kilonovae colored blue. For sufficiently low $Y_e$, the $r$-process extends up to the lanthanides and/or up to the actinides. At higher $Y_e$, the abundances are instead dominated by lighter elements. \textbf{\textit{Center:}} Abundances as a function of the expansion velocity, for fixed $Y_e=0.20$ and $s = 25~k_{\mathrm{B}}/\mathrm{nucleon}$. While the abundances are less sensitive to the velocities than the electron fraction, variations in the velocity still induce order-of-magnitude changes in the abundances of particular elements. \textbf{\textit{Bottom:}} Abundances as a function of entropy, for fixed $Y_e=0.20$ and $v_{\mathrm{exp}} = 0.25c$. As with the velocities, the abundances are not as sensitive to the entropy as the electron fraction, but variations in entropy can nonetheless alter the abundance of certain elements by several order of magnitudes. While the effect of increasing the entropy is similar to that of increasing the expansion velocity around the point $(Y_e = 0.20,~v_{\mathrm{exp}}=0.20c,~s = 25~k_{\mathrm{B}}/\mathrm{nucleon})$, this is not necessarily a general trend or feature of the input abundance patterns. Indeed, the abundances are a complex, non-linear function of $Y_e$, $v_{\mathrm{exp}}$, $s$, and $Z$ (\eg, \citealt{lippuner15}).}\label{fig:abunds_W18}
\end{figure*}

\subsection{Elemental Abundances}\label{ssc:W18abunds}

A useful tracer for the abundances in kilonova ejecta is the electron fraction $Y_e$ (equivalently the proton fraction: the ratio of protons to all baryons). Different physical/morphological components of the kilonova are distinguished by their $Y_e$ and corresponding abundance pattern (\eg, \citealt{wanajo14, just15, mendoza-temis15, wu16}). The electron fraction is determined by the initial composition of the merging
NS(s) and subsequent neutrino emission and absorption via charged-current weak interactions. These microphysical processes are in turn driven by various global processes, with one of the most important being the fate of the merger remnant. In a NS-NS merger, a short or long-lived hypermassive or supermassive NS remnant can produce significant neutrino radiation and raise the electron fraction of a surrounding accretion disk through weak interactions, leading to a higher-$Y_e$ outflow from this disk (\eg, \citealt{metzger14, perego14, lippuner17, siegel18, fahlman18, miller19, nedora21}). If the remnant instead promptly collapses to form a BH, the electron fraction in this outflow may be lower; this is simplified in NS-BH mergers, in which no NS remnant can survive. A sufficiently low electron fraction (\eg, $Y_e \lesssim 0.25$) can enable production of all three observed $r$-process peaks (see \citealt{cowan21}), producing isotopes up to $A \sim 250$, whereas a higher electron fraction ejecta (\eg, $Y_e \gtrsim 0.25$) may be truncated at $A \lesssim 140$ or may induce a shift in the location of these peaks. This in turn leaves an indelible impact on the observed spectrum: a sufficiently low-$Y_e$ ejecta will contain a significant abundance of lanthanides and/or actinides, with millions to billions of blended atomic lines predominantly in the optical/UV, leading to a redder spectrum. A higher-$Y_e$ ejecta will instead be dominated by lighter elements, and yield bluer emission.

The composition of the ejecta also depends on the entropy $s$ and expansion velocity $v_{\mathrm{exp}}$ of the ejecta. At the time of $r$-process nucleosynthesis (first $\sim$ seconds post-merger), a high-entropy (`hot') outflow will remain in an extended ($n, \gamma$) $\leftrightharpoons$ ($\gamma, n$) equilibrium (\ie, equilibrium between neutron captures and photodissociations), while a low-entropy (`cold') outflow will instead quickly $\beta$-decay to form quasi-equilibrium groups of isotopes, leading to differing abundance patterns (\citealt{cowan21, vassh21}). In addition, the expansion timescale can influence the cooling and entropy of the material, and thus the expansion velocity is also important. We therefore employ the full, time-dependent reaction network calculations of \cite{wanajo18}. These calculations were performed for electron fractions $Y_e = (0.01, 0.02, \ldots, 0.50)$, expansion velocities $v_{\mathrm{exp}}/c = (0.05, 0.10, ..., 0.30)$, and specific entropies $s = (10, 15, ..., 35)~k_{\mathrm{B}}/\mathrm{nucleon}$, where $k_{\mathrm{B}}$ is Boltzmann's constant, for a total of $50 \times 6 \times 6 = 1800$ calculations. We treat $Y_e$, $v_{\mathrm{exp}}$, and $s$ directly as fit parameters (\ref{ssc:params}).

To obtain a smoother mapping of the fit parameters to the elemental abundances, we linearly interpolate the abundances for each element, for values $v_{\mathrm{exp}}/c = (0.05, 0.06, ..., 0.30)$ and $s = (10, 11, ..., 35)~k_{\mathrm{B}}/\mathrm{nucleon}$. In Figure~\ref{fig:abunds_W18}, we show the variation in abundances holding two of $(Y_e,~v_{\mathrm{exp}},~s)$ fixed and allowing the other parameter to vary, at $t = 1.4~\mathrm{days}$ post-merger. The abundances are most sensitive to the electron fraction, as expected. However, variations in the expansion velocity and entropy can introduce order of magnitude differences in the abundances for a given electron fraction. We emphasize that these parameters $Y_e,~v_{\mathrm{exp}},~s$~describe the \textit{initial} conditions of the $r$-process, which operates on a timescale of $t \sim$~seconds post-merger, and then sets the initial abundances in the ejecta. These abundances themselves are then time-evolved within the reaction network to produce abundances at times which are relevant to our spectral retrieval, \eg, $t = 1.4$ days. Thus, $v_{\mathrm{exp}}$ should not be directly equated with the velocities in the ejecta at 1.4 days. The calculations in \cite{wanajo18} assume a spherical density profile which is fully equivalent to the power law density in Equation~(\ref{eqn:plaw_dens}) at $t \gg 150~\mathrm{km} / v_{\mathrm{exp}}$, as is the case at $t = 1.4$~days and beyond. However, they further assume that the velocity in the ejecta remains constant over time. $v_{\mathrm{exp}}$ is thus equivalent to the ejecta velocities if and only if the ejecta does not experience any acceleration before the onset of homologous expansion. The ejecta might experience this acceleration if subject to a strong magnetic field from a remnant NS in the first $\sim$ second post-merger (\eg, \citealt{metzger18}), significant $r$-process heating (\eg, \citealt{klion22}), or a Poynting flux-dominated wind from a central engine remnant NS (\citealt{ai22}). 

We note that it is well-documented that uncertainties in nuclear models (nuclear masses, neutron capture rates, $\beta$-decay rates, and treatments of fission) impact the abundances obtained from different reaction networks (\textit{e.g.}, \citealt{eichler15, eichler19, barnes16, barnes21, mumpower16, cote18, zhu21, kullmann22b}). It is thus possible that new reaction network calculations which incorporate, \eg, improved measurements of decay rates or more accurate nuclear mass models, will replace the calculations used here in the future. This motivates our use of a modular framework in \SPARK~in which different inputs can easily be swapped for others.

\subsection{Inference with \textsc{approxposterior}}\label{ssc:inference}

As a first test of \SPARK, we perform inference on the VLT/X-shooter spectrum of GW170817 (\citealt{pian17, smartt17}) at $t=1.4$ days post-merger.\footnote{Acquired through WISeREP (\citealt{yaron12}).} This spectrum spans roughly 3200~\text{\AA} to 24800~\text{\AA}. The relevant parameters~$\theta$ are $\{\log_{10}(L_{\mathrm{outer}}/L_{\odot}),~\log_{10}(\rho_0/\mathrm{g~cm^{-3}}),~v_{\mathrm{inner}},~v_{\mathrm{exp}},\\~Y_e,~s\}$, and we wish to compare synthetic spectra of various $\theta$ to the spectrum of GW170817. However, due to the considerable run times for each radiative transfer run, traditional inference techniques such as a Markov chain Monte Carlo (MCMC) would be prohibitively computationally expensive. At 6 parameters, one would easily require $>10^6$ forward model evaluations (\TARDIS~runs) to perform this MCMC. We therefore instead use \approxposterior~(\citealt{fleming18, fleming20}), in which the posterior distribution is approximated using active learning with a surrogate Gaussian process (GP). We specifically use the technique of Bayesian Active Posterior Estimation (BAPE, \citealt{kandasamy17}). In this technique, the GP surrogate for the posterior is iteratively trained using active learning on points obtained by maximizing an acquisition function.

A summary of the steps involved in \SPARK~is given in Algorithm~\ref{algo:spark}. A more detailed description is as follows. We begin with a training set $T$ of $m_0$ Latin Hypercube-sampled pairs $(\theta_i, L_p (\theta_i))$, $\theta_i \in \mathcal{D}$, where $L_p (\theta)$ is the posterior and $\mathcal{D}$ is the domain of the parameters. We similarly construct a \textit{test} set of $m_{\mathrm{test}}$ Latin Hypercube-sampled points to be used later to assess convergence. We then condition a GP $f(\theta) \equiv \mathcal{GP} (\mu(\theta),~k(\theta, \theta'))$ on $T$, where $\mu(\theta)$ and $k(\theta, \theta')$ are the mean and covariance kernel of the GP, respectively. For the mean, we initialize with $\mu=0$. For the kernel, we employ the squared exponential:

\begin{equation}\label{eqn:covariance_kernel}
    k(\theta_i, \theta_j) = \exp\Big(-\frac{||\theta_i - \theta_j||^2}{2 \ell^2}\Big),   
\end{equation}

\noindent where $\ell$ is a hyperparameter which controls the length scale of correlations in each dimension of parameter space. A small amount of white noise of $\exp(-3)$ is added to the diagonals of the kernel's covariance matrix as a form of regularization (to prevent the GP from overfitting to the potentially complex posterior) and to ensure numerical stability. The value of white noise which was required to prevent over-fitting was determined through trial and error; in the future, it will be useful to regularize the GP with some more robust technique. The solution to this problem is outside the scope of this work, but will be required for future works fitting other spectra.

With the base training set constructed, we transition to active learning, and obtain $m_{\mathrm{active}}$ new points for the training set. These points are selected by maximizing an acquisition (`utility') function. We use the exponentiated variance utility function (Equation (5) of \citealt{kandasamy17}):  

\begin{equation}\label{eqn:EV_utility}
    u_{\mathrm{EV}}(\theta) = \exp\big(2\mu_n(\theta) + \sigma_n^2(\theta)\big)\cdot\big(\exp(\sigma_n^2(\theta)) - 1\big),
\end{equation}
    
\noindent where $\mu_n(\theta)$ and $\sigma_n^2(\theta)$ are the mean and variance of the GP's predictive conditional distribution, respectively. The $\theta^*$ at the maximum of this function maximizes a combination of the approximate posterior density \textit{and} the predictive uncertainty of the GP, ensuring that we sample around the expected peak of the posterior but also explore regions of parameter space where the GP fit is uncertain. This optimization is reinitialized $N_{\mathrm{MinObjRestarts}}$ times to mitigate the impact of local extrema. We then synthesize a spectrum at the optimal $\theta^*$ and compute the posterior $L_p (\theta^*)$, and the new pair $(\theta^*, L_p (\theta^*))$ is added to the training set $T$. For every $N_{\mathrm{optGPEvery}}$ points sampled, the GP is re-conditioned on the augmented training set $T$, and the GP hyperparameters are re-optimized. Optimization of hyperparameters is also reinitialized, $N_{\mathrm{GPRestarts}}$ times. 

Once $m_{\mathrm{active}}$ new points have been sampled, the active learning ends. With a final training set of $m_0 + m_{\mathrm{active}}$ samples, we perform dynamic nested sampling with \texttt{dynesty} (\citealt{speagle20}) to obtain a final approximate posterior distribution. We opt for nested sampling rather than an MCMC due to nested sampling's greater ability to sample complex and potentially multi-modal distributions, and capture the tails of distributions. Since we are currently only interested in the posterior and not the Bayesian evidence, we perform dynamic nested sampling with 100\% weight on the posterior. We use another implementation of nested sampling, \texttt{UltraNest} (\citealt{buchner21}), to verify our posteriors. We observe no significant differences in our tests.

\begin{algorithm}\label{algo:spark}
 \SetAlgoLined
 \textbf{Set} an input domain $\mathcal{D}$ and a prior on the GP surrogate $f(\theta) \approx L_p (\theta)$ \\
 \textbf{Build} initial training set $T$ of $m_0$ pairs ($\theta_i$, $L_p (\theta_i)$), where $\theta_i$ are Latin Hypercube-sampled \\
 \textbf{Build} initial test set of $m_{\mathrm{test}}$ points, Latin Hypercube-sampled \\
 \textbf{Condition} $f(\theta)$ on $T$ and optimize GP hyperparameters \\
    \For{$j=0, 1, ..., m_{\mathrm{active}}/N_{\mathrm{optGPEvery}}$}{
      \For{$i=0, 1, ..., N_{\mathrm{optGPEvery}}$}{
        Find $\theta^*$ which maximizes $u_{\mathrm{EV}}(\theta)$ over $N_{\mathrm{MinObjRestarts}}$ optimizations \\
         Run \TARDIS~on $\theta^*$ \\
         Compare synthetic spectrum to observed, compute $L_p (\theta^*)$ \\
         Append $(\theta^*, L_p (\theta^*))$ to $T$ \\
       }
       \textbf{Condition} $f(\theta)$ on augmented $T$ and determine optimal GP hyperparameters over $N_{\mathrm{GPRestarts}}$ optimizations \\

 }
  \textbf{Sample} final $f(\theta)$ with dynamic nested sampling to obtain final approximate posterior distribution
  \newline
  
\caption{The algorithm used in \SPARK/\approxposterior, from the construction of the initial training set to the final dynamic nested sampling run. Choices for all hyperparameters are included in Table~\ref{tab:SPARK_params} and described in Section~\ref{ssc:inference}.}
\end{algorithm}

Table~\ref{tab:SPARK_params} contains our choices for the \approxposterior~parameters $m_0$, $m_{\mathrm{test}}$, $m_{\mathrm{active}}$, $N_{\mathrm{optGPEvery}}$, $N_{\mathrm{MinObjRestarts}}$, $N_{\mathrm{GPRestarts}}$, the optimizers used for the GP hyperparameters/acquisition function, and other choices relevant to the inference. These choices were initially based on observations presented in~\cite{fleming20}, which used \approxposterior~for an inference problem in five dimensions. We find that these parameters are effective with minor changes when scaled up to six dimensions, except that we require a larger number of points $m_0 = 1500$ in the base training set and let active learning proceed for longer. Table~\ref{tab:SPARK_params} also includes our priors, which are uniform in all dimensions. The prior on the luminosity is approximately centered around the observed bolometric luminosity of the kilonova at $t=1.4$~days (\eg, \citealt{villar17} and references therein). Priors for the density and inner boundary (photospheric) velocity are wide, but span the values estimated for these parameters in other works (\eg, \citealt{kasen17, villar17, watson19, domoto21, gillanders21, gillanders22}). For the electron fraction, expansion velocity, and entropy, we allow the entire range spanned by the reaction network calculations, but we ignore $Y_e > 0.4$.

\begin{deluxetable}{cc}
\tablewidth{0.5\textwidth}
\centering
\tablecaption{Parameters used in the \SPARK~run on the 1.4 day GW170817 kilonova spectrum.}
\tablehead{parameter & value}
\startdata
$m_{\mathrm{test}}$ & 60 \\
$m_0$ & 1500 \\
$m_{\mathrm{new}}$ & 60 \\
$m_{\mathrm{active}}$ & 1140 \\
$N_{\mathrm{optGPEvery}}$ & 10 \\\hline
$N_{\mathrm{MinObjRestarts}}$ & 10 \\
$N_{\mathrm{GPRestarts}}$ & 10 \\\hline
$u_{\mathrm{EV}}$ optimization & \cite{neldermead65} \\
GP optimization & \cite{powell64} \\\hline
priors & \multicolumn{1}{l}{$\log_{10}(\frac{L_{\mathrm{outer}}}{L_{\odot}}) \in [7.6, 8.0]$} \\
  & \multicolumn{1}{l}{$\log_{10}(\frac{\rho_0}{\mathrm{g~cm^{-3}}}) \in [-16.0, -14.0]$} \\
  & \multicolumn{1}{l}{$v_{\mathrm{inner}}/c \in [0.25, 0.34]$} \\
  & \multicolumn{1}{l}{$v_{\mathrm{exp}}/c \in [0.05, 0.30]$} \\
  & \multicolumn{1}{l}{$Y_e \in [0.01, 0.40]$} \\
  & \multicolumn{1}{l}{$s \in [10, 35]~k_{\mathrm{B}}/\mathrm{nucleon}$} \\\hline
likelihood & \multicolumn{1}{l}{\cite{czekala15} global} \\ & \multicolumn{1}{l}{$a_{\mathrm{G}} = 10^{-34}~(\mathrm{erg~s^{-1}~cm^{-2}}$~\text{\AA}${}^{-1})^{2}$}, \\
& \multicolumn{1}{l}{$\ell = 0.025c$} \\
\enddata
\end{deluxetable}\label{tab:SPARK_params}

Finally, to compute likelihoods, we use the full log-likelihood function of \cite{czekala15}:

\begin{equation}
    \ln p (F_{\mathrm{obs}} | \theta) = - \frac{1}{2} \big(\mathrm{R}^{\mathrm{T}} \mathrm{C}^{-1} \mathrm{R} + \ln{\det \mathrm{C}} + N_{\mathrm{pix}} \ln 2\pi \big) ,
\end{equation}

\noindent where $F_{\mathrm{obs}}$ is the observed spectrum, $\mathrm{R}$ is the residual between the observed and \TARDIS~spectrum of parameters $\theta$, $N_{\mathrm{pix}}$ is the number of wavelength bins (pixels) in the spectrum.\footnote{The observed spectrum is re-sampled onto an array of $N_{\mathrm{pix}} = 5000$ logarithmically-spaced wavelength bins, to match the wavelength bins of the synthetic \TARDIS~spectra.} $\mathrm{C}$ is the covariance matrix of the observed spectrum, and can include both the inherent measurement noise and pixel-to-pixel covariances. In the trivial scenario where pixels in the spectrum are uncorrelated, this reduces to the common $\ln p = -\frac{\chi^2}{2} + \mathrm{constant}$. In reality, this is not the case for most spectra, and so \cite{czekala15} construct a more complete $\mathrm{C}$ for the general problem of spectroscopic inference which can include both `global' covariance structure (pixel-to-pixel covariances) and `local' covariance structure arising from, \eg, missing lines in a line list. As in \cite{czekala15}, we adopt a Mat{\'e}rn-3/2 kernel which depends only on the separation between two pixels to represent the global covariance structure (see their equation (11)). The benefits of this approach are a more realistic estimation of the uncertainties on the best-fit parameters and a smoother likelihood / acquisition function $u_{\mathrm{EV}}$ over which it is easier to optimize. After some trial and error, we find that an amplitude term $a_{\mathrm{G}} = 10^{-34}~(\mathrm{erg~s^{-1}~cm^{-2}}$~\text{\AA}${}^{-1})^{2}$ and a correlation length scale $\ell = 0.025c$ for the global covariance lead to an accurate representation of the uncertainties on the observed spectrum. This amplitude is roughly the mean uncertainty on the spectrum, squared, while this correlation length scale is approximately 10\% of the photospheric velocities present in the ejecta at early times (\citealt{watson19}).

At present, we do not employ any local covariance features in the \cite{czekala15} framework, as it is unclear \textit{a priori} which regions of the spectrum suffer most from incomplete atomic data. In the future, determining which parts of the spectrum require such local covariance features to improve the fit could provide insight into where atomic data is most lacking, or, where the model fails more generally.


\begin{figure*}[!ht]
    \includegraphics[width=0.95\textwidth]{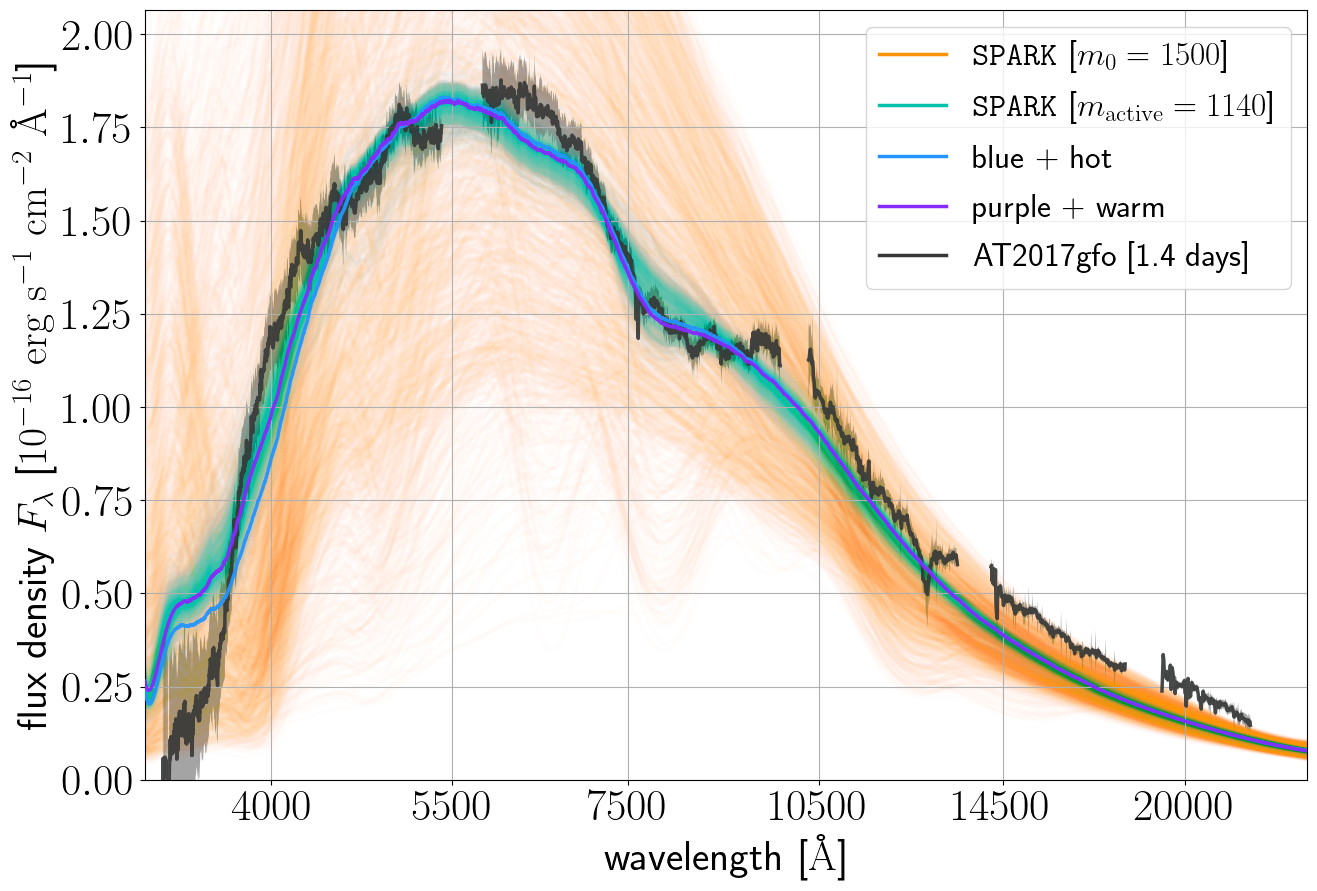}
    \figcaption{\textbf{Set of all synthetic spectra generated during the \SPARK~run}. Orange traces denote the $m_{0} = 1500$ spectra acquired by Latin Hypercube sampling, while turquoise traces show the $m_{\mathrm{active}}=1140$ points where \SPARK~chose to evaluate the forward model using Bayesian Active Posterior Estimation (BAPE) with the acquisition function given by Equation~(\ref{eqn:EV_utility}). We also show the two best-fitting models, `blue+hot' and `purple+warm', which are discussed at length in Sections \ref{sssc:bluehot} and \ref{sssc:purplewarm}. The $1.4$-day spectrum for the GW170817 kilonova, AT2017gfo, is also included. Missing regions in the observed spectrum are regions of either poor sensitivity in the X-shooter spectrograph or telluric features. All fluxes are obtained assuming a fiducial distance of 40 Mpc to the kilonova. }\label{fig:all_TARDIS_runs}
\end{figure*}

\begin{figure}[!ht]
    \centering
    \includegraphics[width=0.47\textwidth]{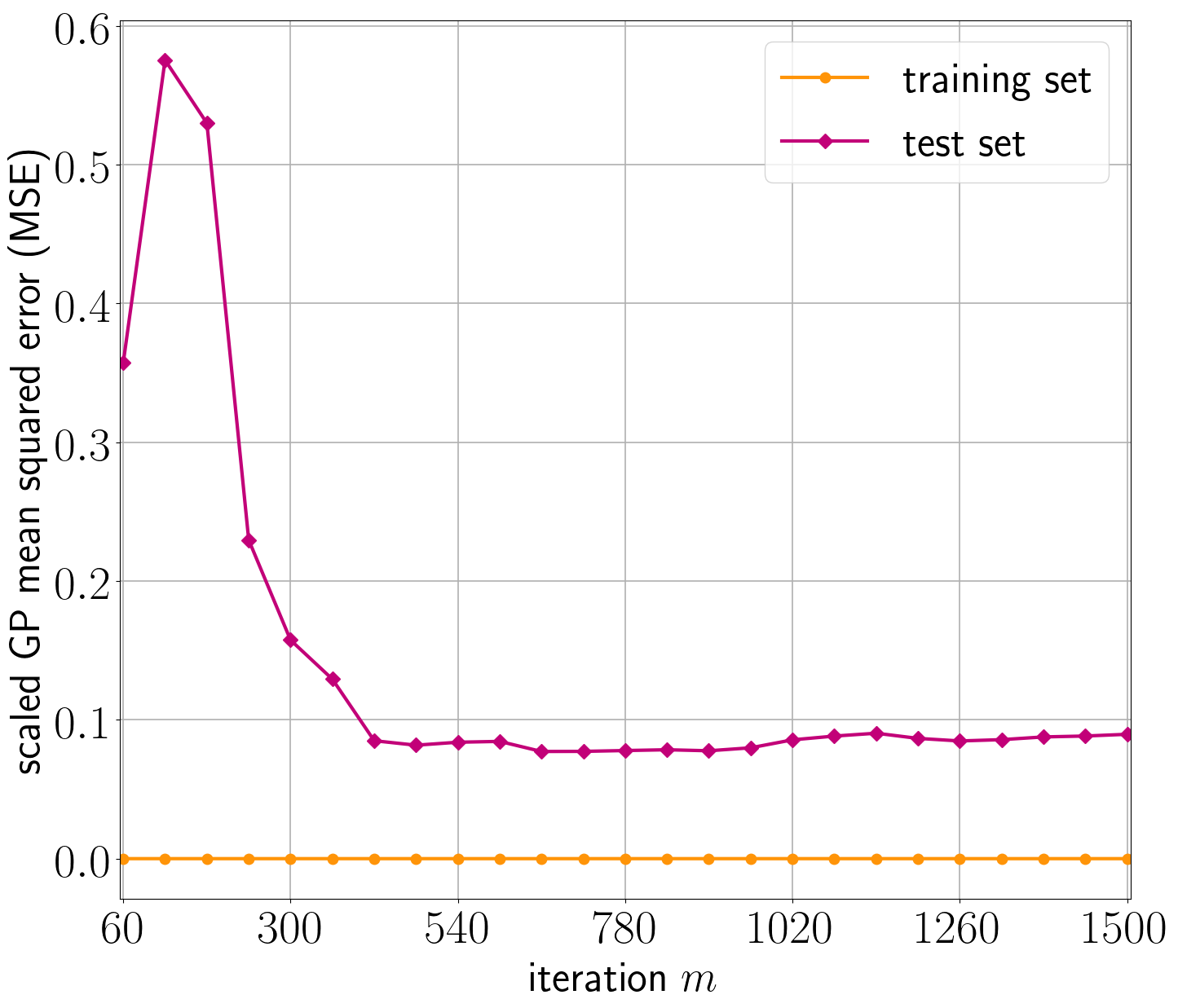}
    \figcaption{\textbf{Scaled mean-squared error (MSE) of the GP's predictions on the test and training sets, over the course of building up the base training set.} The test set contains a fixed number of points $m_{\mathrm{test}} = 60$. The training set grows in size with $m$. The training set error is always low (about $10^9$ times smaller than the test set), as expected, since the GP is conditioned on this set. The test set error gradually decreases and then reaches a plateau, demonstrating that the GP has converged and is accurately reproducing the overall posterior distribution.}\label{fig:MSE}
\end{figure}

\begin{figure}[!ht]
    \centering
    \includegraphics[width=0.47\textwidth]{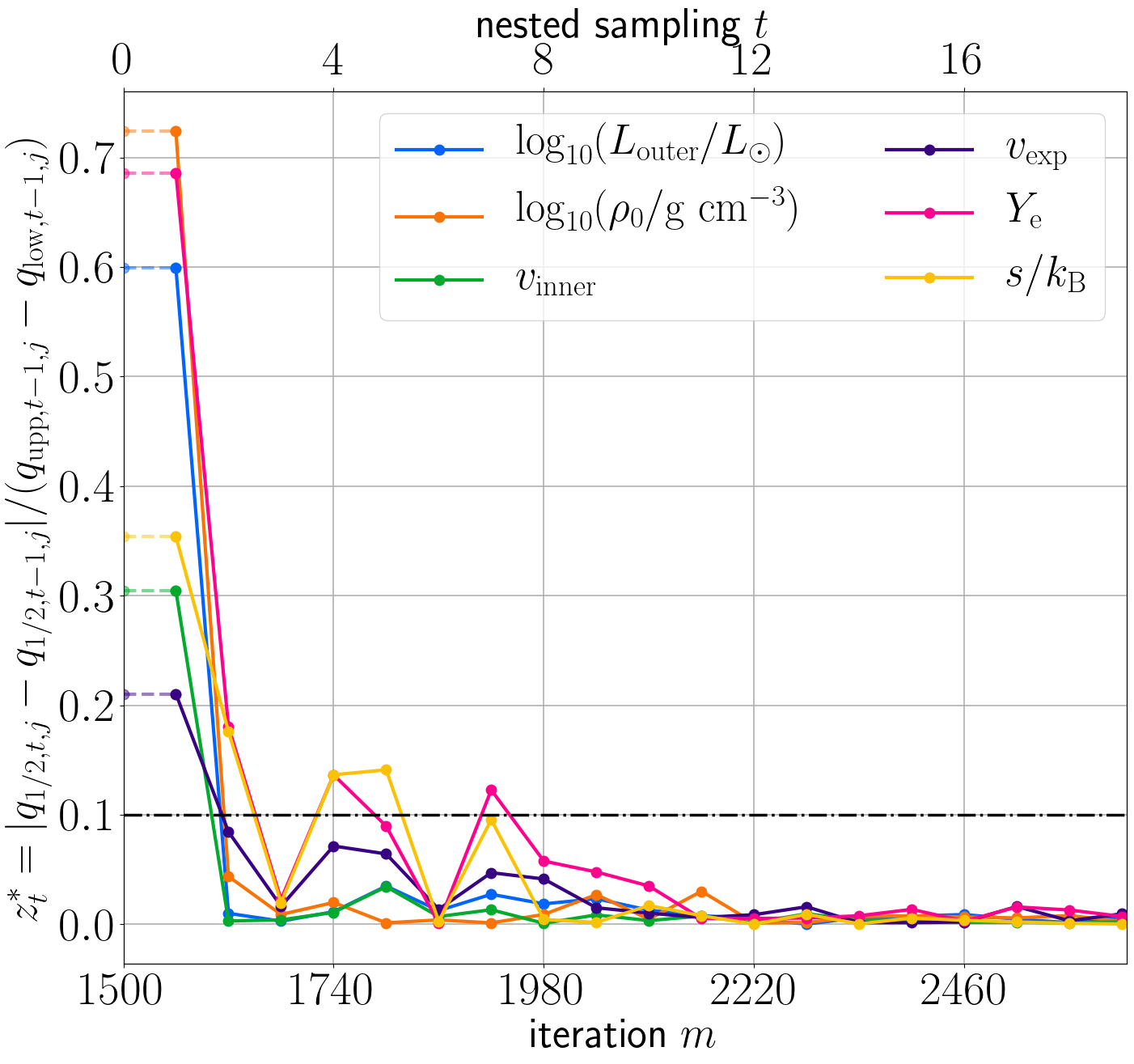}
    \figcaption{\textbf{Convergence diagnostic $z$ for the active learning portion of training.} A new nested sampling run is performed on the re-optimized GP every $m_{\mathrm{new}} = 60$ iterations, and the median, 2.5\%, and 97.5\% quantiles are computed. These are compared at consecutive iterations according to Equation~(\ref{eqn:convergence_z_modified}). The diagnostic $z^{*}_{t,j} \leqslant 0.1$ for all parameters $j$ for 10 consecutive iterations, demonstrating that the posterior has converged.}\label{fig:convergencez}
\end{figure}

\begin{figure*}[!ht]
    \includegraphics[width=0.95\textwidth]{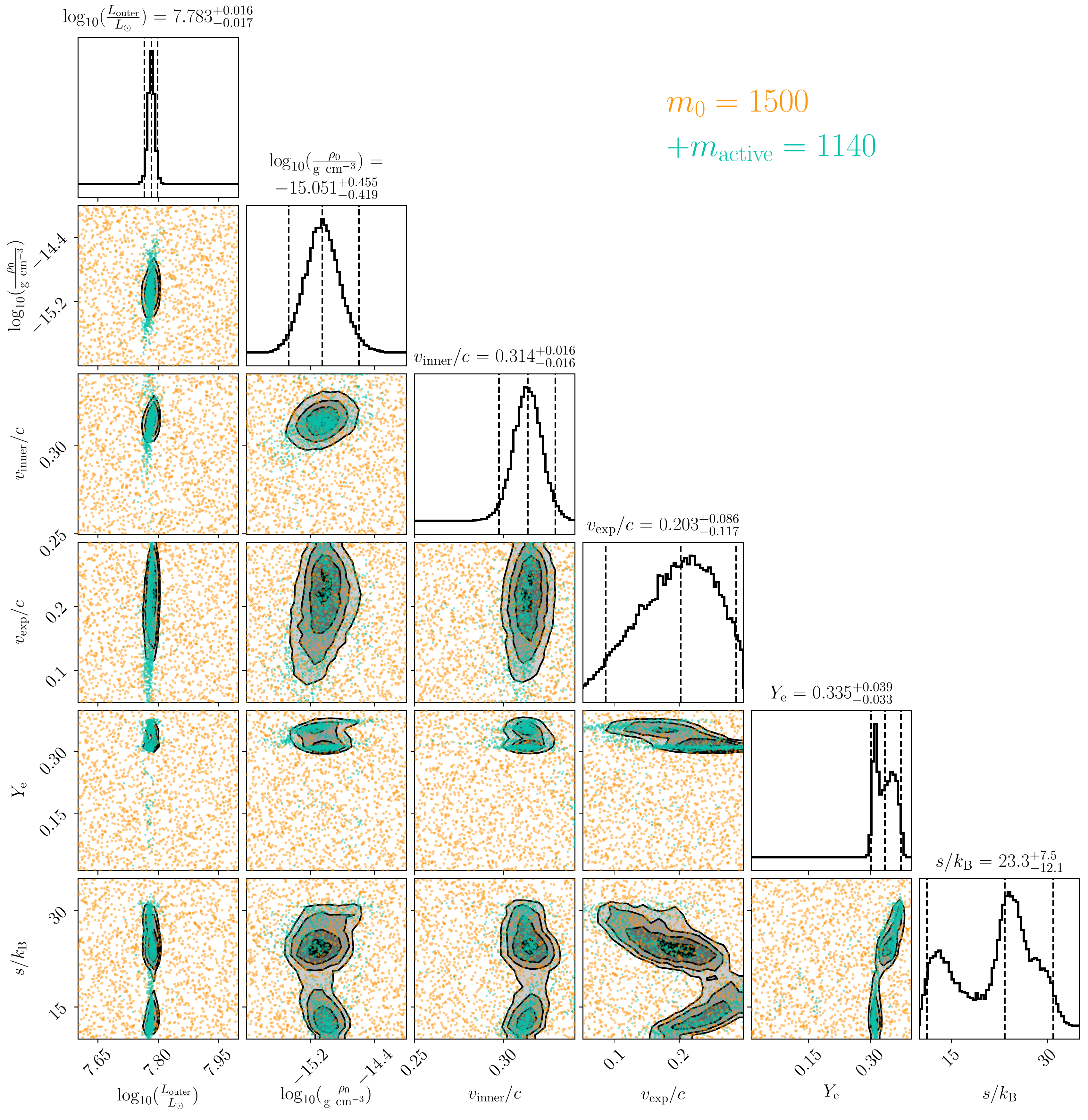}
    \figcaption{\textbf{Corner plot showing the approximate posterior as obtained from dynamic nested sampling}. Dashed lines in the 1D marginalized distributions indicate the 2.5\%, 50\%, and 97.5\% quantiles. Orange points indicate the $m_{0} = 1500$ initial Latin Hypercube-sampled points, while turquoise points indicate the $m_{\mathrm{active}}=1140$ points where \SPARK~chose to evaluate the forward model using Bayesian Active Posterior Estimation (BAPE) with the acquisition function given by Equation~(\ref{eqn:EV_utility}). The BAPE points are grouped into two clusters which are easiest to distinguish in the $Y_e$ and $s$ dimensions. These clusters lie on the two peaks of the bimodal posterior which we recover with nested sampling, demonstrating the ability of BAPE to sample complex, multi-modal posteriors with few forward model evaluations. We highlight the apparent higher electron fraction, higher entropy (`blue + hot') mode of the posterior in Figure~\ref{fig:corner_hotblue}, and the moderate electron fraction, moderate entropy (`purple + warm') mode in Figure~\ref{fig:corner_purplewarm}.}\label{fig:corner}
\end{figure*}

\begin{figure*}[!ht]
    \includegraphics[width=0.95\textwidth]{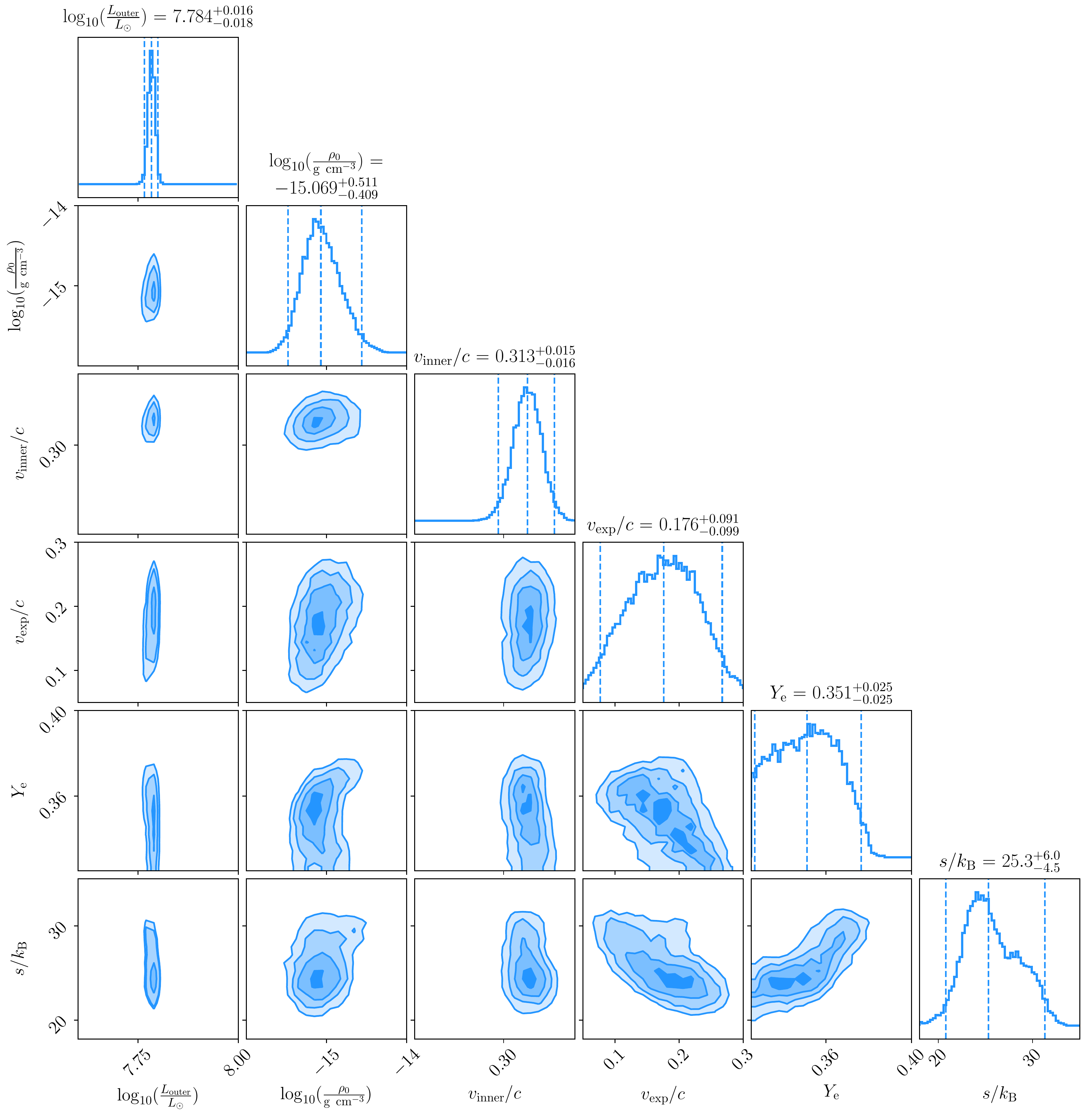}
    \figcaption{\textbf{Corner plot showing the approximate posterior obtained from dynamic nested sampling, zoomed in to the higher electron fraction, higher entropy (`blue + hot') mode of the bimodal posterior shown in Figure~\ref{fig:corner}}. This blue+hot mode has electron fraction $Y_e = 0.351^{+0.025}_{-0.025}$, expansion velocity $v_{\mathrm{exp}}/c = 0.176^{+0.091}_{-0.099}$, and entropy $s = 25.3^{+6.0}_{-4.5}~k_{\mathrm{B}}/\mathrm{nucleon}$. This electron fraction and entropy are both substantially higher than those of the purple+warm mode. The expansion velocity is smaller than that of the purple+warm, but is poorly constrained and is consistent with that of the other mode. While the luminosity and photospheric velocities are nearly identical for both modes, the density (and thus overall ejecta mass above the photosphere; $M_{\mathrm{ej}} > 3.5^{+4.2}_{-3.3}~\times 10^{-5} M_{\odot} = 11.8^{+13.8}_{-11.1} M_{\oplus}$) is slightly smaller for this blue+hot mode. The abundance pattern corresponding to this blue+hot $r$-process is shown in the the top panel of Figure~\ref{fig:infer_abunds}. The spectrum generated by this model is shown in Figure~\ref{fig:all_TARDIS_runs}.}\label{fig:corner_hotblue}
\end{figure*}

\begin{figure*}[!ht]

    \includegraphics[width=0.95\textwidth]{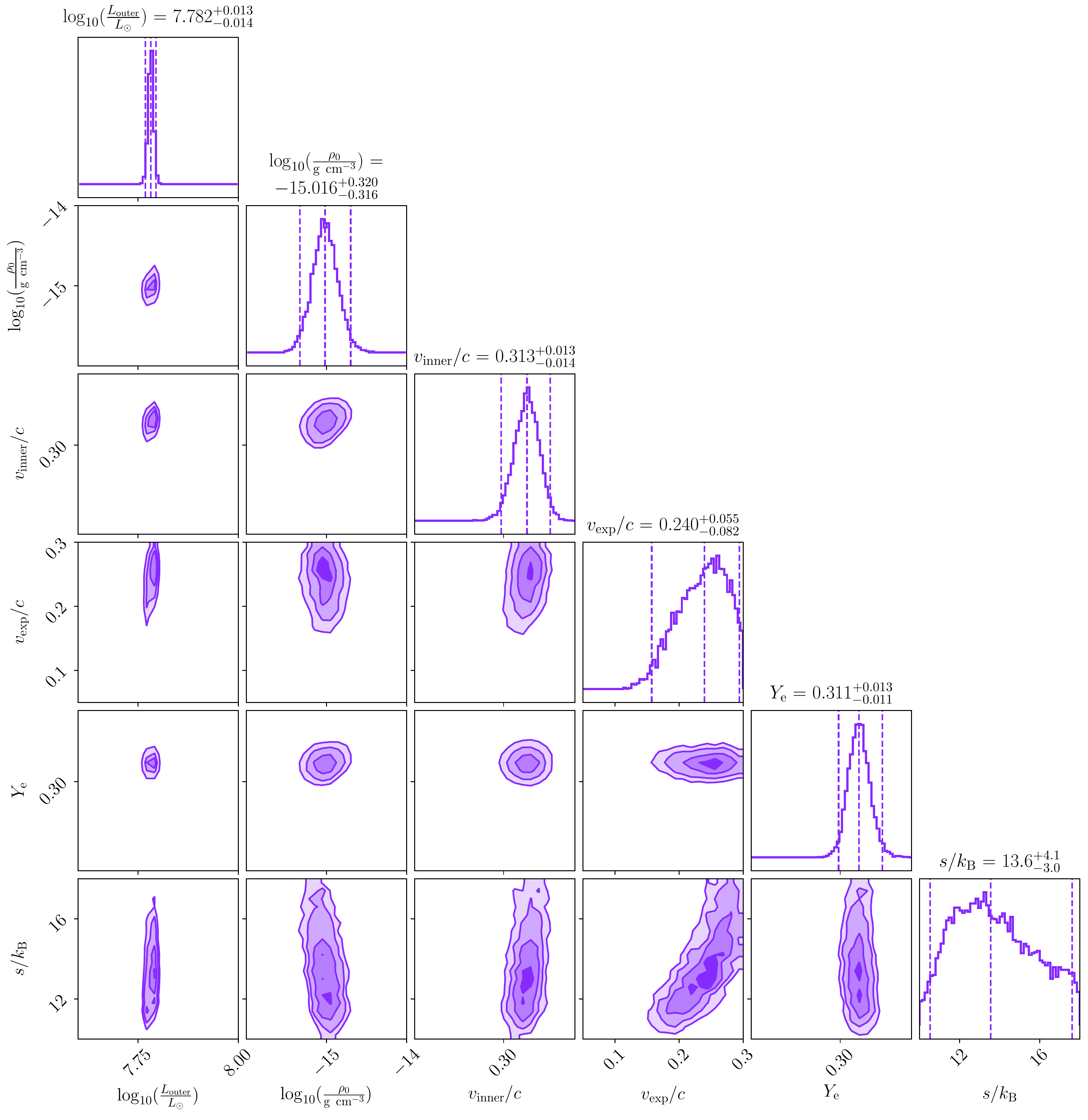}
    \figcaption{\textbf{Corner plot showing the approximate posterior obtained from dynamic nested sampling, zoomed in to the moderate electron fraction, moderate entropy (`purple + warm') mode of the bimodal posterior shown in Figure~\ref{fig:corner}}. In contrast with the blue+hot mode, this purple+warm mode has electron fraction $Y_e = 0.311^{+0.013}_{-0.011}$, and entropy $s = 13.6^{+4.1}_{-3.0}~k_{\mathrm{B}}/\mathrm{nucleon}$, both substantially lower than those of the blue+hot mode. The expansion velocity $v_{\mathrm{exp}}/c = 0.240^{+0.055}_{-0.082}$ here is faster than that of the blue+hot, but is once again poorly constrained. The luminosity and photospheric velocity are nearly identical for both modes, but the density (and thus overall ejecta mass above the photosphere; $M_{\mathrm{ej}} > 4.0^{+2.9}_{-2.9}~\times 10^{-5} M_{\odot} = 13.3^{+9.8}_{-9.7} M_{\oplus}$) is slightly larger for this purple+warm mode. All fit parameters are more tightly-constrained at this mode than in the blue+hot; this is reflected in the abundance pattern corresponding to this purple+warm $r$-process, shown in the bottom panel of Figure~\ref{fig:infer_abunds}. The spectrum corresponding to this model is included in Figure~\ref{fig:all_TARDIS_runs}.}\label{fig:corner_purplewarm}
\end{figure*}

\section{Results}\label{sec:results}

\begin{deluxetable}{ccccc}
\centering
\tablecaption{Best-fit parameters for the GW170817 kilonova at $t=1.4$~days, for the blue+hot and purple+warm models discussed in the text. Upper and lower bounds correspond to 97.5\% and 2.5\% quantiles, respectively. The inner boundary temperature $T_{\mathrm{inner}}$, lower bound on the ejecta mass $M_{\mathrm{ej}}$, and lanthanide mass fraction $X_{\mathrm{lan}}$ are derived parameters, \ie, they are not dimensions in $\theta$-space.}
\tablehead{parameter & \SPARK~blue+hot & \SPARK~purple+warm}
\startdata\tablewidth{1.0\textwidth}
 \vspace{2pt}
$\log_{10}(L_\mathrm{outer}/L_{\odot})$ & $7.784^{+0.016}_{-0.018}$ & $7.782^{+0.013}_{-0.014}$ \\ 
$\log_{10}(\rho_0/\mathrm{g~cm^{-3}})$ & $-15.069^{+0.511}_{-0.409}$ & $-15.016^{+0.320}_{-0.316}$ \\ 
$v_{\mathrm{inner}}/c$ & $0.313^{+0.015}_{-0.016}$ & $0.313^{+0.013}_{-0.014}$ \\
$v_{\mathrm{exp}}/c$ & $0.176^{+0.091}_{-0.099}$ & $0.240^{+0.055}_{-0.082}$ \\
$Y_e$ & $0.351^{+0.025}_{-0.025}$ & $0.311^{+0.013}_{-0.011}$ \\
$s~[k_{\mathrm{B}}/\mathrm{nucleon}]$ & $25.3^{+6.0}_{-4.5}$ & $13.6^{+4.1}_{-3.0}$\\ \hline
$T_{\mathrm{inner}}~[\mathrm{K}]$ & $3962^{+102}_{-109}$ & $3958^{+87}_{-94}$ \\ 
$M_{\mathrm{ej}}~[M_{\odot}]$ & $>3.5^{+4.2}_{-3.3}~\times 10^{-5}$ & $>4.0^{+2.9}_{-2.9}~\times 10^{-5}$ \\ 
 \textcolor{white}{$M_{\mathrm{ej}}$}~$[M_{\oplus}]$ & $>11.8^{+13.8}_{-11.1}$ & $>13.3^{+9.8}_{-9.7}$ \\
 $X_{\mathrm{lan}}$ & $\leqslant 5.6~\times 10^{-9}$ & $1.82^{+26.3}_{-1.79}~\times 10^{-7}$ \\
\enddata
\end{deluxetable}\label{tab:bestfit}

\subsection{Convergence and best fits}\label{ssc:main_results}

X-shooter acquired spectra of the GW170817 kilonova every night beginning at $t=0.5$ days post-merger. The early 1.4-day spectrum shows the most striking, broad absorption feature at $\sim$ $8000$ \text{\AA}, and so we perform inference at this epoch. Moreover, our assumption of LTE in the radiative transfer is most accurate at earlier times. At later times ($\gtrsim 3 - 4$ days), when the ejecta becomes optically thin, this approximation breaks down and \TARDIS~may not be able to reproduce the observed spectra.

Our final inference is performed with 2640 synthetic \TARDIS~spectra: $m_0 = 1500$ which are obtained with Latin Hypercube sampling to obtain a coarse map of the posterior, followed by $m_{\mathrm{active}} = 1140$ which are obtained with BAPE. We present all of these spectra in Figure~\ref{fig:all_TARDIS_runs}, alongside the observed X-shooter spectrum at $t=1.4$~days. We also show our best-fitting models, which we discuss at length below (\ref{sssc:bluehot}, \ref{sssc:purplewarm}) after first assessing the convergence of our algorithm.

We use our test set to quantitatively assess the convergence of the GP surrogate \textit{before} active learning begins. The test set contains $m_{\mathrm{test}} = 60$ points and is also obtained with Latin Hypercube sampling to evenly sample parameter space. Since the GP is not conditioned on these test points, whether the GP accurately predicts the values at these points serves as a test of the GP's ability to capture the entire posterior distribution. Figure~\ref{fig:MSE} shows the mean-squared-error (MSE) of the GP's predictions on the test set as the training set grows. As training progresses, the test set MSE decreases, and eventually reaches a plateau, demonstrating that our training set of $m_0 = 1500$ is adequately large to capture the global properties of the posterior.

Once active learning begins, BAPE attempts to better resolve the peak of the posterior. In this regime, we introduce a convergence diagnostic $z^{*}$:

\begin{equation}
z^{*}_{t,j} = \frac{| q_{1/2,t,j} - q_{1/2,t-1,j} |}{ q_{\mathrm{upp},t-1,j} - q_{\mathrm{low},t-1,j}},
\end{equation}\label{eqn:convergence_z_modified}

\noindent which compares the median ($q_{1/2,t,j}$) and the range between some upper and lower quantiles ($q_{\mathrm{upp},t,j} - q_{\mathrm{low},t,j}$) of the marginal distributions, for all fit parameters $j$, at successive nested sampling runs $t$. This $z^{*}$ is analogous to the Z-score used to measure the distance of some value from the mean of a distribution. A new nested sampling run is performed every time $m_{\mathrm{new}} = 60$ points are added to the active learning set to compute this diagnostic. Convergence is obtained if $z^{*}_{t,j} \leqslant $ some small number $\epsilon$ for all parameters $j$ for multiple consecutive nested sampling runs, indicating that the posterior is stable. In Figure~\ref{fig:convergencez}, we find that $z^{*}_{t,j} \leqslant 0.1$ for 10 consecutive iterations, demonstrating convergence. \footnote{This diagnostic is a modified version of the convergence $z$ presented in \citealt{fleming20}, which used the mean and standard deviation. We use the median and 2.5\%, 97.5\% quantiles instead to account for the possibility of complex, non-Gaussian posteriors, but find similar results using the original convergence $z$.}

One final diagnostic to assess the convergence of the GP is to directly follow the evolution of the optimized GP hyperparameters over the course of training. This diagnostic is instructive both when constructing the base training set and once active learning begins. This diagnostic again confirms that the GP has converged; we show this in Appendix~\ref{app:GP_hparams}.

With the GP converged, we perform a last nested sampling run to finally obtain the approximate posterior distribution.  We show the joint posterior distributions and 1D marginal posteriors for all fit parameters in Figure~\ref{fig:corner}. We also superimpose the points at which we performed radiative transfer during the inference, distinguishing between the Latin Hypercube samples and BAPE-selected active learning samples. 

The active learning samples are grouped into two clusters which are most distinguished in the $Y_e$ and $s$ dimensions. After performing our final nested sampling, we see that these clusters lie over two modes of a distinctly bimodal posterior. Our best-fit parameters for both modes, and associated uncertainties, are listed in Table~\ref{tab:bestfit}. We adopt as best-fit parameters and uncertainties the median and the 2.5\%, 97.5\% quantiles, and keep this convention for the remainder of this work. We separate the two modes using a cut at $s/k_{\mathrm{B}} = 18$, and then cuts of $Y_e \geqslant 0.325$ and $Y_e \leqslant 0.34$, on the samples. These modes are characterized by (1) a larger electron fraction and higher entropy (`blue + hot'), and, (2) a moderate electron fraction and moderate entropy (`purple + warm'). We discuss these modes below.

\subsubsection{Blue, hot model}\label{sssc:bluehot}

We present the posterior, zoomed in to the higher electron fraction and entropy blue+hot mode, in Figure~\ref{fig:corner_hotblue}. For this model, we infer a luminosity of $\log_{10}(L_\mathrm{outer}/L_{\odot}) = 7.784^{+0.016}_{-0.018}$, or equivalently, $L_{\mathrm{outer}} = 2.328^{+0.086}_{-0.096}~\times 10^{41}~\mathrm{erg~s^{-1}}$. Recalling that this outer boundary luminosity is an analog for the inner boundary temperature, we obtain a temperature of $T_{\mathrm{inner}} = 3962^{+102}_{-109}~\mathrm{K}$.  

We infer a density normalization constant of $\log_{10} (\rho_0 / \mathrm{g~cm^{-3}}) = -15.069^{+0.511}_{-0.409}$ or equivalently $\rho_0 = 8.531^{+10.038}_{-8.034}~\times 10^{-16}~\mathrm{g~cm^{-3}}$. For the inner boundary velocity (\ie, photospheric velocity), we find $v_{\mathrm{inner}}/c = 0.313^{+0.015}_{-0.016}$. Combining this density normalization, our inferred inner boundary velocity, and the fixed outer boundary velocity $v_{\mathrm{outer}} = 0.35c$, we find a total mass of $ 3.5^{+4.2}_{-3.3}~\times 10^{-5}~M_{\odot} = 11.8^{+13.8}_{-11.1}~M_{\oplus}$ is contained in our simulation. We emphasize that this is only the mass above the photosphere; most of the ejecta mass should be contained below the photosphere. Indeed, simulations predict a larger total ejecta mass of $10^{-3} - 10^{-1} M_{\odot}$ across a range of NS-NS mergers (\eg, \citealt{fernandez16, shibata19}), while inferences on the light curves suggest a mass of $\lesssim 0.02~M_{\odot}$ for an early, blue component of the GW170817 kilonova (\eg, \citealt{villar17} and references therein). Spectral modelling, in particular at early epochs when the photosphere lies in the outer regions of the total ejecta, cannot be used to infer the total ejecta mass. Nonetheless, we can place a conservative lower limit on the ejecta mass, $M_{\mathrm{ej}} > 3.5^{+4.2}_{-3.3}~\times 10^{-5}~M_{\odot} = 11.8^{+13.8}_{-11.1}~M_{\oplus}$. This mass estimate is dependent on the interplay between our inferred $\rho_0$ and $v_{\mathrm{inner}}$; different $(\rho_0$, $v_{\mathrm{inner}})$ may yield the same photospheric masses. To this end, the uncertainties on the mass estimate are computed by propagating the uncertainties on $\rho_0$ and $v_{\mathrm{inner}}$. We note that the uncertainty in mass is dominated by the much larger uncertainty in $\rho_0$, compared to $v_{\mathrm{inner}}$. Also, there is no evidence for any strong degeneracy between $\rho_0$ and $v_{\mathrm{inner}}$ in our posteriors.

This model is distinguished by the parameters which set the abundance pattern. We infer an electron fraction $Y_{\mathrm{e}} = 0.351^{+0.025}_{-0.025}$, an expansion velocity $v_{\mathrm{exp}}/c = 0.176^{+0.091}_{-0.099}$, and specific entropy $s = 25.3^{+6.0}_{-4.5}~k_{\mathrm{B}}/\mathrm{nucleon}$. These parameters describe an $r$-process with a relatively large electron fraction, which leads to bluer emission, and a higher (hotter) entropy. The expansion velocity is moderate, but relatively poorly constrained compared to the other parameters, considering our initial uniform prior of $v_{\mathrm{exp}}/c \in [0.05, 0.30]$. Indeed, the expansion velocity inferred here is consistent with that of the purple+warm model described in the following section.

We include the spectrum of this blue+hot model in Figure~\ref{fig:all_TARDIS_runs}. In all, the fit captures the general shape of the continuum, as well as the broad absorption feature at $\sim$ $8000$~\AA. It somewhat underestimates the expected emission at the red end of this feature, in the interpretation that this is a P Cygni feature (\eg, \citealt{watson19}). It struggles most with the spectrum at the blue end, $\lesssim$ $ 5000$~\AA.

\subsubsection{Purple, warm model}\label{sssc:purplewarm}

We show the posterior distribution at the purple+warm mode in Figure~\ref{fig:corner_purplewarm}. The inferred luminosity, density, and velocity, and thus the temperature and ejecta mass, are remarkably consistent between the blue+hot mode discussed above and this purple+warm mode. This model is distinct from the previously discussed model in its abundance-setting parameters: we find electron fraction $Y_{\mathrm{e}} = 0.311^{+0.013}_{-0.011}$, expansion velocity $v_{\mathrm{exp}}/c = 0.240^{+0.055}_{-0.082}$, and specific entropy $s = 13.6^{+4.1}_{-3.0}~k_{\mathrm{B}}/\mathrm{nucleon}$. This expansion velocity is faster than that of the blue+hot model--however, we reiterate that the expansion velocity is in general poorly constrained, and that this value is consistent with that of the blue+hot model. The electron fraction and entropy, in contrast, are distinctly lower than those of the previously discussed model. They describe a more moderate electron fraction (hence `purple' emission, being somewhat redder than `blue'), and a lower, but still substantial entropy (hence `warm'), given the ranges spanned by existing kilonova simulations (\eg, \citealt{kawaguchi20}.)  

The spectrum generated by this model is included in Figure~\ref{fig:all_TARDIS_runs}. In all, the quality of the fit is similar to that of the blue+hot model. The most notable difference is that the purple+warm slightly improves the fit to the blue end at $\lesssim$ $5000$~\AA. This is due to the differing abundance patterns of these two modes, which we discuss in the following section (\ref{ssc:bestfit-abunds}).

All parameters are systematically more tightly constrained in this model. This is not surprising given that the $Y_e$ peak of this mode is much sharper (see Figure~\ref{fig:corner}). This has important consequences on the inferred abundances and the features in the emergent spectra, as presented in Sections~\ref{ssc:bestfit-abunds} and~\ref{ssc:spectral_features}.

\subsection{Inferred abundance patterns}\label{ssc:bestfit-abunds}

\begin{figure*}[!ht]
    \includegraphics[width=0.98\textwidth]{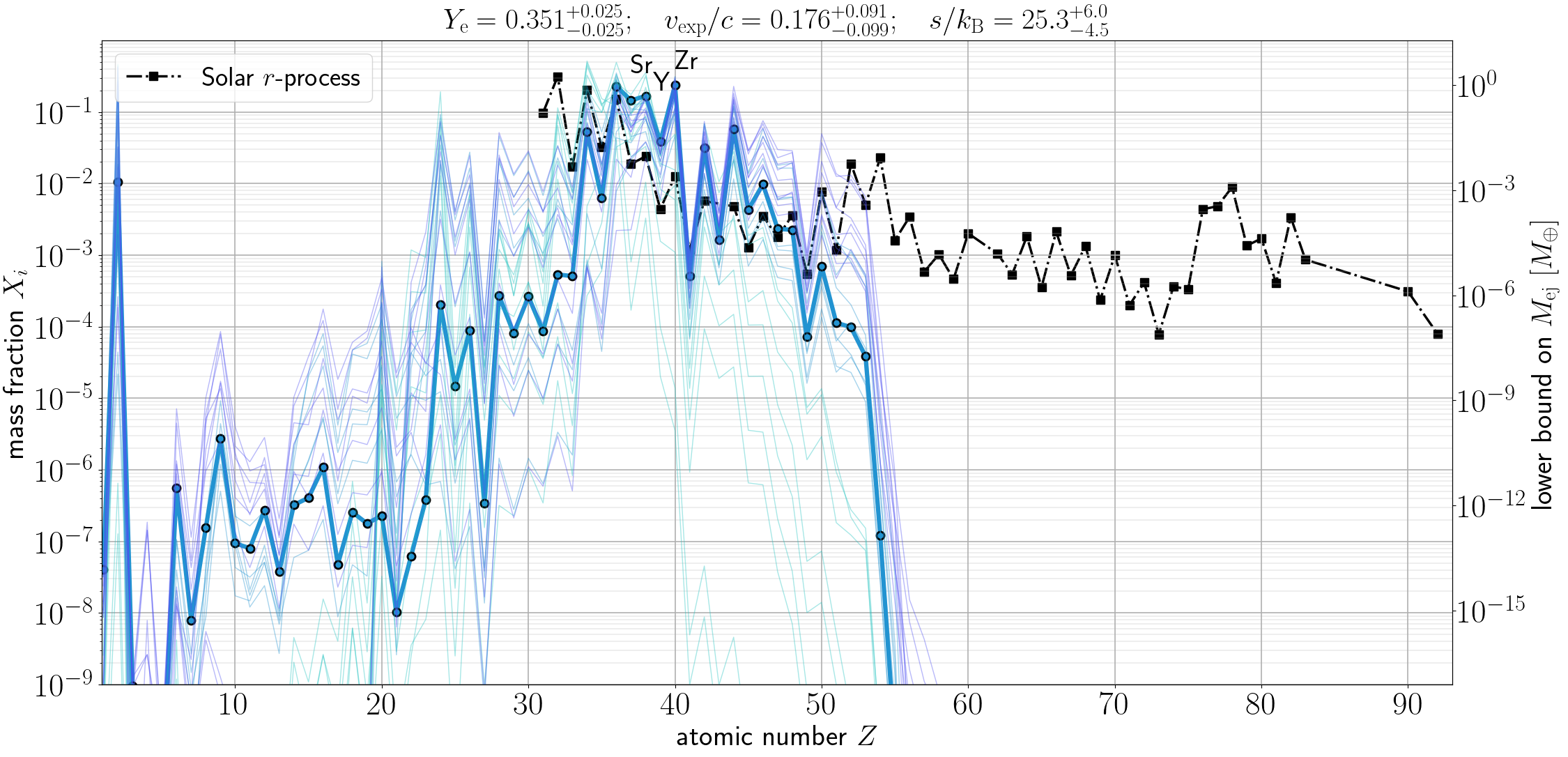}
    \includegraphics[width=0.98\textwidth]{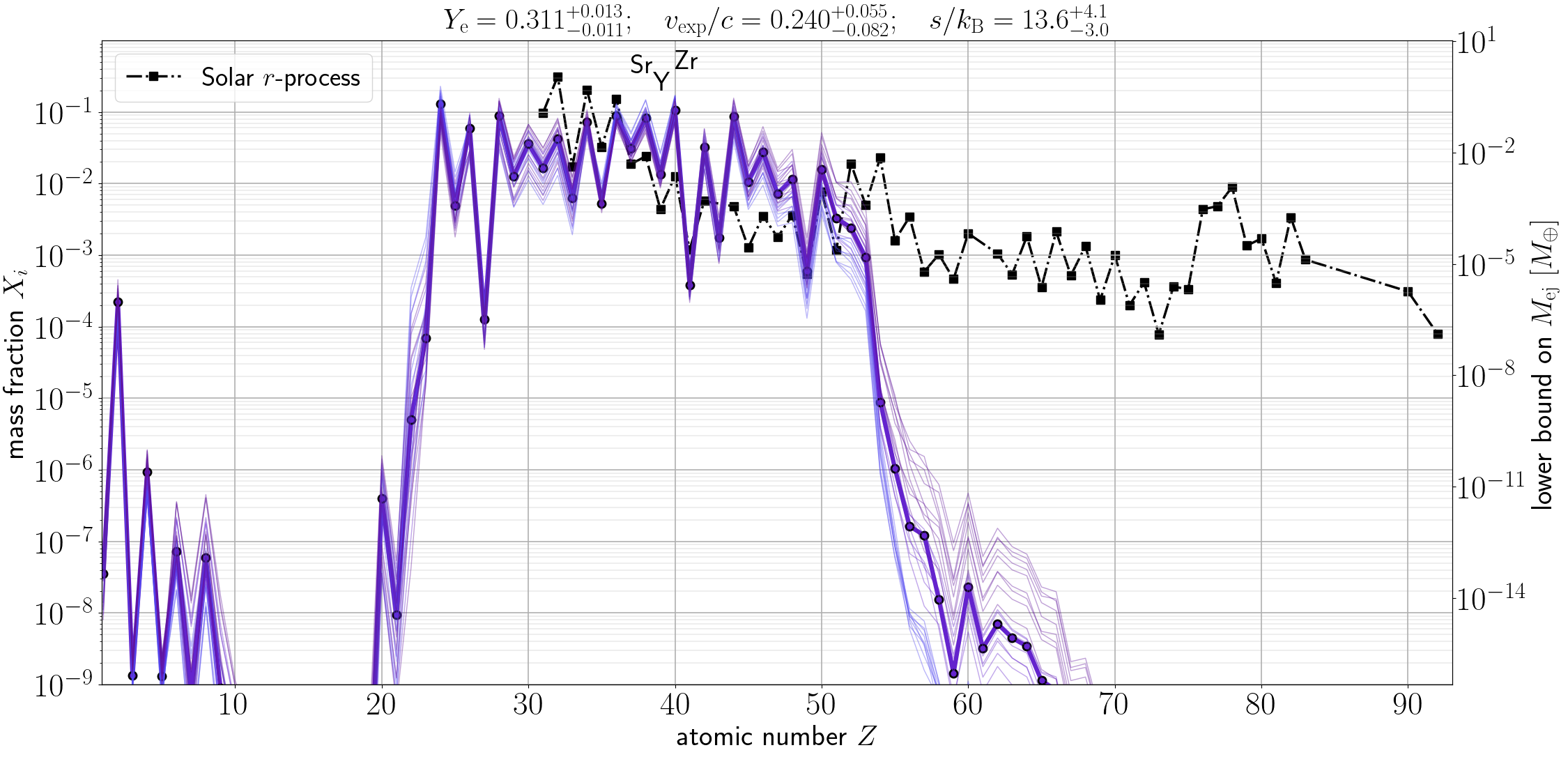}
    \figcaption{\textbf{Elemental mass fractions for the blue+hot and purple+warm best-fit models.} The blue+hot model (top) has electron fraction $Y_e = 0.351^{+0.025}_{-0.025}$, expansion velocity $v_{\mathrm{exp}}/c = 0.176^{+0.091}_{-0.099}$ and entropy $s = 25.3^{+6.0}_{-4.5}~k_{\mathrm{B}}/\mathrm{nucleon}$, while the purple+warm model (bottom) has $Y_e = 0.311^{+0.013}_{-0.011}$, $v_{\mathrm{exp}}/c = 0.240^{+0.055}_{-0.082}$, and $s = 13.6^{+4.1}_{-3.0}~k_{\mathrm{B}}/\mathrm{nucleon}$. The solid lines indicate the best fit, while the semi-translucent lines reflect how uncertainties in electron fraction, entropy, and expansion velocity of the ejecta lead to uncertainties in the abundances. The smaller scatter of lines in the purple+warm abundances reflects the fact that all fit parameters are more tightly constrained in this model. Lines are color-coded by their electron fraction, \ie, redder lines for a lower electron fraction and bluer for higher. Both abundance patterns are very lanthanide-poor: the blue+hot model has an upper bound of $X_{\mathrm{lan}} \leqslant 5.7~\times 10^{-9}$, but has a much smaller best-fit $X_{\mathrm{lan}} \sim 10^{-15}$. The purple+warm model marginally produces some lanthanides, but still has a low lanthanide fraction $X_{\mathrm{lan}} = 1.82^{+26.3}_{-1.79} \times 10^{-7}$. Also shown are the Solar $r$-process mass fractions for elements $Z\geqslant31$, obtained using the Solar System data of \cite{lodders09} subtracted by the $s$-process fractions of \cite{bisterzo14}. The best-fit abundance patterns are both clearly non-Solar. We also include a mass axis, in Earth masses, based on the inferred lower bound on the ejecta mass, $M_{\mathrm{ej}} > 3.5^{+4.2}_{-3.3}~\times 10^{-5}~M_{\odot} = 11.8^{+13.8}_{-11.1}~M_{\oplus}$ for the blue+hot model and $M_{\mathrm{ej}} > 4.0^{+2.9}_{-2.9}~\times 10^{-5}~M_{\odot} = 13.3^{+9.8}_{-9.7}~M_{\oplus}$ for the purple+warm. We highlight three elements of interest, strontium (${}_{38}$Sr), yttrium (${}_{39}$Y), and zirconium (${}_{40}$Zr), in both panels.}\label{fig:infer_abunds}
\end{figure*}

The two models discussed in the previous section are characterized by different abundance patterns due to their differences in electron fraction $Y_e$, expansion velocity $v_{\mathrm{exp}}$, and entropy $s$. We present the complete abundance patterns of both models for the inferred $Y_e$, $v_{\mathrm{exp}}$, $s$ in Figure~\ref{fig:infer_abunds}. The abundance patterns broadly have similar properties: they have large abundances at the first $r$-process peak and quickly drop off as we approach the lanthanides. However, the purple+warm model evidently has a larger abundance of the iron group elements ($Z \sim 26-30$) and of the lanthanides. Moreover, because the purple+warm model systematically imposes tighter constraints on all parameters, the scatter in the allowed abundance patterns is smaller. Indeed, the abundances of the first $r$-process peak elements strontium ${}_{38}$Sr, yttrium ${}_{39}$Y, and zirconium ${}_{40}$Zr, the importance of which is discussed in Section \ref{ssc:spectral_features}, are all constrained to within a factor of 2 in the purple+warm abundance pattern.

Both abundance patterns show a scarcity of lanthanides and heavier elements. We place a conservative upper limit $X_{\mathrm{lan}} \leqslant 5.6 \times 10^{-9}$ for the blue+hot model, but note that the best-fit lanthanide fraction for this model is substantially lower at $X_{\mathrm{lan}} \sim 10^{-15}$. For the purple+warm, the lanthanide fraction is better constrained, but still small: $X_{\mathrm{lan}} = 1.82^{+26.3}_{-1.79} \times 10^{-7}$. Notably, this lanthanide fraction is 5 orders of magnitude smaller than that of the Solar system ($X_{\mathrm{lan,\odot}} \approx 10^{-1.4}$; \citealt{ji19}). Beyond this single quantity, we can further compare our complete abundance pattern to that of the Solar system, as computed by \cite{lodders09} with the $s$-process subtraction of \cite{bisterzo14}. Both models' abundances are evidently non-Solar. This indicates that the early, blue component of the kilonova which is visible at $1.4$ days can not account for the Solar $r$-process abundances on its own. However, a later, redder component which is buried under the photosphere at these early times might be characterized by an abundance pattern which contains heavier elements and is closer to Solar. 

A subtle, but important difference between the blue+hot model and the purple+warm is the differences in the best-fit abundances of Sr, Y, and Zr. The blue+hot model has best-fit abundances of Sr, Y, Zr which are a factor of $2 \times$, $3 \times$, and $2 \times$ larger than the purple+warm, respectively, albeit the uncertainties on the blue+hot model's abundances are large. In the following section, we examine the impact of these elements on the emergent spectrum, and how these differences in abundances might lead to the observed differences at the blue end ($\lesssim 4500$~\AA) of the spectrum.

\begin{figure*}[!ht]
    \includegraphics[width=0.985\textwidth]{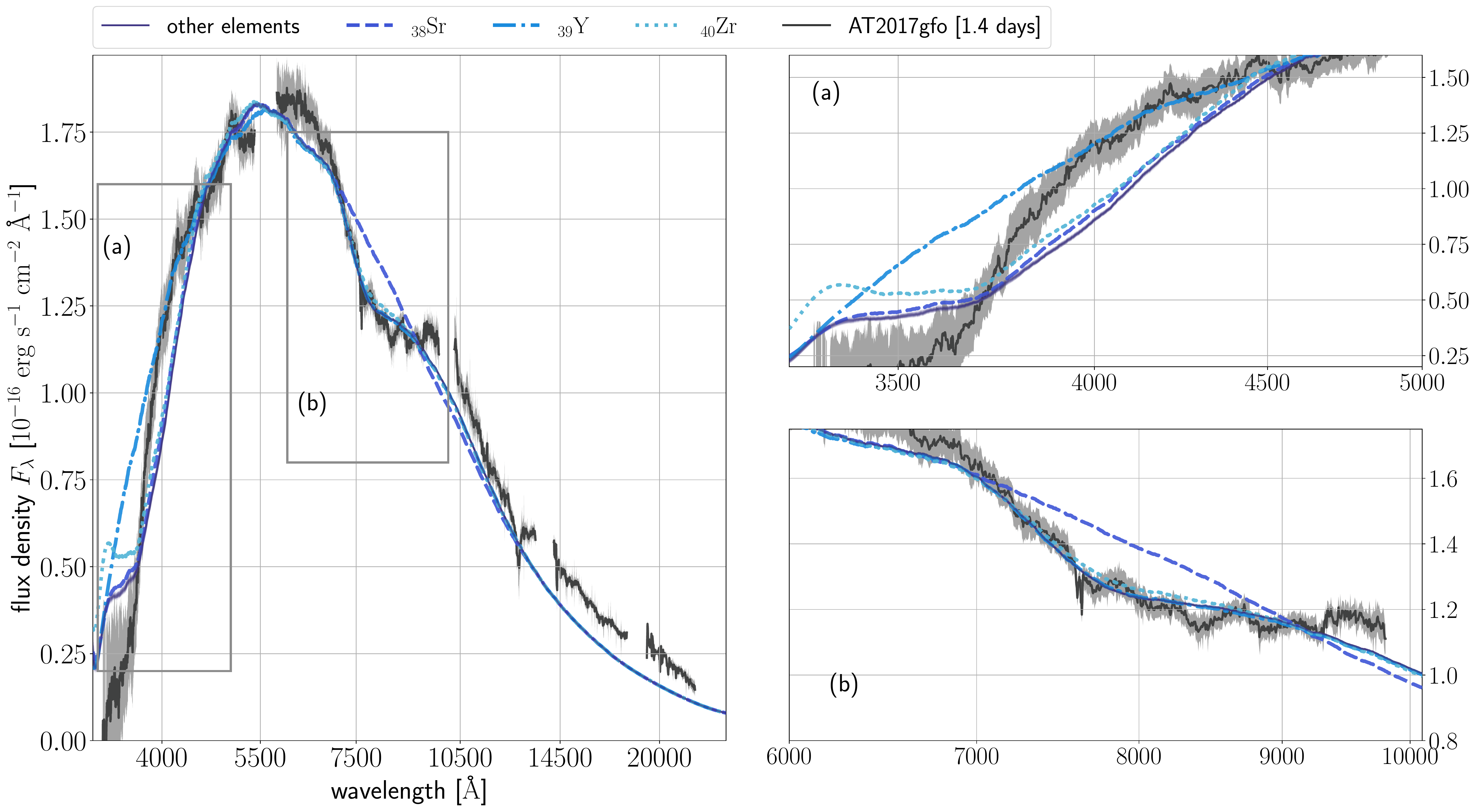}
    \figcaption{\textbf{Leave-one-out spectra, using the best-fit parameters from the blue+hot model}. The synthetic spectra demonstrate the effect of omitting a particular element by transferring its abundance to the filler element He (which is not expected to contribute to the emission significantly) and re-computing the radiative transfer. These leave-one-out spectra are then compared the observed spectrum. Only three elements contribute substantial absorption. We highlight the $3000 - 5000$~\AA~and $6000-10000$~\AA~regions in insets \textbf{(a)} and \textbf{(b)}. In inset \textbf{(b)}, at $\sim$8000~\AA, we see strong evidence that the absorption is produced by strontium ${}_{38}$Sr, as has been found in other works (\citealt{watson19, domoto21, gillanders22}). In inset \textbf{(a)}, we see tentative evidence for some combination of absorption at $\lesssim$4500~\AA~produced by yttrium ${}_{39}$Y, and absorption at $\lesssim$3600~\AA~produced by zirconium ${}_{40}$Zr. If these elements are responsible, the absorption from these elements is overestimated from $\sim$ $3600 - 4500$~\AA~and underestimated at $\lesssim$3600~\AA. While this region is near the edge of the X-shooter sensitivity, the systematic under- and over-estimations are significant.}\label{fig:leave_out_bluehot}
\end{figure*}

\begin{figure*}[!ht]
    \includegraphics[width=0.985\textwidth]{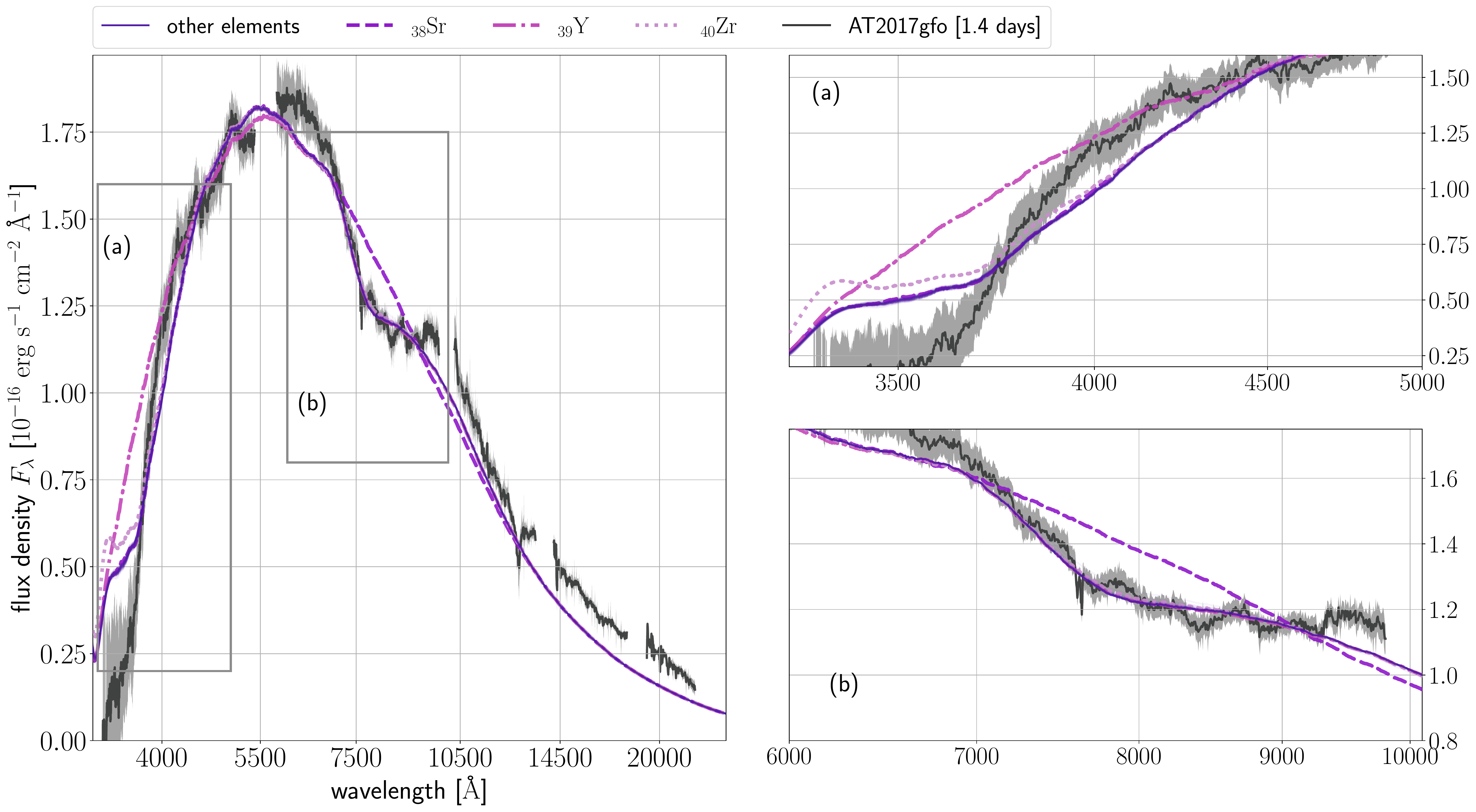}
    \figcaption{\textbf{Leave-one-out spectra for the purple+warm model}. As in Figure~\ref{fig:leave_out_bluehot}, we see strong evidence for absorption from strontium ${}_{38}$Sr at $\sim$8000~\AA~(inset \textbf{(b)}), and tentative evidence for absorption from some combination of yttrium ${}_{39}$Y and zirconium ${}_{40}$Zr at $\lesssim$4500~\AA~(inset \textbf{(a)}). The similarities between the leave-one-out spectra of the blue+hot model and the purple+warm model suggest that, at $1.4~\mathrm{days}$, the absorption in the spectrum is indeed dominated by these elements from the first peak of the $r$-process, and not others such as the iron group elements (present in large amounts) or the lanthanides (present in modest amounts) in the abundance pattern of the purple+warm model. Compared to the blue+hot model, the over- and under-estimation of the absorption from Y and Zr at $\lesssim$4500~\AA~is less severe.}\label{fig:leave_out_purplewarm}
\end{figure*}

\subsection{Spectral features}\label{ssc:spectral_features}

To assess the contribution of individual elements to the emergent spectrum, we produce `leave-one-out' spectra. Specifically, we take a single element and set its abundance to 0, replacing it by increasing the abundance of the filler element helium, an element which should not have a large impact on the emergent spectrum at ealytimes when the approximation of LTE is valid (\citealt{perego22}). For a given set of best-fit parameters, we then iteratively perform radiative transfer, leaving each element out, and determine which elements' presence substantially alters the best-fitting spectrum.

In Figure~\ref{fig:leave_out_bluehot}, we show the leave-one-out spectra for the blue+hot model. All elements with a relative abundance $>$ $10^{-15}$ are tested. At $\sim$8000~\AA, we see clear evidence that the absorption is due to the presence of strontium ${}_{38}$Sr. This agrees with other works (\citealt{watson19, domoto21, gillanders22}) which have similarly attributed this absorption to Sr. No other element contributes substantial absorption from $\sim$$7000 - 9000$~\AA.

At shorter wavelengths $\lesssim$4500~\AA, there is evidence for absorption from some combination of yttrium ${}_{40}$Y and zirconium ${}_{40}$Zr. In particular, Y contributes absorption over this entire range, while Zr contributes some smaller amount of absorption at $\lesssim$3600~\AA. However, this absorption is too strong from $\sim$ $3600 - 4500$~\AA, and too weak at $\lesssim$3600~\AA. While we note that this region of the spectrum is at the edge of the X-shooter spectrograph's sensitivity, the systematic over- and under-estimation are still significant. The reason for the overestimated absorption at $3600 - 4500$~\AA~may be that the abundance of Y is overestimated in this blue+hot model. For the underestimated absorption at $\lesssim$3600~\AA, this may be point to either an underestimation of the Zr in the model, or, incompleteness of the line lists for Y and/or Zr.

In Figure~\ref{fig:leave_out_purplewarm}, we see the leave-one-out spectra for the purple+warm model. Because the abundances of this model extend to heavier elements, we perform our leave-one-out analysis for a larger set of elements. Nonetheless, the elements which generate the prominent absorption features in the spectrum are still Sr, Y, and Zr. The $\sim$8000~\AA~absorption by Sr is once again recovered, as is the absorption by Y and Zr at shorter wavelengths. In contrast with the blue+hot model, the purple+warm is better able to capture the absorption from predominantly Y at $\sim$ $3600-4500$~\AA. The purple+warm model has best-fit abundances of Sr, Y, Zr which are a factor of $2 \times$, $3 \times$, and $2 \times$ smaller than the blue+hot. We thus see that in the purple+warm model, the smaller abundance of Y leads to less absorption at $\sim$ $3600 - 4500$~\AA~and a better fit. This reiterates the importance of this element and strengthens the association between this feature and this element. The absorption from Zr at $\lesssim 3600$~\AA~is also weaker in the purple+warm model due to its smaller abundance, but less noticeably so. 

We can further quantitatively verify the importance of the mentioned elements, and determine which particular ionization states are responsible for the absorption, using the Spectral element DEComposition (SDEC) tool available in \TARDIS. In this technique, at the end of a radiative transfer simulation, all escaping photon packets are labelled by the species they last interacted with (or, an electron scattering event or escaping without interaction). We can then assess which species dominate radiation-matter interactions and thus contribute the greatest absorption. In our purple+warm model, we find that photon packets interact predominantly with singly-ionized species, and we confirm the importance of Sr, Y, and Zr in the emergent spectrum. At $7000 - 9000$~\AA, 84\% of line interactions occur with \ion{Sr}{2}, 12\% with \ion{Y}{2}, and 4\% with \ion{Zr}{2}. At $\leqslant 4500$~\AA, 4\% of interactions occur with \ion{Sr}{2}, 29\% with \ion{Y}{2}, and 58\% with \ion{Zr}{2}. Singly-ionized iron group elements (chromium ${}_{24}$\ion{Cr}{2}, manganese ${}_{25}$\ion{Mn}{2}, iron ${}_{26}$\ion{Fe}{2}) contribute at these shorter wavelengths as well, but are sub-dominant, contributing at the level of 1\% - 4\%. All neutral or doubly-ionized species contribute $< 0.1\%$ to radiation-matter interactions, indicating that singly-ionized species dominate the opacity of the ejecta at early times. The statistics of the line interactions in the blue+hot spectrum are similar, affirming that singly-ionized Sr, Y, and Zr are responsible for the absorption in the model spectra.

In all, the association of \ion{Sr}{2} to the $\sim$8000~\AA~feature is strong, and in agreement with other works (\citealt{watson19,domoto21,gillanders22,domoto22}). The association with \ion{Y}{2} and \ion{Zr}{2} at $\lesssim$4500~\AA~is less secure due to our models' inability to adequately fit the spectrum at these wavelengths. We nevertheless tentatively associate these features with Y and Zr, for multiple reasons: (1) the effect of leaving out these elements, in particular Y, is substantial in both models, (2) no other elements contribute substantial absorption in this region of the spectrum, (3) the smaller abundances of Y and Zr in the purple+warm model compared to the blue+hot lead to less absorption, as would be expected if these elements were responsible, and (4) other works using different methods (\citealt{gillanders22, domoto22}) have similarly attributed these features to these elements.

\subsection{Computational  feasibility}\label{ssc:computing}

\begin{figure}[!ht]
    \centering
    \includegraphics[width=0.49\textwidth]{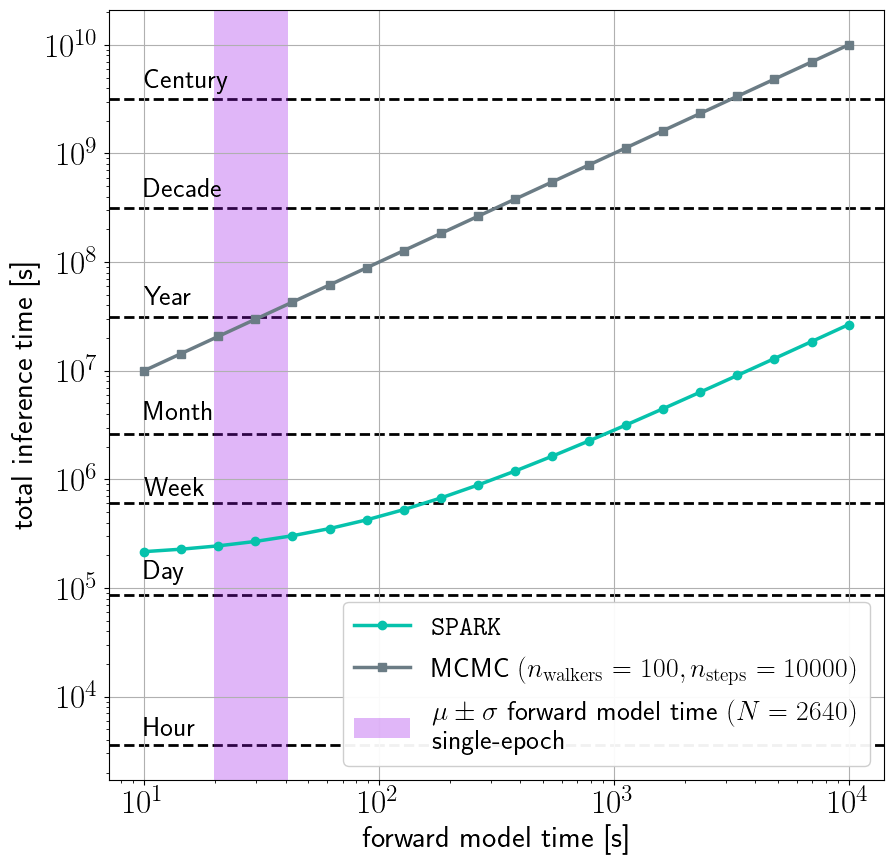}
    \figcaption{\textbf{Comparison of run times for the entire inference with \SPARK~versus an MCMC.} The purple band indicates the mean forward model run time of $31 \pm 10~\mathrm{s}$ to generate a single synthetic spectrum at a single epoch, averaged across all 2640 forward model evaluations. The variations in forward model time arise from two factors: (1) as the abundances in the ejecta vary, so do the opacities and the number of matter-radiation interactions, and (2) the inner edge of the computational grid, $v_{\mathrm{inner}}$, is itself a fit parameter. While inference with an MCMC of $n_{\mathrm{samples}}= 10^2 \times 10^4 = 10^6$ (reasonable given the 6D parameter space) would take as prohibitively long as a year, the time required by \SPARK~is more reasonable at $\lesssim$1~week.  We note that while an MCMC can be arbitrarily parallelized, the degree of parallelization needed to be competitive with \SPARK~is currently unreachable. Moreover, because \TARDIS~itself benefits from large parallelization, any increases in parallelization will also decrease the run time for \SPARK.}\label{fig:mcmc_compare}
\end{figure}

Beyond the goal of inferring the abundances and identifying spectral features, our goal in this work is also to make spectral retrieval of kilonovae computationally feasible. Despite the small set of just 2640 points in a 6D parameter space, \SPARK~converges to a good fit to the observed spectrum. In contrast, a MCMC approach to this 6D inference problem might require as many as $10^2$ walkers $\times~10^4$ steps $=10^6$ forward model evaluations, and this number could be larger for a complex, multi-modal posterior. This would be prohibitively computationally expensive. To demonstrate this, Figure~\ref{fig:mcmc_compare} presents a comparison between the total time to inference for \SPARK~versus a standard MCMC, as a function of the time required for a single forward model evaluation. Across all 2640 runs, we observe an average $\pm$ standard deviation forward model evaluation time of $31 \pm 10$ s.\footnote{All \TARDIS~runs are performed on a node of 48 cores, with $8~\mathrm{GiB} \sim 8.6~\mathrm{GB}$ memory per core. We refer the reader to \href{https://docs.alliancecan.ca/wiki/Narval/en}{https://docs.alliancecan.ca/wiki/Narval/en} for details of the nodes.} This average forward model time leads to a total inference time of $\lesssim$~1 week for \SPARK, compared to as long as a year for a standard MCMC. As our model increases in complexity and we incorporate both multi-epoch and multi-component inference into \SPARK, these forward model evaluation times will only increase and this discrepancy will become more severe. For example, a simple 2-epoch fit would double the forward model time by requiring twice as many radiative transfer simulations. 

We note that, while our initial $m_0$ samples can be obtained in parallel, our active learning approach is necessarily sequential. In contrast, MCMC can be arbitrarily parallelized---one could thus argue that an MCMC, if highly parallelized, would have a run time comparable to \SPARK. However, our method enables an almost $1000\times$ decrease in the number of forward model evaluations needed compared to an MCMC. Such a large amount of parallelization is currently impractical. Furthermore, since \TARDIS~itself is a MC radiative transfer code and can be arbitrarily parallelized, any increase in parallelization also accelerates \SPARK. We also note that while other sampling techniques such as nested sampling can increase the sampling efficiency and reduce the number of required forward model evaluations, this typically leads to a reduction by a factor of $\sim$10$\times$ at most. Thus, these faster sampling techniques are similarly slower than \SPARK.


\section{Discussion}\label{sec:disco}

\subsection{Interpretation of the bimodal posterior}\label{ssc:bimodal}

Spectral synthesis is a highly non-linear process, and thus it is not surprising that our inferred posterior might be complex or multi-modal. However, the physical meaning of this bimodality merits some discussion. We find that an $r$-process with higher electron fraction and entropy can generate a qualitatively similar spectrum to that of a moderate electron fraction and entropy. We further find that the differences between these two models' spectra arise primarily from differing abundances of just three elements: Sr, Y, and Zr. If the spectrum at this epoch is indeed insensitive to all but three elements, this may allow for some degeneracy in the abundance patterns which can adequately reproduce the observed spectrum. This degeneracy may be worsened by the fact that the elemental abundances obtained from reaction network calculations are highly non-linear functions of $Y_e$, $v_{\mathrm{exp}}$, and $s$ (\eg, \citealt{wanajo14, lippuner15, wanajo18}). At later epochs, when more elements (\eg, the lanthanides) may also contribute to the emergent spectrum, it may be possible to break these degeneracies.  

We caution that this bimodality should not be interpreted as evidence for multiple ejecta components. Our model is single-component and assumes a uniform abundance. This approach is likely most valid at early times, when the emission may be dominated by that of a single component (\eg, \citealt{kasen17}). Furthermore, the two models presented here yield similar spectra. A true multi-component analysis would require a fully multi-component model. Evidence for multiple components might also be stronger at later epochs, when the photosphere recedes into the ejecta, allowing the observer to potentially peer through the faster, bluer ejecta to see a slower, redder component.

\subsection{Consistency with other studies}\label{ssc:comparison}

Our inferred temperatures, velocities, and ejecta mass are broadly consistent with those of other works which have studied the spectra (and light curves) of the GW170817 kilonova, but we find some differences. 

Our inferred temperature of $T_{\mathrm{inner}} = 3960~\mathrm{K}$ (the same in both models) is in agreement with that of \cite{watson19}. It is somewhat cooler than that of \cite{gillanders22}, but we note that their spectrum has undergone a different calibration which raises the overall flux. We obtain a lower limit on the ejecta mass of $\gtrsim$$3 \times 10^{-5}~M_{\odot}$, which is well below the $\lesssim$$0.02 M_{\odot}$ inferred from photometric analyses for a putative early, blue component (\eg, \citealt{drout17, villar17} and references therein). Similarly, our inferred photospheric velocity of $v_{\mathrm{inner}} = 0.31c$ is broadly consistent with photometric analyses which inferred an ejecta velocity of $\sim$ $0.3c$ for this early, blue component (\citealt{villar17} and references therein), and consistent with the blackbody expansion velocity measured from the spectra in \cite{watson19}. It is slightly larger than the $v_{\mathrm{min}} = 0.28c$ in \cite{gillanders22}, but they note some arbitrariness in the choice of these and other parameters. Finally, our inferred lower bound on the ejecta mass is a factor $\sim$$6 \times$ and a factor $\sim$$2 \times$~smaller than the `$Y_e-0.37 \mathrm{a}$' and `$1^{\mathrm{st}}$~$r$-peak' models of \cite{gillanders22}, respectively. This is because (1) our inferred $\rho_0$ is smaller, and (2) our photosphere is at $v = 0.31c$, compared to their $v_{\mathrm{min}} = 0.28c$. 

We can also compare our inferred abundances to those of other works. In Tables~\ref{tab:ion_abunds_bluehot} and \ref{tab:ion_abunds_purplewarm} (Appendix~\ref{app:ion_abunds}), we list the mass fractions, lower bound on the ejecta mass, and mass assuming some fiducial total mass, for all elements in the ejecta. The larger uncertainties on our abundances in the blue+hot model preclude any useful comparison with those of other works. In contrast, the abundances in the purple+warm are tightly constrained. In particular, we infer lower bounds on the ejecta masses of the important ions \ion{Sr}{2}, \ion{Y}{2}, and \ion{Zr}{2} as $1.85^{+1.14}_{-0.98} \times 10^{-7}~M_{\odot}$, $2.35^{+1.14}_{-0.98} \times 10^{-7}~M_{\odot}$, and $2.45^{+1.14}_{-0.98} \times 10^{-6}~M_{\odot}$, respectively. For \ion{Sr}{2}, this mass is within a factor of $\sim$ $15 \%$ of the $Y_e -0.37$a model of \cite{gillanders22}, and within a factor of $2 \times$ the $1^{\mathrm{st}}$~$r$-peak model of the same work. The inferred mass of \ion{Y}{2} is similarly within a factor of $\sim$ $3 - 4 \times$ that of the $Y_e -0.37$a model, and the inferred mass of \ion{Zr}{2} is within a factor of $\sim$ $2 - 3 \times$ that of the $1^{\mathrm{st}}$~$r$-peak model. The agreement among these ions' masses reiterates their importance in shaping the emergent spectrum.

It is also useful to compute the abundance ratios of certain elements. \cite{domoto21} note that calcium ${}_{20}$Ca and Sr are co-produced under many $r$-process conditions due to their similar electronic structures. They find that \ion{Ca}{2} lines will appear in the spectrum unless $X_{\mathrm{Ca}} / X_{\mathrm{Sr}} \lesssim 0.002$, and note the absence of \ion{Ca}{2} lines in the observed kilonova spectrum. We do not see \ion{Ca}{2} absorption in our model spectra, and indeed we recover $X_{\mathrm{Ca}} / X_{\mathrm{Sr}} \sim 10^{-5}$ in both our blue+hot and purple+warm models. \cite{domoto21} further show that, for velocities $v/c \sim 0.15 - 0.20$, this small ratio $X_{\mathrm{Ca}} / X_{\mathrm{Sr}} < 0.002$ can be obtained with either (1) $s/k_{\mathrm{B}} \gtrsim 25$ and $Y_e \sim 0.35-0.45$, or (2) $s/k_{\mathrm{B}} \sim 10$ and $Y_e \lesssim 0.35$ (see their Figure 10). These constraints match our blue+hot and purple+warm models, respectively.

We emphasize that we are probing only the outermost $3 \times 10^{-5}~M_{\odot}$ of approximately $0.02~M_{\odot}$ of ejecta which was produced during the merger according to other studies, \ie, less than 1\% of the total ejecta. Thus, we do not claim that our inferred mass fractions (in particular the lanthanide mass fraction $X_{\mathrm{lan}}$), $Y_e$, $v_{\mathrm{exp}}$ or $s$ describe all of the ejecta produced during the merger. The inferred parameters describe the line-forming region of the ejecta which (remarkably, given its small mass) produces the prominent spectral features in the early, 1.4-day spectrum.

\subsection{Physical origin of the early emission}\label{ssc:physical_origin}

By directly inferring the electron fraction $Y_e$, expansion velocity $v_{\mathrm{exp}}$, and entropy $s$, we are able to constrain the fundamental conditions of the $r$-process which generate the early-time ejecta in the line-forming region. In particular, we can determine whether our inferred values match those of any of the various components expected from a merger. 

In the purple+warm model, we infer a moderate electron fraction of $Y_e = 0.311^{+0.013}_{-0.011}$ and moderate entropy of $s/k_{\mathrm{B}} = 13.6^{+4.1}_{-3.0}$. These parameters are comparable to those of the simulations of \cite{fujibayashi20}, who study the post-merger mass ejection from low-mass NS+NS systems (total mass $\sim 2.5 M_{\odot}$; compare to the total mass of $2.74 M_{\odot}$ inferred from GWs for GW170817; \citealt{abbott17c}) in full GRMHD with neutrino radiation and viscosity. They find a robust $Y_e \sim 0.32 - 0.34$ and $s/k_{\mathrm{B}} \sim 15-19$ in the viscously-driven outflows from the accretion disks which form around the merger remnant in their models. However, they find much smaller (mean) expansion velocities, $v/c \sim 0.09-0.11$, compared to our $v_{\mathrm{exp}}/c = 0.240^{+0.055}_{-0.082}$. This discrepancy might arise from the fact that the early spectra probe only the higher-velocity component of the velocity distribution, whereas the values reported in, \eg, \cite{fujibayashi20} are the averages over the entire (multi-component) distributions. Indeed, \cite{fahlman18} find that viscous hydrodynamic simulations of the disk outflows cannot produce ejecta with average velocities greater than $\sim 0.15c$. One solution to this discrepancy might be the presence of a strong magnetic field around a remnant NS (\citealt{metzger18, fujibayashi20}). Within a strong magnetic field, mass ejection is accelerated, raising the velocities in the ejecta. This also has the effect of lowering the electron fraction, as the faster ejecta suffer less neutrino irradiation from the remnant NS. Weaker neutrino heating could also result in smaller entropies in the ejecta. The inclusion of a strong magnetic field could thus raise the expansion velocity, lower the electron fraction, and lower the entropy of the models of \cite{fujibayashi20}, bringing all three of these quantities into closer agreement with our inferred values. \cite{ciolfi20a} and \cite{ciolfi20b} similarly argue that the relatively high velocity $v/c \gtrsim 0.20$ (and large mass $0.015 - 0.025 M_{\odot}$) of the early kilonova, reported across the literature, could be reproduced by a magnetically-accelerated wind from a metastable NS remnant. 

For the blue+hot model, the higher inferred entropy may be explained by significant shock heating in the ejecta. Indeed, the larger electron fraction $Y_e = 0.351^{+0.025}_{-0.025}$, expansion velocity $v_{\mathrm{exp}}/c = 0.176^{+0.091}_{-0.099}$, and high entropy $s/k_{\mathrm{B}} = 25.3^{+6.0}_{-4.5}$ of this blue+hot model are consistent with the shocked, polar dynamical ejecta which originates from the collisional interface of the two NSs during the merger, subject to strong neutrino heating. Including a proper treatment of neutrino irradiation, \cite{kullmann22a} and \cite{just22} find a strong angular dependence for $Y_e$ in the dynamical ejecta. In particular, across multiple NS equations of state and binary mass ratios, they observe electron fractions $Y_e \sim 0.28 - 0.34$ and velocities $v/c \sim 0.22 - 0.28$ in the polar ejecta, consistent with our blue+hot model. \footnote{These $Y_e$ and $v$ are also consistent with the purple+warm model; note, however, that the higher entropies in such shock-heated ejecta are inconsistent with the entropy of this model.} Furthermore, a smaller expansion velocity in their models would have the effect of further increasing neutrino irradiation on the dynamical ejecta, raising its electron fraction and entropy. This would bring these models into greater agreement with our blue+hot model. The need for a strong neutrino flux on the dynamical ejecta could also be satisfied by a short-lived NS remnant.

\subsection{Impact of incomplete atomic line lists}\label{ssc:incomplete_lines}

During spectral synthesis, we have conservatively used only the observed lines from VALD when constructing our line list. This excludes any semi-empirical lines acquired by calibrating theoretically calculated lines to observations as well as any purely theoretical lines. It is well-established that the observed line lists are incomplete, and this may be the reason for our difficulty in fitting the blue end of the spectrum.   

If the observed absorption at $\lesssim$4500~\AA~is indeed solely from \ion{Y}{2} and \ion{Zr}{2}, the fact that we do not adequately fit this portion of the spectrum may suggest that our line lists for these elements are incomplete. For example, missing transitions in the $\lesssim$3600~\AA~range might absorb radiation and re-emit in the $3600 - 4500$~\AA~range, simultaneously solving the problem of underestimated absorption in the former range and overestimated absorption in the latter. However, we note that \cite{gillanders22} observe the same difficulty with fitting this region of the spectrum, even after their inclusion of the semi-empirical extended Kurucz \texttt{ATOMS} lines (\citealt{kurucz18}) for the ions \mbox{Sr\,\textsc{i}--\textsc{iii}}, \mbox{Y\,\textsc{i}--\textsc{ii}} and \mbox{Zr\,\textsc{i}--\textsc{iii}}, which greatly outnumber the observed lines. It is thus possible that the lists are incomplete for other elements which are present in the ejecta. In particular, the lanthanides should have many lines in the UV and optical in this wavelength range (\eg, \citealt{tanaka20}), but very few observed lines exist for these elements. We infer the presence of very few lanthanides in our fits, but this may result from the incompleteness of the line lists for these elements. This reiterates the importance of obtaining complete line lists for these elements in the future. 

Finally, we also note that our fits do not fully capture the IR spectrum at $\gtrsim$10,000~\AA~(nor do those of \citealt{gillanders22}). Some of the emission at these wavelengths may be emission which would be re-processed from the bluer end of the spectrum if the line lists were more complete. Alternatively, the temperature of the the blackbody-like continuum which underlies the fits might be biased by their inability to fully capture the blue end. In either case, more complete line lists might remedy these issues.

\subsection{Alternative approaches to modelling}\label{ssc:alternative_approaches}

In \SPARK, we model the spectra of the kilonova, with the goal of inferring the elemental abundance patterns and identifying spectral features. This is distinct from performing inference on light curves, for two reasons: (1) modelling the spectra is crucial for inferring individual abundances, and (2) simulations which primarily yield light curves may be relatively computationally inexpensive, and other inference schemes may be used. Indeed, the 3D, time-dependent radiative transfer code \texttt{POSSIS} (\citealt{bulla19}) has been used extensively to model the light curves, and different inference techniques have been employed due to its greater speed compared to, \eg, \TARDIS. \cite{almualla21} use \texttt{POSSIS} to generate a grid of synthetic kilonova light curves. They then use a neural network to interpolate over this grid, constructing a surrogate model capable of producing synthetic light curves, and perform inference on the observed light curves. \cite{ristic22} similarly use \texttt{POSSIS} to generate a grid of light curves, but instead use adaptive learning to select the location of some of their simulations and construct their surrogate using a GP. Finally, \cite{lukosiute22} use three existing sets of spectra (\citealt{kasen17, dietrich20, anand21}; the latter two generated with \texttt{POSSIS}) and construct a conditional Variational Auto-Encoder (cVAE), \texttt{KilonovaNet}, which acts as a surrogate for these spectral datasets. They convolve the output of their spectral emulator with standard photometric filters to produce light curves and perform inference. 

These studies take advantage of the speed of \texttt{POSSIS} and inference techniques which are suited to this speed to study the light curves of the GW170817 kilonova and recover its macroscopic properties: total ejecta masses, velocities, geometries, and morphologies. However, \texttt{POSSIS} is faster than \TARDIS~because it takes as input a grid of wavelength- and time-dependent opacities, rather than computing expansion opacities (which depend on the number density of a given ion in the plasma) during the simulation itself. Thus, these studies of the light curves do not constrain the elemental abundance patterns (beyond the lanthanide fraction, \citealt{lukosiute22}), nor the fundamental conditions of the $r$-process.

In contrast, \SPARK~is not sensitive to some overall properties of the kilonova (such as total ejecta masses or observing angle), but can extract the element-by-element abundance patterns, constrain $Y_e$, $v_{\mathrm{exp}}$, and $s$ with uncertainties, and identify spectral features. This inference is only possible due to the direct connection between the abundances and opacities present in \TARDIS. These trade-offs highlight the crucial importance of performing inference on both light curves and spectra using different forward models, and applying different inference schemes depending on the ensuing computational cost of the forward model and the dimensionality of the problem. A variety of complementary approaches are required to fully characterize the kilonova, from the macroscopic to the microscopic. \SPARK~contributes uniquely in this context by performing the first inference of the elemental abundance pattern using a kilonova spectrum.


\section{Conclusions}\label{sec:conclusions}

We introduce \SPARK, a modular inference engine for spectral retrieval of kilonovae. We employ approximate posterior estimation with active learning to solve an inference problem which has been computationally intractable to date. Crucially, our inference approach allows us to estimate uncertainties in key kilonova parameters. With \SPARK, we are able to obtain the complete element-by-element abundance pattern of a kilonova, with uncertainties, given a single observed spectrum. 

The abundance pattern for the GW170817 kilonova in the early, optically-thick phase at $t=1.4$~days is exceptionally lanthanide-poor. The ejecta is dominated by elements from the first $r$-process peak, including strontium ${}_{38}$Sr, yttrium ${}_{39}$Y, and zirconium ${}_{40}$Zr. However, the observed spectrum is well-fit by two different models: a blue+hot model with higher electron fraction and entropy, and, a purple+warm model with more moderate electron fraction and entropy. 

Adoption of our purple+warm model would suggest that we observe the viscously-driven outflows from the remnant accretion disk of a neutron star + neutron star merger at these early times. Our relatively high inferred expansion velocity for this model may hint at the presence of a strong magnetic field around a remnant metastable neutron star, which would accelerate these outflows. Alternatively, the blue+hot model may be evidence for a highly shocked dynamical component which is squeezed at the collisional interface of the two neutron stars and preferentially ejected along the poles of the merger. A strong neutrino flux from a remnant metastable neutron star could support the high electron fraction and entropy of this component. Our purple+warm and blue+hot models thus both suggest the presence of a remnant metastable neutron star following the merger.

In addition to inferring the complete abundance pattern, we also use leave-one-out spectra to identify signatures of specific elements in the spectrum of the kilonova. We recover evidence for absorption from \ion{Sr}{2}, \ion{Y}{2}, and \ion{Zr}{2} in the ejecta. 

In this work, we have explored single-component ejecta and have studied only the 1.4-day spectrum of the GW170817 kilonova. In the future, we will adapt \SPARK~to handle multi-component, stratified ejecta, which will enable us to capture important effects like lanthanide curtaining, if present. Our inference approach also lends itself to performing Bayesian model comparison, which will allow us to quantitatively determine whether additional component(s) are needed to accurately model the kilonova. There is also information in the temporal evolution of the spectrum, and so in future work we will jointly fit multiple epochs of data by self-consistently evolving the relevant fit parameters. As the ejecta expands and the photosphere recedes deeper into the ejecta at later times, we may be able to identify different components in the ejecta. This naturally combines itself with multi-component analyses.

Given the modularity of our approach, it is easy to swap out reaction network calculations or atomic line lists for some other set of calculations or line lists. We will thus also explore the effect of incorporating theoretical lines (generated by \AUTOSTRUCTURE; \citealt{badnell16}) which have been calibrated to observations in future work. 

Beyond extending our analysis of the GW170817 kilonova, we also anticipate applying \SPARK~to any future kilonovae discovered by follow-up of gravitational wave sources, in the next LIGO-Virgo-KAGRA Observing Run 4 (O4) and beyond.

\acknowledgments

NV works in Tiohti{\'a}:ke / Mooniyang, also known as Montr{\'e}al, which lies on the unceded land of the Haudenosaunee and Anishinaabeg nations. This work made use of high-performance computing resources in Tiohti{\'a}:ke / Mooniyang and in Burnaby, British Columbia, the unceded land of the Coast Salish peoples, including the Tsleil-Waututh, Kwikwetlem, Squamish, and Musqueam nations. We acknowledge the ongoing struggle of Indigenous peoples on this land, and elsewhere on Turtle Island, and hope for a future marked by true reconciliation. 

This work made extensive use of the \href{https://docs.alliancecan.ca/wiki/Cedar}{\texttt{Cedar}} and \href{https://docs.alliancecan.ca/wiki/Narval/en}{\texttt{Narval}} clusters of the \href{https://alliancecan.ca/en}{Digital Research Alliance of Canada} at Simon Fraser University (with regional partner \href{https://www.westgrid.ca/}{WestGrid}) and the {\'E}cole de technologie sup{\'e}rieure, respectively. We thank the support staff of Calcul Qu{\'e}bec in particular for their assistance at various steps in this project. We also thank Nikolas Provatas, Victor Ionescu, and Bart Odelman.

We thank Shinya Wanajo for kindly sharing their reaction network calculations. We thank Jessica Birky and David Fleming for useful discussions on approximate Bayesian inference and the use of \href{https://dflemin3.github.io/approxposterior/index.html}{\approxposterior}.

This work has made use of the \href{http://vald.astro.uu.se/~vald/php/vald.php}{Vienna Atomic Line Database (VALD)}, operated at Uppsala University, the Institute of Astronomy RAS in Moscow, and the University of Vienna. We thank Nikolai Piskunov and Eric Stempels for help in obtaining the VALD data.

This research made use of \href{https://tardis-sn.github.io/tardis/index.html}{\TARDIS}, a community-developed software package for spectral synthesis in supernovae (\citealt{kerzendorf14}). The development of \TARDIS~received support from the Google Summer of Code initiative and from the European Space Agency (ESA)'s Summer of Code in Space program. \TARDIS~makes extensive use of \href{https://docs.astropy.org/en/stable/}{\texttt{astropy}}. We thank Andrew Fullard, Wolfgang Kerzendorf, and the entire \TARDIS~development team for their assistance and their commitment to the development and maintenance of the code. 

N.V. acknowledges funding from the Bob Wares Science Innovation Prospectors Fund and the Murata Family Fellowship. J.J.R.\ and D.H.\ acknowledge support from the Canada Research Chairs (CRC) program, the NSERC Discovery Grant program, the FRQNT Nouveaux Chercheurs Grant program, and the Canadian Institute for Advanced Research (CIFAR). J.J.R.\ acknowledges funding from the Canada Foundation for Innovation (CFI), and the Qu\'{e}bec Ministère de l’\'{E}conomie et de l’Innovation. M.R.D. acknowledges support from the NSERC through grant RGPIN-2019-06186, the Canada Research Chairs Program, the Canadian Institute for Advanced Research (CIFAR), and the Dunlap Institute at the University of Toronto. R.F. acknowledges support from NSERC Discovery Grant RGPIN-2022-03463. N.R.B acknowledges funding by STFC (UK) through the University of Strathclyde
UK APAP Network grant ST/V000863/1.

We thank the anonymous reviewer for their helpful feedback, questions, and comments, which have strengthened this study.
\newline

\software{
\href{https://dflemin3.github.io/approxposterior/index.html}{\approxposterior}: \cite{fleming18};
\href{https://docs.astropy.org/en/stable/}{\texttt{astropy}}: \cite{astropy18};
\href{https://cmasher.readthedocs.io/}{\texttt{cmasher}}: \cite{velden20};
\href{https://corner.readthedocs.io/en/latest/index.html}{\texttt{corner}}: \cite{foreman-mackey16};
\href{https://dynesty.readthedocs.io/en/latest/index.html}{\texttt{dynesty}}: 
\cite{speagle20};
\href{https://emcee.readthedocs.io/en/stable/}{\texttt{emcee}}: \cite{foreman-mackey13}; 
\href{https://george.readthedocs.io/en/latest/}{\texttt{george}}: \cite{ambikasaran15};
\href{https://tardis-sn.github.io/tardis/index.html}{\TARDIS}: \cite{kerzendorf14};
\href{https://johannesbuchner.github.io/UltraNest/}{\texttt{UltraNest}}: \cite{buchner21}} 

\clearpage

\bibliographystyle{apj}

\appendix{}

\section{Line list details}\label{app:linelist}

We provide a breakdown of our line list in Table~\ref{tab:linecounts}, which details the number of lines for each ion. We only list elements which are synthesized in part by the $r$-process ($Z \geqslant 31$), but note that we use 1,476,338 additional VALD lines for $Z \leqslant 30$, dominated by iron-group elements. For these elements with $Z \leqslant 30$, we acquire lines as available; for some doubly-ionized iron-group elements, no lines are present.

\section{Gaussian Process Hyperparameters}\label{app:GP_hparams}

\begin{figure}[!ht]
    \centering
    \includegraphics[width=0.47\textwidth]{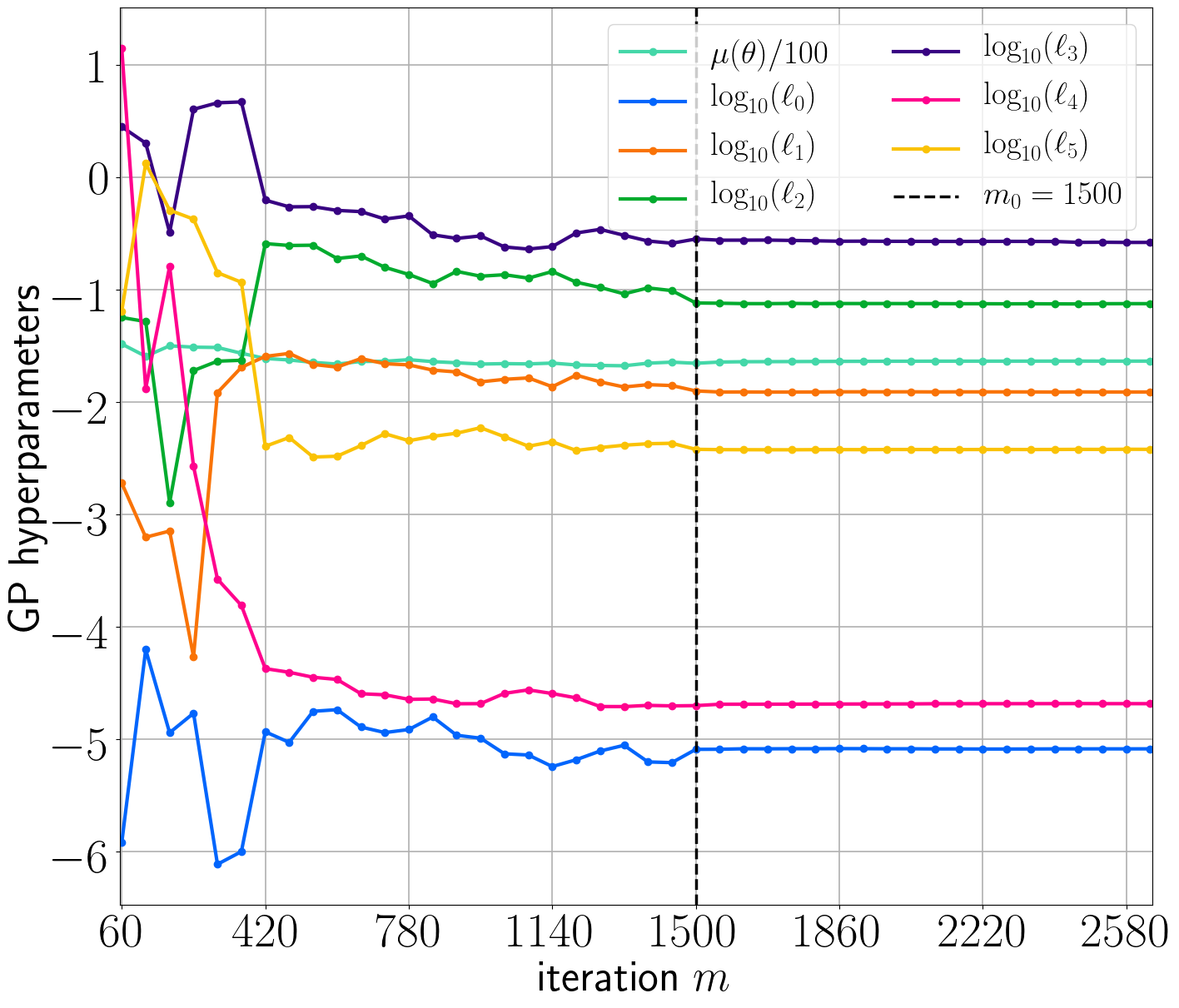}
    \figcaption{\textbf{Hyperparameters of the GP over the course of training.} Hyperparameters include the mean value and covariance kernel length scales of the GP, $\ell_j$, in each dimension $j$ of $\theta$-space. The mean value is normalized by $10^2$ for legibility. The hyperparameters rapidly converge and then remain stable during construction of the base training set, indicating that the optimal hyperparameters have been found. Once active learning begins at $m = 1500$, the hyperparameters remain stable.}\label{fig:hparams}
\end{figure}

To further assess the convergence of the \SPARK~run, we can plot the hyperparameters of the GP over the course of training. In Figure~\ref{fig:hparams}, we see this evolution for the mean $\mu(\theta)$ of the GP and the scale length hyperparameters $\ell_j$ in each dimension $j$ of $\theta$-space. This diagnostic is instructive both when constructing the base training set and during active learning. As with the GP test set error, during the construction of the base training set, the hyperparameters eventually converge to stable values, indicating that the GP has captured the global properties of the posterior. Once active learning begins at $m = 1500$, the hyperparameters remain stable.

\section{Detailed Ion Abundances}\label{app:ion_abunds}

In Tables~\ref{tab:ion_abunds_bluehot} and \ref{tab:ion_abunds_purplewarm}, we provide mass fractions and uncertainties for all of the relevant ions in the ejecta in our best-fit blue+hot and purple+warm models. The majority of elements are singly ionized. We also provide the total mass of each ion contained in our simulation, and, the mass of each ion under the assumption of some fiducial total ejecta mass of $0.02~M_{\odot}$ and uniform composition above and below the photosphere.

\startlongtable
\begin{deluxetable}{cc|c}
\tablewidth{\textwidth}
\centering
\tablecaption{Number of lines for each ion, for elements with $Z \geqslant 31$. For elements with $Z<31$, an additional 1,476,338 lines (dominated by the iron group) are used.}\label{tab:linecounts} 
\tablehead{\colhead{$Z$} & \colhead{ion} & \colhead{$N_{\mathrm{lines}}$}} 
\startdata
1-30 & H - Zn \ion{}{1}~-\ion{}{3}\tablenotemark{a} & 1,476,338 \\\hline
31 & \ion{Ga}{1} & 41 \\
 & \ion{Ga}{2} & 194 \\
 & \ion{Ga}{3} & 2 \\\hline
32 & \ion{Ge}{1} & 64 \\
 & \ion{Ge}{2} & 22 \\\hline
33 & \ion{As}{1} & 110 \\\hline
34 & \ion{Se}{1} & 8 \\\hline
37 & \ion{Rb}{1} & 143 \\\hline
38 & \ion{Sr}{1} & 110 \\
 & \ion{Sr}{2} & 695 \\\hline
39 & \ion{Y}{1} & 5,365 \\
 & \ion{Y}{2} & 7,753 \\
 & \ion{Y}{3} & 39 \\\hline
40 & \ion{Zr}{1} & 699 \\
 & \ion{Zr}{2} & 542 \\
 & \ion{Zr}{3} & 359 \\\hline
41 & \ion{Nb}{1} & 1,044 \\
 & \ion{Nb}{2} & 2,988 \\
 & \ion{Nb}{3} & 76 \\\hline
42 & \ion{Mo}{1} & 2,892 \\
 & \ion{Mo}{2} & 328 \\\hline
43 & \ion{Tc}{1} & 52 \\\hline
44 & \ion{Ru}{1} & 1,028 \\
 & \ion{Ru}{2} & 127  \\\hline
45 & \ion{Rh}{1} & 440 \\
 & \ion{Rh}{2} & 94 \\\hline
46 & \ion{Pd}{1} & 76 \\
 & \ion{Pd}{2} & 13 \\\hline
47 & \ion{Ag}{1} & 15 \\
 & \ion{Ag}{2} & 10 \\\hline
48 & \ion{Cd}{1} & 20  \\
 & \ion{Cd}{2} & 4 \\\hline
49 & \ion{In}{1} & 24 \\
 & \ion{In}{2} & 16 \\\hline
50 & \ion{Sn}{1} & 59 \\
 & \ion{Sn}{2} & 22 \\\hline
51 & \ion{Sb}{1} & 73 \\\hline
52 & \ion{Te}{1} & 18 \\\hline
54 & \ion{Xe}{2} & 33 \\\hline
55 & \ion{Cs}{1} & 150 \\\hline
56 & \ion{Ba}{1} & 125 \\
 & \ion{Ba}{2} & 244 \\\hline
57 & \ion{La}{1} & 267 \\
 & \ion{La}{2} & 3,938 \\
 & \ion{La}{3} & 131 \\\hline
58 & \ion{Ce}{1} & 903 \\
 & \ion{Ce}{2} & 16,014 \\
 & \ion{Ce}{3} & 3,023 \\\hline
59 & \ion{Pr}{1} & 127 \\
 & \ion{Pr}{2} & 7,723 \\
 & \ion{Pr}{3} & 1,151 \\\hline
60 & \ion{Nd}{1} & 281 \\
 & \ion{Nd}{2} & 1,279 \\
 & \ion{Nd}{3} & 71 \\\hline
62 & \ion{Sm}{1} & 520 \\
 & \ion{Sm}{2} & 1,334 \\
 & \ion{Sm}{3} & 49 \\\hline
63 & \ion{Eu}{1} & 352 \\
 & \ion{Eu}{2} & 871 \\
 & \ion{Eu}{3} & 1,150 \\\hline
64 & \ion{Gd}{1} & 620 \\
 & \ion{Gd}{2} & 963 \\
 & \ion{Gd}{3} & 52 \\\hline
65 & \ion{Tb}{1} & 3 \\
 & \ion{Tb}{2} & 1,822 \\
 & \ion{Tb}{3} & 80 \\\hline
66 & \ion{Dy}{1} & 834 \\
 & \ion{Dy}{2} & 897 \\
 & \ion{Dy}{3} & 1,337 \\\hline
67 & \ion{Ho}{1} & 711 \\
 & \ion{Ho}{2} & 496 \\
 & \ion{Ho}{3} & 1,309 \\\hline
68 & \ion{Er}{1} & 365 \\
 & \ion{Er}{2} & 775 \\
 & \ion{Er}{3} & 1,308 \\\hline
69 & \ion{Tm}{1} & 532 \\
 & \ion{Tm}{2} & 7,919 \\
 & \ion{Tm}{3} & 1,479 \\\hline
70 & \ion{Yb}{1} & 83 \\
 & \ion{Yb}{2} & 6,794  \\
 & \ion{Yb}{3} & 271 \\\hline
71 & \ion{Lu}{1} & 247 \\
 & \ion{Lu}{2} & 125 \\
 & \ion{Lu}{3} & 59 \\\hline
72 & \ion{Hf}{1} & 430 \\
 & \ion{Hf}{2} & 434 \\\hline
73 & \ion{Ta}{1} & 11,470 \\
 & \ion{Ta}{2} & 3,992 \\\hline
74 & \ion{W}{1} & 1,062 \\
 & \ion{W}{2} & 221 \\\hline
75 & \ion{Re}{1} & 772 \\
 & \ion{Re}{2} & 47 \\\hline
76 & \ion{Os}{1} & 891 \\
 & \ion{Os}{2} & 47 \\\hline
77 & \ion{Ir}{1} & 501 \\
 & \ion{Ir}{2} & 28 \\\hline
78 & \ion{Pt}{1} & 215 \\
 & \ion{Pt}{2} & 119 \\
 & \ion{Pt}{3} & 666 \\\hline
79 & \ion{Au}{1} & 61 \\
 & \ion{Au}{2} & 498 \\
 & \ion{Au}{3} & 175 \\\hline
80 & \ion{Hg}{1} & 27 \\
 & \ion{Hg}{2} & 446 \\
 & \ion{Hg}{3} & 42 \\\hline
81 & \ion{Tl}{1} & 22 \\\hline
82 & \ion{Pb}{1} & 38 \\
 & \ion{Pb}{2} & 60 \\\hline
83 & \ion{Bi}{1} & 29 \\
 & \ion{Bi}{2} & 14 \\\hline
90 & \ion{Th}{1} & 700 \\
 & \ion{Th}{2} & 1,448 \\
 & \ion{Th}{3} & 903 \\\hline
92 & \ion{U}{1} & 561 \\
 & \ion{U}{2} & 622 \\\hline
 & \textbf{TOTAL} & \textbf{1,597,376}  \\
\enddata
\tablenotetext{a}{Lines acquired as available: for some doubly-ionized iron-group elements, lines are not available.}
\end{deluxetable}

\startlongtable
\begin{deluxetable*}{cc|c|c|c}
\tablewidth{\textwidth}
\centering
\tablecaption{Mass fractions and masses for all ions in the blue+hot model with a mass fraction $\mathrm{X}_i \geqslant 10^{-9}$. Columns include the atomic number $Z$, the name of the ion, the mass fraction $\mathrm{X}_i$ of the ion, and the mass of the ion in our simulation, using our inferred lower bound on the ejecta mass $M_{\mathrm{ej}} > 3.5^{+4.2}_{-3.3} \times 10^{-5}~M_{\odot}$. We also include the mass of the ion assuming some fiducial total ejecta mass of $M_{\mathrm{ej}} = 0.02~M_{\odot}$ and a uniform composition above and below the photosphere. We highlight ions of interest: \ion{Sr}{2}, \ion{Y}{2}, and \ion{Zr}{2}. }\label{tab:ion_abunds_bluehot} 
\tablehead{\colhead{$Z$} & \colhead{ion} & \colhead{mass fraction $\mathrm{X}_i$} & \colhead{lower bound $M_{\mathrm{ej,i}}~[M_{\odot}]$} & \colhead{fiducial $M_{0.02,\mathrm{i}}$} } 
\startdata
1 & \ion{H}{1} & ${5.01}^{+0.04}_{-1.00} \times 10^{-5}$ & ${1.75}^{+1.20}_{-1.37} \times 10^{-9}$ & ${1.00}^{+0.04}_{-1.00} \times 10^{-6}$ \\
 & \ion{H}{2} & ${4.75}^{+0.04}_{-1.00} \times 10^{-7}$ & ${1.66}^{+1.20}_{-1.37} \times 10^{-11}$ & ${9.51}^{+0.04}_{-1.00} \times 10^{-9}$ \\\hline
2 & \ion{He}{1} & ${8.36}^{+4.03}_{-1.00} \times 10^{-1}$ & ${2.93}^{+4.21}_{-1.37} \times 10^{-5}$ & ${1.67}^{+4.03}_{-1.00} \times 10^{-2}$ \\\hline
3 & \ion{Li}{2} & ${2.49}^{+0.55}_{-1.00} \times 10^{-8}$ & ${8.72}^{+1.32}_{-1.37} \times 10^{-13}$ & ${4.98}^{+0.55}_{-1.00} \times 10^{-10}$ \\\hline
4 & \ion{Be}{2} & ${8.53}^{+0.55}_{-1.00} \times 10^{-9}$ & ${2.98}^{+1.32}_{-1.37} \times 10^{-13}$ & ${1.71}^{+0.55}_{-1.00} \times 10^{-10}$ \\\hline
6 & \ion{C}{1} & ${3.49}^{+356.60}_{-1.00} \times 10^{-7}$ & ${1.22}^{+356.60}_{-1.37} \times 10^{-11}$ & ${6.97}^{+356.60}_{-1.00} \times 10^{-9}$ \\
 & \ion{C}{2} & ${4.63}^{+356.60}_{-1.00} \times 10^{-6}$ & ${1.62}^{+356.60}_{-1.37} \times 10^{-10}$ & ${9.26}^{+356.60}_{-1.00} \times 10^{-8}$ \\\hline
7 & \ion{N}{1} & ${5.16}^{+356.60}_{-1.00} \times 10^{-8}$ & ${1.81}^{+356.60}_{-1.37} \times 10^{-12}$ & ${1.03}^{+356.60}_{-1.00} \times 10^{-9}$ \\\hline
8 & \ion{O}{1} & ${7.72}^{+12.24}_{-1.00} \times 10^{-7}$ & ${2.70}^{+12.30}_{-1.37} \times 10^{-11}$ & ${1.54}^{+12.24}_{-1.00} \times 10^{-8}$ \\
 & \ion{O}{2} & ${6.30}^{+12.24}_{-1.00} \times 10^{-9}$ & ${2.20}^{+12.30}_{-1.37} \times 10^{-13}$ & ${1.26}^{+12.24}_{-1.00} \times 10^{-10}$ \\\hline
9 & \ion{F}{1} & ${9.59}^{+14.04}_{-1.00} \times 10^{-6}$ & ${3.36}^{+14.09}_{-1.37} \times 10^{-10}$ & ${1.92}^{+14.04}_{-1.00} \times 10^{-7}$ \\\hline
10 & \ion{Ne}{1} & ${2.94}^{+14.04}_{-1.00} \times 10^{-7}$ & ${1.03}^{+14.09}_{-1.37} \times 10^{-11}$ & ${5.88}^{+14.04}_{-1.00} \times 10^{-9}$ \\\hline
11 & \ion{Na}{2} & ${1.93}^{+33.94}_{-1.00} \times 10^{-7}$ & ${6.76}^{+33.96}_{-1.37} \times 10^{-12}$ & ${3.86}^{+33.94}_{-1.00} \times 10^{-9}$ \\\hline
12 & \ion{Mg}{2} & ${5.89}^{+33.94}_{-1.00} \times 10^{-7}$ & ${2.06}^{+33.96}_{-1.37} \times 10^{-11}$ & ${1.18}^{+33.94}_{-1.00} \times 10^{-8}$ \\\hline
13 & \ion{Al}{2} & ${6.60}^{+12.79}_{-1.00} \times 10^{-8}$ & ${2.31}^{+12.85}_{-1.37} \times 10^{-12}$ & ${1.32}^{+12.79}_{-1.00} \times 10^{-9}$ \\\hline
14 & \ion{Si}{2} & ${5.21}^{+12.79}_{-1.00} \times 10^{-7}$ & ${1.82}^{+12.85}_{-1.37} \times 10^{-11}$ & ${1.04}^{+12.79}_{-1.00} \times 10^{-8}$ \\\hline
15 & \ion{P}{1} & ${1.34}^{+12.69}_{-1.00} \times 10^{-9}$ & ${4.68}^{+12.74}_{-1.37} \times 10^{-14}$ & ${2.68}^{+12.69}_{-1.00} \times 10^{-11}$ \\
 & \ion{P}{2} & ${5.39}^{+6.06}_{-1.00} \times 10^{-7}$ & ${1.89}^{+6.18}_{-1.37} \times 10^{-11}$ & ${1.08}^{+6.06}_{-1.00} \times 10^{-8}$ \\\hline
16 & \ion{S}{1} & ${1.07}^{+6.06}_{-1.00} \times 10^{-8}$ & ${3.74}^{+6.18}_{-1.37} \times 10^{-13}$ & ${2.14}^{+6.06}_{-1.00} \times 10^{-10}$ \\
 & \ion{S}{2} & ${1.34}^{+3.52}_{-1.00} \times 10^{-6}$ & ${4.70}^{+3.72}_{-1.37} \times 10^{-11}$ & ${2.68}^{+3.52}_{-1.00} \times 10^{-8}$ \\\hline
17 & \ion{Cl}{1} & ${4.09}^{+3.52}_{-1.00} \times 10^{-8}$ & ${1.43}^{+3.72}_{-1.37} \times 10^{-12}$ & ${8.17}^{+3.52}_{-1.00} \times 10^{-10}$ \\
 & \ion{Cl}{2} & ${7.55}^{+3.52}_{-1.00} \times 10^{-9}$ & ${2.64}^{+3.72}_{-1.37} \times 10^{-13}$ & ${1.51}^{+3.52}_{-1.00} \times 10^{-10}$ \\\hline
18 & \ion{Ar}{1} & ${2.05}^{+1.41}_{-1.00} \times 10^{-7}$ & ${7.16}^{+1.85}_{-1.37} \times 10^{-12}$ & ${4.09}^{+1.41}_{-1.00} \times 10^{-9}$ \\\hline
19 & \ion{K}{2} & ${1.48}^{+1.41}_{-1.00} \times 10^{-7}$ & ${5.17}^{+1.85}_{-1.37} \times 10^{-12}$ & ${2.96}^{+1.41}_{-1.00} \times 10^{-9}$ \\\hline
20 & \ion{Ca}{2} & ${7.18}^{+1.41}_{-1.00} \times 10^{-8}$ & ${2.51}^{+1.85}_{-1.37} \times 10^{-12}$ & ${1.44}^{+1.41}_{-1.00} \times 10^{-9}$ \\
 & \ion{Ca}{3} & ${1.10}^{+5.15}_{-1.00} \times 10^{-7}$ & ${3.86}^{+5.29}_{-1.37} \times 10^{-12}$ & ${2.21}^{+5.15}_{-1.00} \times 10^{-9}$ \\\hline
21 & \ion{Sc}{2} & ${6.07}^{+5.15}_{-1.00} \times 10^{-9}$ & ${2.13}^{+5.29}_{-1.37} \times 10^{-13}$ & ${1.21}^{+5.15}_{-1.00} \times 10^{-10}$ \\\hline
22 & \ion{Ti}{2} & ${3.45}^{+5.15}_{-1.00} \times 10^{-8}$ & ${1.21}^{+5.29}_{-1.37} \times 10^{-12}$ & ${6.89}^{+5.15}_{-1.00} \times 10^{-10}$ \\\hline
23 & \ion{V}{2} & ${1.88}^{+17.02}_{-1.00} \times 10^{-7}$ & ${6.59}^{+17.06}_{-1.37} \times 10^{-12}$ & ${3.76}^{+17.02}_{-1.00} \times 10^{-9}$ \\\hline
24 & \ion{Cr}{2} & ${9.65}^{+17.02}_{-1.00} \times 10^{-5}$ & ${3.38}^{+17.06}_{-1.37} \times 10^{-9}$ & ${1.93}^{+17.02}_{-1.00} \times 10^{-6}$ \\
 & \ion{Cr}{3} & ${1.16}^{+20.41}_{-1.00} \times 10^{-9}$ & ${4.04}^{+20.45}_{-1.37} \times 10^{-14}$ & ${2.31}^{+20.41}_{-1.00} \times 10^{-11}$ \\\hline
25 & \ion{Mn}{2} & ${6.20}^{+20.41}_{-1.00} \times 10^{-6}$ & ${2.17}^{+20.45}_{-1.37} \times 10^{-10}$ & ${1.24}^{+20.41}_{-1.00} \times 10^{-7}$ \\\hline
26 & \ion{Fe}{2} & ${3.66}^{+46.99}_{-1.00} \times 10^{-5}$ & ${1.28}^{+47.01}_{-1.37} \times 10^{-9}$ & ${7.32}^{+46.99}_{-1.00} \times 10^{-7}$ \\\hline
27 & \ion{Co}{2} & ${1.24}^{+46.99}_{-1.00} \times 10^{-7}$ & ${4.35}^{+47.01}_{-1.37} \times 10^{-12}$ & ${2.48}^{+46.99}_{-1.00} \times 10^{-9}$ \\\hline
28 & \ion{Ni}{2} & ${1.00}^{+34.30}_{-1.00} \times 10^{-4}$ & ${3.51}^{+34.32}_{-1.37} \times 10^{-9}$ & ${2.00}^{+34.30}_{-1.00} \times 10^{-6}$ \\\hline
29 & \ion{Cu}{2} & ${2.54}^{+34.30}_{-1.00} \times 10^{-5}$ & ${8.90}^{+34.32}_{-1.37} \times 10^{-10}$ & ${5.09}^{+34.30}_{-1.00} \times 10^{-7}$ \\\hline
30 & \ion{Zn}{1} & ${7.42}^{+73.25}_{-1.00} \times 10^{-9}$ & ${2.60}^{+73.26}_{-1.37} \times 10^{-13}$ & ${1.48}^{+73.25}_{-1.00} \times 10^{-10}$ \\
 & \ion{Zn}{2} & ${7.86}^{+3698.81}_{-0.86} \times 10^{-5}$ & ${2.75}^{+3698.81}_{-1.28} \times 10^{-9}$ & ${1.57}^{+3698.81}_{-0.86} \times 10^{-6}$ \\\hline
31 & \ion{Ga}{2} & ${2.27}^{+3698.81}_{-0.86} \times 10^{-5}$ & ${7.93}^{+3698.81}_{-1.28} \times 10^{-10}$ & ${4.53}^{+3698.81}_{-0.86} \times 10^{-7}$ \\\hline
32 & \ion{Ge}{2} & ${1.30}^{+3698.81}_{-0.86} \times 10^{-4}$ & ${4.54}^{+3698.81}_{-1.28} \times 10^{-9}$ & ${2.59}^{+3698.81}_{-0.86} \times 10^{-6}$ \\\hline
33 & \ion{As}{1} & ${2.95}^{+1118.47}_{-1.00} \times 10^{-7}$ & ${1.03}^{+1118.47}_{-1.37} \times 10^{-11}$ & ${5.91}^{+1118.47}_{-1.00} \times 10^{-9}$ \\
 & \ion{As}{2} & ${1.15}^{+1118.47}_{-1.00} \times 10^{-4}$ & ${4.03}^{+1118.47}_{-1.37} \times 10^{-9}$ & ${2.30}^{+1118.47}_{-1.00} \times 10^{-6}$ \\\hline
34 & \ion{Se}{1} & ${1.01}^{+1118.47}_{-1.00} \times 10^{-5}$ & ${3.53}^{+1118.47}_{-1.37} \times 10^{-10}$ & ${2.02}^{+1118.47}_{-1.00} \times 10^{-7}$ \\
& \ion{Se}{2} & ${1.07}^{+18627.73}_{-0.98} \times 10^{-2}$ & ${3.76}^{+18627.73}_{-1.36} \times 10^{-7}$ & ${2.15}^{+18627.73}_{-0.98} \times 10^{-4}$ \\\hline
35 & \ion{Br}{1} & ${2.14}^{+18627.73}_{-0.98} \times 10^{-4}$ & ${7.48}^{+18627.73}_{-1.36} \times 10^{-9}$ & ${4.28}^{+18627.73}_{-0.98} \times 10^{-6}$ \\
& \ion{Br}{2} & ${1.04}^{+18627.73}_{-0.98} \times 10^{-3}$ & ${3.64}^{+18627.73}_{-1.36} \times 10^{-8}$ & ${2.08}^{+18627.73}_{-0.98} \times 10^{-5}$ \\\hline
36 & \ion{Kr}{1} & ${4.06}^{+5859.54}_{-0.99} \times 10^{-2}$ & ${1.42}^{+5859.54}_{-1.37} \times 10^{-6}$ & ${8.12}^{+5859.54}_{-0.99} \times 10^{-4}$ \\
& \ion{Kr}{2} & ${9.31}^{+5859.54}_{-0.99} \times 10^{-4}$ & ${3.26}^{+5859.54}_{-1.37} \times 10^{-8}$ & ${1.86}^{+5859.54}_{-0.99} \times 10^{-5}$ \\\hline
37 & \ion{Rb}{2} & ${2.53}^{+5859.54}_{-0.99} \times 10^{-2}$ & ${8.84}^{+5859.54}_{-1.37} \times 10^{-7}$ & ${5.05}^{+5859.54}_{-0.99} \times 10^{-4}$ \\\hline
38 & \textbf{\ion{Sr}{2}} & $\mathbf{{1.35}^{+920.70}_{-1.00} \times 10^{-3}}$ & $\mathbf{{4.72}^{+920.70}_{-1.37} \times 10^{-8}}$ & $\mathbf{{2.70}^{+920.70}_{-1.00} \times 10^{-5}}$ \\
& \ion{Sr}{3} & ${2.65}^{+920.70}_{-1.00} \times 10^{-2}$ & ${9.29}^{+920.70}_{-1.37} \times 10^{-7}$ & ${5.31}^{+920.70}_{-1.00} \times 10^{-4}$ \\\hline
39 & \textbf{\ion{Y}{2}} & $\mathbf{{3.55}^{+920.70}_{-1.00} \times 10^{-3}}$ & $\mathbf{{1.24}^{+920.70}_{-1.37} \times 10^{-7}}$ & $\mathbf{{7.10}^{+920.70}_{-1.00} \times 10^{-5}}$ \\
& \ion{Y}{3} & ${2.68}^{+298.76}_{-1.00} \times 10^{-3}$ & ${9.37}^{+298.76}_{-1.37} \times 10^{-8}$ & ${5.35}^{+298.76}_{-1.00} \times 10^{-5}$ \\\hline
40 & \ion{Zr}{1} & ${1.23}^{+298.76}_{-1.00} \times 10^{-9}$ & ${4.31}^{+298.76}_{-1.37} \times 10^{-14}$ & ${2.46}^{+298.76}_{-1.00} \times 10^{-11}$ \\
& \textbf{\ion{Zr}{2}} & $\mathbf{ {3.50}^{+298.76}_{-1.00} \times 10^{-2} }$ & $\mathbf{{1.23}^{+298.76}_{-1.37} \times 10^{-6}}$ & $\mathbf{ {7.01}^{+298.76}_{-1.00} \times 10^{-4} }$ \\
& \ion{Zr}{3} & ${1.24}^{+318.33}_{-1.00} \times 10^{-3}$ & ${4.33}^{+318.34}_{-1.37} \times 10^{-8}$ & ${2.48}^{+318.33}_{-1.00} \times 10^{-5}$ \\\hline
41 & \ion{Nb}{2} & ${7.49}^{+318.33}_{-1.00} \times 10^{-5}$ & ${2.62}^{+318.34}_{-1.37} \times 10^{-9}$ & ${1.50}^{+318.33}_{-1.00} \times 10^{-6}$ \\
& \ion{Nb}{3} & ${5.13}^{+318.33}_{-1.00} \times 10^{-8}$ & ${1.80}^{+318.34}_{-1.37} \times 10^{-12}$ & ${1.03}^{+318.33}_{-1.00} \times 10^{-9}$ \\\hline
42 & \ion{Mo}{2} & ${4.40}^{+456.69}_{-1.00} \times 10^{-3}$ & ${1.54}^{+456.69}_{-1.37} \times 10^{-7}$ & ${8.80}^{+456.69}_{-1.00} \times 10^{-5}$ \\
& \ion{Mo}{3} & ${5.93}^{+456.69}_{-1.00} \times 10^{-9}$ & ${2.08}^{+456.69}_{-1.37} \times 10^{-13}$ & ${1.19}^{+456.69}_{-1.00} \times 10^{-10}$ \\\hline
43 & \ion{Tc}{2} & ${2.18}^{+456.69}_{-1.00} \times 10^{-4}$ & ${7.64}^{+456.69}_{-1.37} \times 10^{-9}$ & ${4.37}^{+456.69}_{-1.00} \times 10^{-6}$ \\
& \ion{Tc}{3} & ${2.50}^{+203.20}_{-1.00} \times 10^{-8}$ & ${8.74}^{+203.20}_{-1.37} \times 10^{-13}$ & ${4.99}^{+203.20}_{-1.00} \times 10^{-10}$ \\\hline
44 & \ion{Ru}{1} & ${5.54}^{+203.20}_{-1.00} \times 10^{-9}$ & ${1.94}^{+203.20}_{-1.37} \times 10^{-13}$ & ${1.11}^{+203.20}_{-1.00} \times 10^{-10}$ \\
& \ion{Ru}{2} & ${7.20}^{+203.20}_{-1.00} \times 10^{-3}$ & ${2.52}^{+203.20}_{-1.37} \times 10^{-7}$ & ${1.44}^{+203.20}_{-1.00} \times 10^{-4}$ \\
& \ion{Ru}{3} & ${5.07}^{+179.83}_{-1.00} \times 10^{-9}$ & ${1.77}^{+179.83}_{-1.37} \times 10^{-13}$ & ${1.01}^{+179.83}_{-1.00} \times 10^{-10}$ \\\hline
45 & \ion{Rh}{2} & ${5.23}^{+179.83}_{-1.00} \times 10^{-4}$ & ${1.83}^{+179.83}_{-1.37} \times 10^{-8}$ & ${1.05}^{+179.83}_{-1.00} \times 10^{-5}$ \\\hline
46 & \ion{Pd}{1} & ${2.55}^{+115.22}_{-1.00} \times 10^{-9}$ & ${8.91}^{+115.23}_{-1.37} \times 10^{-14}$ & ${5.09}^{+115.22}_{-1.00} \times 10^{-11}$ \\
& \ion{Pd}{2} & ${1.10}^{+115.22}_{-1.00} \times 10^{-3}$ & ${3.84}^{+115.23}_{-1.37} \times 10^{-8}$ & ${2.19}^{+115.22}_{-1.00} \times 10^{-5}$ \\\hline
47 & \ion{Ag}{2} & ${2.55}^{+115.22}_{-1.00} \times 10^{-4}$ & ${8.94}^{+115.23}_{-1.37} \times 10^{-9}$ & ${5.11}^{+115.22}_{-1.00} \times 10^{-6}$ \\\hline
48 & \ion{Cd}{1} & ${6.49}^{+94.70}_{-1.00} \times 10^{-9}$ & ${2.27}^{+94.71}_{-1.37} \times 10^{-13}$ & ${1.30}^{+94.70}_{-1.00} \times 10^{-10}$ \\
& \ion{Cd}{2} & ${2.27}^{+94.70}_{-1.00} \times 10^{-4}$ & ${7.95}^{+94.71}_{-1.37} \times 10^{-9}$ & ${4.54}^{+94.70}_{-1.00} \times 10^{-6}$ \\\hline
49 & \ion{In}{2} & ${6.99}^{+208.53}_{-1.00} \times 10^{-6}$ & ${2.45}^{+208.54}_{-1.37} \times 10^{-10}$ & ${1.40}^{+208.53}_{-1.00} \times 10^{-7}$ \\\hline
50 & \ion{Sn}{2} & ${6.25}^{+208.53}_{-1.00} \times 10^{-5}$ & ${2.19}^{+208.54}_{-1.37} \times 10^{-9}$ & ${1.25}^{+208.53}_{-1.00} \times 10^{-6}$ \\
& \ion{Sn}{3} & ${1.91}^{+208.53}_{-1.00} \times 10^{-8}$ & ${6.70}^{+208.54}_{-1.37} \times 10^{-13}$ & ${3.83}^{+208.53}_{-1.00} \times 10^{-10}$ \\\hline
51 & \ion{Sb}{2} & ${9.76}^{+38.01}_{-1.00} \times 10^{-6}$ & ${3.42}^{+38.03}_{-1.37} \times 10^{-10}$ & ${1.95}^{+38.01}_{-1.00} \times 10^{-7}$ \\\hline
52 & \ion{Te}{2} & ${7.71}^{+38.01}_{-1.00} \times 10^{-6}$ & ${2.70}^{+38.03}_{-1.37} \times 10^{-10}$ & ${1.54}^{+38.01}_{-1.00} \times 10^{-7}$ \\\hline
53 & \ion{I}{1} & ${1.07}^{+38.01}_{-1.00} \times 10^{-8}$ & ${3.74}^{+38.03}_{-1.37} \times 10^{-13}$ & ${2.14}^{+38.01}_{-1.00} \times 10^{-10}$ \\
& \ion{I}{2} & ${3.03}^{+9.27}_{-1.00} \times 10^{-6}$ & ${1.06}^{+9.35}_{-1.37} \times 10^{-10}$ & ${6.05}^{+9.27}_{-1.00} \times 10^{-8}$ \\\hline
54 & \ion{Xe}{1} & ${1.27}^{+9.27}_{-1.00} \times 10^{-9}$ & ${4.43}^{+9.35}_{-1.37} \times 10^{-14}$ & ${2.53}^{+9.27}_{-1.00} \times 10^{-11}$ \\
& \ion{Xe}{2} & ${7.68}^{+24.70}_{-1.00} \times 10^{-9}$ & ${2.69}^{+24.73}_{-1.37} \times 10^{-13}$ & ${1.54}^{+24.70}_{-1.00} \times 10^{-10}$
\enddata
\end{deluxetable*}

\startlongtable
\begin{deluxetable*}{cc|c|c|c}
\tablewidth{\textwidth}
\centering
\tablecaption{Same as Table~\ref{tab:ion_abunds_bluehot}, for the purple+warm best-fitting model. The lower bound on the total ejecta mass, which is used to compute the mass of each ion in the simulation, is $M_{\mathrm{ej}} > 4.0^{+2.9}_{-2.9} \times 10^{-5}~M_{\odot}$ for this model. We highlight ions of interest: \ion{Sr}{2}, \ion{Y}{2}, and \ion{Zr}{2}. }\label{tab:ion_abunds_purplewarm} 
\tablehead{\colhead{$Z$} & \colhead{ion} & \colhead{mass fraction $\mathrm{X}_i$} & \colhead{lower bound $M_{\mathrm{ej,i}}~[M_{\odot}]$} & \colhead{fiducial $M_{0.02,\mathrm{i}}$} } 
\startdata
1 & \ion{H}{1} & ${1.69}^{+1.93}_{-0.78} \times 10^{-4}$ & ${6.74}^{+2.06}_{-1.07} \times 10^{-9}$ & ${3.37}^{+1.93}_{-0.78} \times 10^{-6}$ \\
& \ion{H}{2} & ${8.27}^{+1.93}_{-0.78} \times 10^{-7}$ & ${3.31}^{+2.06}_{-1.07} \times 10^{-11}$ & ${1.65}^{+1.93}_{-0.78} \times 10^{-8}$ \\\hline
2 & \ion{He}{1} & ${6.79}^{+1.20}_{-0.50} \times 10^{-2}$ & ${2.72}^{+1.40}_{-0.88} \times 10^{-6}$ & ${1.36}^{+1.20}_{-0.50} \times 10^{-3}$ \\\hline
3 & \ion{Li}{2} & ${1.34}^{+0.90}_{-0.50} \times 10^{-7}$ & ${5.34}^{+1.16}_{-0.88} \times 10^{-12}$ & ${2.67}^{+0.90}_{-0.50} \times 10^{-9}$ \\\hline
4 & \ion{Be}{1} & ${8.10}^{+0.90}_{-0.50} \times 10^{-9}$ & ${3.24}^{+1.16}_{-0.88} \times 10^{-13}$ & ${1.62}^{+0.90}_{-0.50} \times 10^{-10}$ \\
& \ion{Be}{2} & ${5.56}^{+1.22}_{-0.58} \times 10^{-5}$ & ${2.23}^{+1.42}_{-0.93} \times 10^{-9}$ & ${1.11}^{+1.22}_{-0.58} \times 10^{-6}$ \\\hline
5 & \ion{B}{2} & ${5.45}^{+1.22}_{-0.58} \times 10^{-8}$ & ${2.18}^{+1.42}_{-0.93} \times 10^{-12}$ & ${1.09}^{+1.22}_{-0.58} \times 10^{-9}$ \\\hline
6 & \ion{C}{1} & ${3.06}^{+1.22}_{-0.58} \times 10^{-7}$ & ${1.22}^{+1.42}_{-0.93} \times 10^{-11}$ & ${6.11}^{+1.22}_{-0.58} \times 10^{-9}$ \\
& \ion{C}{2} & ${2.12}^{+1.39}_{-0.66} \times 10^{-6}$ & ${8.47}^{+1.57}_{-0.98} \times 10^{-11}$ & ${4.23}^{+1.39}_{-0.66} \times 10^{-8}$ \\\hline
7 & \ion{N}{1} & ${2.01}^{+1.39}_{-0.66} \times 10^{-8}$ & ${8.02}^{+1.57}_{-0.98} \times 10^{-13}$ & ${4.01}^{+1.39}_{-0.66} \times 10^{-10}$ \\\hline
8 & \ion{O}{1} & ${1.11}^{+4.42}_{-0.74} \times 10^{-6}$ & ${4.45}^{+4.48}_{-1.03} \times 10^{-11}$ & ${2.23}^{+4.42}_{-0.74} \times 10^{-8}$ \\
& \ion{O}{2} & ${4.70}^{+4.42}_{-0.74} \times 10^{-9}$ & ${1.88}^{+4.48}_{-1.03} \times 10^{-13}$ & ${9.40}^{+4.42}_{-0.74} \times 10^{-11}$ \\\hline
9 & \ion{F}{1} & ${1.20}^{+18.89}_{-0.89} \times 10^{-8}$ & ${4.78}^{+18.91}_{-1.15} \times 10^{-13}$ & ${2.39}^{+18.89}_{-0.89} \times 10^{-10}$ \\\hline
20 & \ion{Ca}{2} & ${6.75}^{+18.89}_{-0.89} \times 10^{-7}$ & ${2.70}^{+18.91}_{-1.15} \times 10^{-11}$ & ${1.35}^{+18.89}_{-0.89} \times 10^{-8}$ \\
& \ion{Ca}{3} & ${5.41}^{+6.35}_{-0.81} \times 10^{-7}$ & ${2.16}^{+6.39}_{-1.08} \times 10^{-11}$ & ${1.08}^{+6.35}_{-0.81} \times 10^{-8}$ \\\hline
21 & \ion{Sc}{2} & ${2.17}^{+6.35}_{-0.81} \times 10^{-8}$ & ${8.70}^{+6.39}_{-1.08} \times 10^{-13}$ & ${4.35}^{+6.35}_{-0.81} \times 10^{-10}$ \\\hline
22 & \ion{Ti}{2} & ${1.05}^{+17.53}_{-0.88} \times 10^{-5}$ & ${4.21}^{+17.54}_{-1.14} \times 10^{-10}$ & ${2.11}^{+17.53}_{-0.88} \times 10^{-7}$ \\
& \ion{Ti}{3} & ${4.27}^{+17.53}_{-0.88} \times 10^{-8}$ & ${1.71}^{+17.54}_{-1.14} \times 10^{-12}$ & ${8.53}^{+17.53}_{-0.88} \times 10^{-10}$ \\\hline
23 & \ion{V}{2} & ${1.28}^{+2.72}_{-0.92} \times 10^{-4}$ & ${5.14}^{+2.81}_{-1.17} \times 10^{-9}$ & ${2.57}^{+2.72}_{-0.92} \times 10^{-6}$ \\
& \ion{V}{3} & ${2.58}^{+2.72}_{-0.92} \times 10^{-8}$ & ${1.03}^{+2.81}_{-1.17} \times 10^{-12}$ & ${5.17}^{+2.72}_{-0.92} \times 10^{-10}$ \\\hline
24 & \ion{Cr}{1} & ${4.43}^{+2.72}_{-0.92} \times 10^{-8}$ & ${1.77}^{+2.81}_{-1.17} \times 10^{-12}$ & ${8.85}^{+2.72}_{-0.92} \times 10^{-10}$ \\
& \ion{Cr}{2} & ${2.33}^{+4.41}_{-0.94} \times 10^{-1}$ & ${9.31}^{+4.47}_{-1.19} \times 10^{-6}$ & ${4.65}^{+4.41}_{-0.94} \times 10^{-3}$ \\
& \ion{Cr}{3} & ${1.43}^{+4.41}_{-0.94} \times 10^{-6}$ & ${5.71}^{+4.47}_{-1.19} \times 10^{-11}$ & ${2.85}^{+4.41}_{-0.94} \times 10^{-8}$ \\\hline
25 & \ion{Mn}{1} & ${6.86}^{+4.41}_{-0.94} \times 10^{-9}$ & ${2.75}^{+4.47}_{-1.19} \times 10^{-13}$ & ${1.37}^{+4.41}_{-0.94} \times 10^{-10}$ \\
& \ion{Mn}{2} & ${7.89}^{+62.39}_{-0.99} \times 10^{-3}$ & ${3.16}^{+62.39}_{-1.23} \times 10^{-7}$ & ${1.58}^{+62.39}_{-0.99} \times 10^{-4}$ \\
& \ion{Mn}{3} & ${1.44}^{+62.39}_{-0.99} \times 10^{-7}$ & ${5.76}^{+62.39}_{-1.23} \times 10^{-12}$ & ${2.88}^{+62.39}_{-0.99} \times 10^{-9}$ \\\hline
26 & \ion{Fe}{1} & ${2.45}^{+62.39}_{-0.99} \times 10^{-7}$ & ${9.78}^{+62.39}_{-1.23} \times 10^{-12}$ & ${4.89}^{+62.39}_{-0.99} \times 10^{-9}$ \\
& \ion{Fe}{2} & ${9.32}^{+22.49}_{-0.76} \times 10^{-2}$ & ${3.73}^{+22.51}_{-1.05} \times 10^{-6}$ & ${1.86}^{+22.49}_{-0.76} \times 10^{-3}$ \\
& \ion{Fe}{3} & ${2.17}^{+22.49}_{-0.76} \times 10^{-7}$ & ${8.66}^{+22.51}_{-1.05} \times 10^{-12}$ & ${4.33}^{+22.49}_{-0.76} \times 10^{-9}$ \\\hline
27 & \ion{Co}{1} & ${1.12}^{+22.49}_{-0.76} \times 10^{-9}$ & ${4.48}^{+22.51}_{-1.05} \times 10^{-14}$ & ${2.24}^{+22.49}_{-0.76} \times 10^{-11}$ \\
& \ion{Co}{2} & ${1.77}^{+0.60}_{-0.56} \times 10^{-4}$ & ${7.07}^{+0.94}_{-0.92} \times 10^{-9}$ & ${3.54}^{+0.60}_{-0.56} \times 10^{-6}$ \\\hline
28 & \ion{Ni}{1} & ${7.32}^{+0.60}_{-0.56} \times 10^{-7}$ & ${2.93}^{+0.94}_{-0.92} \times 10^{-11}$ & ${1.46}^{+0.60}_{-0.56} \times 10^{-8}$ \\
& \ion{Ni}{2} & ${1.24}^{+0.60}_{-0.56} \times 10^{-1}$ & ${4.98}^{+0.94}_{-0.92} \times 10^{-6}$ & ${2.49}^{+0.60}_{-0.56} \times 10^{-3}$ \\
& \ion{Ni}{3} & ${2.40}^{+1.69}_{-0.58} \times 10^{-9}$ & ${9.59}^{+1.84}_{-0.93} \times 10^{-14}$ & ${4.79}^{+1.69}_{-0.58} \times 10^{-11}$ \\\hline
29 & \ion{Cu}{1} & ${8.23}^{+1.69}_{-0.58} \times 10^{-8}$ & ${3.29}^{+1.84}_{-0.93} \times 10^{-12}$ & ${1.65}^{+1.69}_{-0.58} \times 10^{-9}$ \\
& \ion{Cu}{2} & ${1.51}^{+1.69}_{-0.58} \times 10^{-2}$ & ${6.03}^{+1.84}_{-0.93} \times 10^{-7}$ & ${3.02}^{+1.69}_{-0.58} \times 10^{-4}$ \\\hline
30 & \ion{Zn}{1} & ${7.30}^{+0.53}_{-0.60} \times 10^{-6}$ & ${2.92}^{+0.90}_{-0.94} \times 10^{-10}$ & ${1.46}^{+0.53}_{-0.60} \times 10^{-7}$ \\
& \ion{Zn}{2} & ${4.06}^{+0.53}_{-0.60} \times 10^{-2}$ & ${1.62}^{+0.90}_{-0.94} \times 10^{-6}$ & ${8.12}^{+0.53}_{-0.60} \times 10^{-4}$ \\\hline
31 & \ion{Ga}{1} & ${1.16}^{+0.53}_{-0.60} \times 10^{-9}$ & ${4.62}^{+0.90}_{-0.94} \times 10^{-14}$ & ${2.31}^{+0.53}_{-0.60} \times 10^{-11}$ \\
& \ion{Ga}{2} & ${1.64}^{+0.43}_{-0.56} \times 10^{-2}$ & ${6.58}^{+0.84}_{-0.92} \times 10^{-7}$ & ${3.29}^{+0.43}_{-0.56} \times 10^{-4}$ \\\hline
32 & \ion{Ge}{1} & ${2.66}^{+0.43}_{-0.56} \times 10^{-7}$ & ${1.07}^{+0.84}_{-0.92} \times 10^{-11}$ & ${5.33}^{+0.43}_{-0.56} \times 10^{-9}$ \\
& \ion{Ge}{2} & ${3.88}^{+0.43}_{-0.56} \times 10^{-2}$ & ${1.55}^{+0.84}_{-0.92} \times 10^{-6}$ & ${7.76}^{+0.43}_{-0.56} \times 10^{-4}$ \\
& \ion{Ge}{3} & ${8.71}^{+0.67}_{-0.63} \times 10^{-8}$ & ${3.49}^{+0.98}_{-0.96} \times 10^{-12}$ & ${1.74}^{+0.67}_{-0.63} \times 10^{-9}$ \\\hline
33 & \ion{As}{1} & ${2.67}^{+0.67}_{-0.63} \times 10^{-5}$ & ${1.07}^{+0.98}_{-0.96} \times 10^{-9}$ & ${5.33}^{+0.67}_{-0.63} \times 10^{-7}$ \\
& \ion{As}{2} & ${5.45}^{+0.67}_{-0.63} \times 10^{-3}$ & ${2.18}^{+0.98}_{-0.96} \times 10^{-7}$ & ${1.09}^{+0.67}_{-0.63} \times 10^{-4}$ \\\hline
34 & \ion{Se}{1} & ${1.00}^{+0.80}_{-0.59} \times 10^{-4}$ & ${4.01}^{+1.08}_{-0.93} \times 10^{-9}$ & ${2.01}^{+0.80}_{-0.59} \times 10^{-6}$ \\
& \ion{Se}{2} & ${5.61}^{+0.80}_{-0.59} \times 10^{-2}$ & ${2.24}^{+1.08}_{-0.93} \times 10^{-6}$ & ${1.12}^{+0.80}_{-0.59} \times 10^{-3}$ \\\hline
35 & \ion{Br}{1} & ${1.14}^{+0.82}_{-0.65} \times 10^{-3}$ & ${4.57}^{+1.09}_{-0.97} \times 10^{-8}$ & ${2.28}^{+0.82}_{-0.65} \times 10^{-5}$ \\
& \ion{Br}{2} & ${2.89}^{+0.82}_{-0.65} \times 10^{-3}$ & ${1.16}^{+1.09}_{-0.97} \times 10^{-7}$ & ${5.78}^{+0.82}_{-0.65} \times 10^{-5}$ \\\hline
36 & \ion{Kr}{1} & ${6.12}^{+0.82}_{-0.65} \times 10^{-2}$ & ${2.45}^{+1.09}_{-0.97} \times 10^{-6}$ & ${1.22}^{+0.82}_{-0.65} \times 10^{-3}$ \\
& \ion{Kr}{2} & ${7.25}^{+0.84}_{-0.65} \times 10^{-4}$ & ${2.90}^{+1.11}_{-0.98} \times 10^{-8}$ & ${1.45}^{+0.84}_{-0.65} \times 10^{-5}$ \\\hline
37 & \ion{Rb}{2} & ${2.07}^{+0.84}_{-0.65} \times 10^{-2}$ & ${8.29}^{+1.11}_{-0.98} \times 10^{-7}$ & ${4.14}^{+0.84}_{-0.65} \times 10^{-4}$ \\\hline
38 & \textbf{\ion{Sr}{2}} & $\mathbf{{4.62}^{+0.88}_{-0.66} \times 10^{-3}}$ & $\mathbf{{1.85}^{+1.14}_{-0.98} \times 10^{-7}}$ & $\mathbf{{9.25}^{+0.88}_{-0.66} \times 10^{-5}}$ \\
& \ion{Sr}{3} & ${4.75}^{+0.88}_{-0.66} \times 10^{-2}$ & ${1.90}^{+1.14}_{-0.98} \times 10^{-6}$ & ${9.51}^{+0.88}_{-0.66} \times 10^{-4}$ \\\hline
39 & \textbf{\ion{Y}{2}} & $\mathbf{{5.88}^{+0.88}_{-0.66} \times 10^{-3}}$ & $\mathbf{{2.35}^{+1.14}_{-0.98} \times 10^{-7}}$ & $\mathbf{{1.18}^{+0.88}_{-0.66} \times 10^{-4}}$ \\
& \ion{Y}{3} & ${2.31}^{+0.89}_{-0.66} \times 10^{-3}$ & ${9.23}^{+1.15}_{-0.98} \times 10^{-8}$ & ${4.62}^{+0.89}_{-0.66} \times 10^{-5}$ \\\hline
40 & \ion{Zr}{1} & ${4.06}^{+0.89}_{-0.66} \times 10^{-9}$ & ${1.63}^{+1.15}_{-0.98} \times 10^{-13}$ & ${8.13}^{+0.89}_{-0.66} \times 10^{-11}$ \\
& \textbf{\ion{Zr}{2}} & $\mathbf{{6.13}^{+0.89}_{-0.66} \times 10^{-2}}$ & $\mathbf{{2.45}^{+1.15}_{-0.98} \times 10^{-6}}$ & $\mathbf{{1.23}^{+0.89}_{-0.66} \times 10^{-3}}$ \\
& \ion{Zr}{3} & ${1.12}^{+0.41}_{-0.30} \times 10^{-3}$ & ${4.49}^{+0.83}_{-0.79} \times 10^{-8}$ & ${2.25}^{+0.41}_{-0.30} \times 10^{-5}$ \\\hline
41 & \ion{Nb}{2} & ${2.15}^{+0.41}_{-0.30} \times 10^{-4}$ & ${8.61}^{+0.83}_{-0.79} \times 10^{-9}$ & ${4.30}^{+0.41}_{-0.30} \times 10^{-6}$ \\
& \ion{Nb}{3} & ${7.62}^{+1.01}_{-0.15} \times 10^{-8}$ & ${3.05}^{+1.24}_{-0.74} \times 10^{-12}$ & ${1.52}^{+1.01}_{-0.15} \times 10^{-9}$ \\\hline
42 & \ion{Mo}{1} & ${7.13}^{+1.01}_{-0.15} \times 10^{-9}$ & ${2.85}^{+1.24}_{-0.74} \times 10^{-13}$ & ${1.43}^{+1.01}_{-0.15} \times 10^{-10}$ \\
& \ion{Mo}{2} & ${1.71}^{+0.59}_{-0.35} \times 10^{-2}$ & ${6.84}^{+0.93}_{-0.80} \times 10^{-7}$ & ${3.42}^{+0.59}_{-0.35} \times 10^{-4}$ \\
& \ion{Mo}{3} & ${1.18}^{+0.59}_{-0.35} \times 10^{-8}$ & ${4.73}^{+0.93}_{-0.80} \times 10^{-13}$ & ${2.37}^{+0.59}_{-0.35} \times 10^{-10}$ \\\hline
43 & \ion{Tc}{2} & ${8.75}^{+0.87}_{-0.38} \times 10^{-4}$ & ${3.50}^{+1.14}_{-0.82} \times 10^{-8}$ & ${1.75}^{+0.87}_{-0.38} \times 10^{-5}$ \\
& \ion{Tc}{3} & ${5.15}^{+0.96}_{-0.41} \times 10^{-8}$ & ${2.06}^{+1.20}_{-0.84} \times 10^{-12}$ & ${1.03}^{+0.96}_{-0.41} \times 10^{-9}$ \\\hline
44 & \ion{Ru}{1} & ${6.01}^{+0.96}_{-0.41} \times 10^{-8}$ & ${2.40}^{+1.20}_{-0.84} \times 10^{-12}$ & ${1.20}^{+0.96}_{-0.41} \times 10^{-9}$ \\
& \ion{Ru}{2} & ${4.13}^{+1.20}_{-0.37} \times 10^{-2}$ & ${1.65}^{+1.41}_{-0.82} \times 10^{-6}$ & ${8.26}^{+1.20}_{-0.37} \times 10^{-4}$ \\
& \ion{Ru}{3} & ${1.49}^{+1.20}_{-0.37} \times 10^{-8}$ & ${5.95}^{+1.41}_{-0.82} \times 10^{-13}$ & ${2.98}^{+1.20}_{-0.37} \times 10^{-10}$ \\\hline
45 & \ion{Rh}{1} & ${8.74}^{+1.20}_{-0.37} \times 10^{-9}$ & ${3.50}^{+1.41}_{-0.82} \times 10^{-13}$ & ${1.75}^{+1.20}_{-0.37} \times 10^{-10}$ \\
& \ion{Rh}{2} & ${4.86}^{+0.79}_{-0.37} \times 10^{-3}$ & ${1.95}^{+1.07}_{-0.82} \times 10^{-7}$ & ${9.73}^{+0.79}_{-0.37} \times 10^{-5}$ \\\hline
46 & \ion{Pd}{1} & ${5.19}^{+0.79}_{-0.37} \times 10^{-8}$ & ${2.08}^{+1.07}_{-0.82} \times 10^{-12}$ & ${1.04}^{+0.79}_{-0.37} \times 10^{-9}$ \\
& \ion{Pd}{2} & ${1.18}^{+0.79}_{-0.37} \times 10^{-2}$ & ${4.72}^{+1.07}_{-0.82} \times 10^{-7}$ & ${2.36}^{+0.79}_{-0.37} \times 10^{-4}$ \\\hline
47 & \ion{Ag}{1} & ${9.45}^{+0.96}_{-0.46} \times 10^{-9}$ & ${3.78}^{+1.21}_{-0.86} \times 10^{-13}$ & ${1.89}^{+0.96}_{-0.46} \times 10^{-10}$ \\
& \ion{Ag}{2} & ${3.00}^{+0.96}_{-0.46} \times 10^{-3}$ & ${1.20}^{+1.21}_{-0.86} \times 10^{-7}$ & ${6.00}^{+0.96}_{-0.46} \times 10^{-5}$ \\\hline
48 & \ion{Cd}{1} & ${2.43}^{+0.96}_{-0.46} \times 10^{-7}$ & ${9.72}^{+1.21}_{-0.86} \times 10^{-12}$ & ${4.86}^{+0.96}_{-0.46} \times 10^{-9}$ \\
& \ion{Cd}{2} & ${4.47}^{+1.07}_{-0.55} \times 10^{-3}$ & ${1.79}^{+1.29}_{-0.91} \times 10^{-7}$ & ${8.94}^{+1.07}_{-0.55} \times 10^{-5}$ \\
& \ion{Cd}{3} & ${1.12}^{+1.07}_{-0.55} \times 10^{-9}$ & ${4.47}^{+1.29}_{-0.91} \times 10^{-14}$ & ${2.24}^{+1.07}_{-0.55} \times 10^{-11}$ \\\hline
49 & \ion{In}{2} & ${2.22}^{+1.07}_{-0.55} \times 10^{-4}$ & ${8.87}^{+1.29}_{-0.91} \times 10^{-9}$ & ${4.43}^{+1.07}_{-0.55} \times 10^{-6}$ \\\hline
50 & \ion{Sn}{1} & ${6.40}^{+1.26}_{-0.58} \times 10^{-9}$ & ${2.56}^{+1.46}_{-0.93} \times 10^{-13}$ & ${1.28}^{+1.26}_{-0.58} \times 10^{-10}$ \\
& \ion{Sn}{2} & ${5.45}^{+1.26}_{-0.58} \times 10^{-3}$ & ${2.18}^{+1.46}_{-0.93} \times 10^{-7}$ & ${1.09}^{+1.26}_{-0.58} \times 10^{-4}$ \\
& \ion{Sn}{3} & ${8.61}^{+1.26}_{-0.58} \times 10^{-7}$ & ${3.44}^{+1.46}_{-0.93} \times 10^{-11}$ & ${1.72}^{+1.26}_{-0.58} \times 10^{-8}$ \\\hline
51 & \ion{Sb}{1} & ${1.67}^{+1.08}_{-0.59} \times 10^{-7}$ & ${6.66}^{+1.30}_{-0.94} \times 10^{-12}$ & ${3.33}^{+1.08}_{-0.59} \times 10^{-9}$ \\
& \ion{Sb}{2} & ${1.08}^{+1.08}_{-0.59} \times 10^{-3}$ & ${4.31}^{+1.30}_{-0.94} \times 10^{-8}$ & ${2.16}^{+1.08}_{-0.59} \times 10^{-5}$ \\
& \ion{Sb}{3} & ${2.51}^{+1.08}_{-0.59} \times 10^{-9}$ & ${1.00}^{+1.30}_{-0.94} \times 10^{-13}$ & ${5.01}^{+1.08}_{-0.59} \times 10^{-11}$ \\\hline
52 & \ion{Te}{1} & ${1.19}^{+1.13}_{-0.63} \times 10^{-7}$ & ${4.77}^{+1.34}_{-0.96} \times 10^{-12}$ & ${2.38}^{+1.13}_{-0.63} \times 10^{-9}$ \\
& \ion{Te}{2} & ${7.22}^{+1.13}_{-0.63} \times 10^{-4}$ & ${2.89}^{+1.34}_{-0.96} \times 10^{-8}$ & ${1.44}^{+1.13}_{-0.63} \times 10^{-5}$ \\\hline
53 & \ion{I}{1} & ${1.88}^{+1.13}_{-0.63} \times 10^{-6}$ & ${7.51}^{+1.34}_{-0.96} \times 10^{-11}$ & ${3.76}^{+1.13}_{-0.63} \times 10^{-8}$ \\
& \ion{I}{2} & ${2.78}^{+1.62}_{-0.69} \times 10^{-4}$ & ${1.11}^{+1.77}_{-1.00} \times 10^{-8}$ & ${5.56}^{+1.62}_{-0.69} \times 10^{-6}$ \\\hline
54 & \ion{Xe}{1} & ${6.01}^{+1.62}_{-0.69} \times 10^{-7}$ & ${2.40}^{+1.77}_{-1.00} \times 10^{-11}$ & ${1.20}^{+1.62}_{-0.69} \times 10^{-8}$ \\
& \ion{Xe}{2} & ${1.89}^{+1.65}_{-0.68} \times 10^{-6}$ & ${7.57}^{+1.80}_{-0.99} \times 10^{-11}$ & ${3.79}^{+1.65}_{-0.68} \times 10^{-8}$ \\\hline
55 & \ion{Cs}{2} & ${2.87}^{+1.65}_{-0.68} \times 10^{-7}$ & ${1.15}^{+1.80}_{-0.99} \times 10^{-11}$ & ${5.74}^{+1.65}_{-0.68} \times 10^{-9}$ \\\hline
56 & \ion{Ba}{3} & ${4.16}^{+1.97}_{-0.78} \times 10^{-8}$ & ${1.66}^{+2.10}_{-1.06} \times 10^{-12}$ & ${8.32}^{+1.97}_{-0.78} \times 10^{-10}$ \\\hline
57 & \ion{La}{2} & ${6.19}^{+1.97}_{-0.78} \times 10^{-9}$ & ${2.47}^{+2.10}_{-1.06} \times 10^{-13}$ & ${1.24}^{+1.97}_{-0.78} \times 10^{-10}$ \\
& \ion{La}{3} & ${2.43}^{+1.97}_{-0.78} \times 10^{-8}$ & ${9.74}^{+2.10}_{-1.06} \times 10^{-13}$ & ${4.87}^{+1.97}_{-0.78} \times 10^{-10}$ \\\hline
58 & \ion{Ce}{3} & ${3.24}^{+2.36}_{-0.80} \times 10^{-9}$ & ${1.30}^{+2.47}_{-1.08} \times 10^{-13}$ & ${6.48}^{+2.36}_{-0.80} \times 10^{-11}$ \\\hline
60 & \ion{Nd}{3} & ${5.09}^{+2.36}_{-0.80} \times 10^{-9}$ & ${2.04}^{+2.47}_{-1.08} \times 10^{-13}$ & ${1.02}^{+2.36}_{-0.80} \times 10^{-10}$ \\\hline
62 & \ion{Sm}{3} & ${1.37}^{+2.36}_{-0.80} \times 10^{-9}$ & ${5.46}^{+2.47}_{-1.08} \times 10^{-14}$ & ${2.73}^{+2.36}_{-0.80} \times 10^{-11}$
\enddata
\end{deluxetable*}

 \hfill 


\end{document}